\documentclass[aps,prd,showpacs,twocolumn,amsmath,10pt,superscriptaddress,floatfix,nofootinbib]{revtex4-1}

\usepackage{epsfig,amssymb,amsfonts,amsmath,mathtools,bm,color,xcolor,graphicx,braket,adjustbox,esint,upgreek}

\newcommand{\be}{\begin{equation}}
\newcommand{\ee}{\end{equation}}
\newcommand{\bea}{\begin{eqnarray}}
\newcommand{\eea}{\end{eqnarray}}
\newcommand{\beas}{\begin{eqnarray*}}
\newcommand{\eeas}{\end{eqnarray*}}
\newcommand{\ds}{\displaystyle}

\newcommand{\tr}{\mbox{Tr}}

\newcommand{\gev}{\mbox{GeV}}

\newcommand{\zb}{Z_b^{\pm}(10610)}
\newcommand{\zbp}{Z_b^{\pm}(10650)}

\def\vec#1{\boldsymbol{#1}}

\newcommand{\Hb}{\bar{H}}

\newcommand{\nn}{\nonumber} 
\newcommand{\lrbd}{\overset{\leftrightarrow}{\partial}}
\newcommand{\veep}{{\bm \epsilon}}
\newcommand{\vep}{{\bm p}}

\newcommand{\veq}{{\bm q}}

\newcommand{\ven}{{\bm n}}

\newcommand{\veA}{{\bm A}}

\newcommand{\vetau}{{\bm \tau}}

\newcommand{\eq}{Eq.~\eqref}

\newcommand{\fpi}{f_{\pi}}

\newcommand{\Xint}{\fint}

\graphicspath{{Figs.dir/}}

\begin{document}

\title{Spin partners $W_{bJ}$ from the line shapes of the $Z_b(10610)$ and $Z_b(10650)$}

\author{V. Baru}
\affiliation{Helmholtz-Institut f\"ur Strahlen- und Kernphysik and Bethe Center for Theoretical Physics, Universit\"at Bonn, D-53115 Bonn, Germany}
\affiliation{Institute for Theoretical and Experimental Physics, B. Cheremushkinskaya 25, 117218 Moscow, Russia}
\affiliation{P.N. Lebedev Physical Institute of the Russian Academy of Sciences, 119991, Leninskiy Prospect 53, Moscow, Russia}

\author{E. Epelbaum}
\affiliation{Ruhr University Bochum, Faculty of Physics and Astronomy,
 Institute for Theoretical Physics II, D-44780 Bochum, Germany}

\author{A. A. Filin}
\affiliation{Ruhr University Bochum, Faculty of Physics and Astronomy,
 Institute for Theoretical Physics II, D-44780 Bochum, Germany}

\author{C. Hanhart}
\affiliation{Forschungszentrum J\"ulich, Institute for Advanced Simulation, Institut f\"ur Kernphysik and
J\"ulich Center for Hadron Physics, D-52425 J\"ulich, Germany}

\author{A. V. Nefediev}
\affiliation{P.N. Lebedev Physical Institute of the Russian Academy of Sciences, 119991, Leninskiy Prospect 53, Moscow, Russia}
\affiliation{National Research Nuclear University MEPhI, 115409, Kashirskoe highway 31, Moscow, Russia}
\affiliation{Moscow Institute of Physics and Technology, 141700, Institutsky lane 9, Dolgoprudny, Moscow Region, Russia}

\author{Q. Wang}
\affiliation{Helmholtz-Institut f\"ur Strahlen- und Kernphysik and Bethe Center for Theoretical Physics, Universit\"at Bonn, D-53115 Bonn, Germany}

\begin{abstract}

In a recent paper Phys.Rev. D {\bf 98}, 074023 (2018), the most up-to-date experimental data for all measured production and decay channels of the bottomonium-like states $Z_b(10610)$ and 
$Z_b(10650)$ 
were analysed in a field-theoretical coupled-channel approach which respects analyticity and unitarity and incorporates both the pion exchange as well as a short-ranged potential nonperturbatively. 
All parameters of the interaction were fixed directly from data, and pole positions for both $Z_b$ states were determined. 
In this work we employ the same approach to predict in a parameter-free way the pole positions and the line shapes in the elastic and inelastic channels of the (still to be discovered) spin partners 
of the $Z_b$ states. They are conventionally referred to as $W_{bJ}$'s
with the quantum numbers $J^{PC}=J^{++}$ ($J=0,1,2$). 
It is demonstrated that the results of our most advanced pionful fit, which gives the best $\chi^2/{\rm d.o.f.}$ for the data in the $Z_b$ channels, are consistent with all $W_{bJ}$ states being 
above-threshold resonances which manifest themselves as well pronounced hump structures in the line shapes. On the contrary, in the pionless approach, all $W_{bJ}$'s are virtual states which can be 
seen as enhanced threshold cusps in the inelastic line shapes. Since the two above scenarios provide different imprints on the observables, the role of the 
one-pion exchange in the $B^{(*)}\bar{B}^{(*)}$ systems can be inferred from the once available experimental data directly.
\end{abstract}



\maketitle

\section{Introduction}
\label{sec:introduction}

Heavy-quark spin symmetry (HQSS) is an approximate symmetry of QCD. It states that in the limit of an infinite mass of the heavy quark its spin is conserved by the strong 
interactions. As one departs from this limit, corrections scale as $\Lambda_{\rm QCD}/M_Q$, where $\Lambda_{\rm QCD}\simeq 200$ MeV denotes the intrinsic mass scale of QCD 
and $M_Q$ is the mass of the heavy quark. Stated differently, in the heavy-quark limit hadronic interactions do not depend on the heavy quark spin orientation, and hadronic states can be 
classified by the quantum numbers of the light degrees of freedom. This implies that, for a state with a given heavy quark spin observed experimentally, its spin partner 
states with different heavy quark spins but the same light quark cloud should exist as well. 
This pattern of spin partners is well established amongst the states with the assumed $\bar{Q}Q$ structure below the open-flavour threshold. For example, in the $b$-sector, for each radial 
excitation number $n$, one identifies $\eta_b(nS)$, with the heavy
quark-antiquark spin 0, as a spin partner of the $\Upsilon(nS)$, with the heavy quark-antiquark spin 1, and $h_b(nP)$, with the heavy
quark-antiquark spin 0, as a spin partner of the $\chi_{bJ}(nP)$ ($J=0,1,2$), with the heavy quark-antiquark spin 1. 
Meanwhile, the present experimental situation does not allow one to
reliably identify such spin partners for the confirmed exotic states
in the $b$-sector above the open-flavour threshold, namely, for the $Z_b(10610)$ 
and $Z_b(10650)$ states. However, the high luminosity and high 
statistics Belle-II experiment which has just started to operate \cite{Kou:2018nap} should be able to provide us with new information on these 
exciting states. 

This work contains a set of model-independent predictions based on
HQSS for both pole parameters and line shapes of the charged $J^{PC}=J^{++}$ ($J=0,1,2$) 
molecular states in the spectrum of bottomonium.
The parameters of the formalism were fixed in an earlier analysis of the most recent experimental information available for the $1^{+-}$ bottomonium-like states $\zb$ and
$\zbp$ (for simplicity often referred to as $Z_b$ and $Z_b'$, respectively). They were
observed by the Belle Collaboration as peaks
in the invariant mass distributions of the $\Upsilon(nS)\pi^\pm$ ($n=1,2,3$) and $h_b(mP)\pi^\pm$ ($m=1,2$) subsystems in the dipion transitions from 
the vector bottomonium 
$\Upsilon(10860)$ \cite{Belle:2011aa} and later confirmed in the elastic 
$B^{(*)}\bar{B}^{*}$ channels \cite{Collaboration:2011gja,Adachi:2012cx,Garmash:2015rfd}. Being isovectors, the $\zb$ and $\zbp$ cannot be conventional $\bar{b}b$ mesons as their minimal 
possible quark 
content is four-quark while their proximity to the $B\bar{B}^*$ and $B^*\bar{B}^*$ thresholds, respectively, provides a strong support for their
molecular interpretation --- see, for example, recent review articles \cite{Guo:2017jvc,Bondar:2016hva}.
In particular, the interference pattern in their inelastic decay channels $\Upsilon(nS)\pi$ and $h_b(mP)\pi$ can be explained naturally in the framework of the molecular 
picture \cite{Bondar:2011ev}. 
For a competing tetraquark interpretation claimed to be also compatible with the data see 
Refs.~\cite{Ali:2011ug,Esposito:2014rxa,Maiani:2017kyi}. 

Since the mass of the $b$ quark is very large compared with the typical QCD scale, $M_b\gg\Lambda_{\rm QCD}$, the constraints from the heavy-quark spin 
symmetry should be very accurate for bottomonium-like systems, including the 
$\zb$ and $\zbp$.
The wave functions of the $Z_b$ states in the molecular picture can be written as \cite{Bondar:2011ev}
\begin{eqnarray}
\label{Zbmolecule}
|Z_b\rangle&=&-\frac{1}{\sqrt 2}\Bigl[(1^-_{b \bar b}\otimes 0^-_{q \bar q})_{S=1}+
(0^-_{b \bar b}\otimes 1^-_{q \bar q})_{S=1}\Bigr]\label{Zb1},\\ \label{Zb'molecule}
|Z_b'\rangle&=&\frac{1}{\sqrt{2}}\Bigl[(1^-_{b \bar b}\otimes 0^-_{q \bar
q})_{S=1}-(0^-_{b \bar b}\otimes 1^-_{q \bar q})_{S=1}\Bigr], \label{Zb2}
\end{eqnarray}
where $S^{P}_{q \bar q}$ denotes the wave function of the light $q\bar{q}$ pair with the total spin $S$ and parity $P$, and $S^{P}_{b \bar b}$ means the same for the $b\bar{b}$ pair. Based on this 
structure four additional isovector sibling 
states $W_{bJ}$ are predicted to exist 
\cite{Bondar:2011ev,Voloshin:2011qa,Mehen:2011yh,Bondar:2016hva,Baru:2017gwo} with the quantum numbers $J^{PC}=J^{++}$ ($J=0,1,2$) and with the 
wave functions
\begin{eqnarray}
|W_{b0}\rangle&=&\frac12\Bigl[\sqrt{3} (1_{b\bar{b}}\otimes 1_{q\bar{q}})_{S=0}
-(0_{b\bar{b}}\otimes 0_{q\bar{q}})_{S=0}\Bigr],\label{Wb0}\\
|W_{b0}'\rangle&=&\frac12\Bigl[(1_{b\bar{b}}\otimes 1_{q\bar{q}})_{S=0}+
\sqrt{3}(0_{b\bar{b}}\otimes 0_{q\bar{q}})_{S=0}\Bigr],\label{Wb0pr}\\[1mm]
|W_{b1}\rangle&=&(1_{b\bar{b}}\otimes 1_{q\bar{q}})_{S=1},\label{Wb1}\\[2mm]
|W_{b2}\rangle&=&(1_{b\bar{b}}\otimes 1_{q\bar{q}})_{S=2}.\label{Wb2}
\end{eqnarray}
Thus, using that the low-energy interaction and the transition potentials between the elastic and inelastic channels in the $Z_b$'s fixed from the existing experimental data
 can be uniquely translated into the $W_{bJ}$ sector using HQSS constraints,
 the theoretical description of the spin partner states $W_{bJ}$ appears to be rather straightforward. 
 
 In Ref.~\cite{Wang:2018jlv} an effective field theory (EFT) approach to the $Z_b(10610)$ and $Z_b(10650)$ was developed and employed in the 
 analysis of the experimental data on the line shapes of these states in the elastic and inelastic channels. The approach is formulated based on an 
 effective Lagrangian consistent with both chiral and heavy-quark
spin symmetry of QCD. The key features of the approach and the central findings of Ref.~\cite{Wang:2018jlv}
 can be summarised as follows (see Ref.~\cite{Epelbaum:2008ga} for a review of a similar chiral EFT
approach in few-nucleon systems):
\begin{itemize}
\item The EFT is constructed employing the so-called Weinberg counting~\cite{Weinberg:1990rz}, proposed originally
to treat few-nucleon systems. The potential is constructed to a given order in $Q/\Lambda_h$ (here $Q$
denotes the soft scales of the given problem and $\Lambda_h\approx 1$ GeV represents the hard scale of the chiral EFT)
and then resummed nonperturbatively employing the Lippmann-Schwinger equation.
Accordingly, at leading order the potential contains two momentum-independent, ${\cal O}(Q^0)$, contact interactions and the
 pion exchange.
 \item Simultaneously with the chiral EFT expansion, the potential is expanded around the spin symmetry limit. 
 At leading order in chiral EFT this calls for the inclusion of the $B^*$-$B$ 
 mass difference together with all interaction vertices constructed in line with HQSS.
 Since $\Lambda_{\rm QCD}/M_b\approx 0.04\ll 1$, subleading HQSS violating contributions, which would lead to additional terms in the potential,
 are expected to play a minor role. This expectation was confirmed in Ref.~\cite{Wang:2018jlv}, where it was shown that the data in the $1^{+-}$ 
 channel were essentially consistent with HQSS constraints imposed on the potential.
\item The binding momenta, the pion mass and the momentum scale generated by the splitting between 
the $B\bar B^*$ and $B^*\bar B^*$ thresholds ($\delta=m_*-m\approx 45$~MeV with $m$ ($m^*$) denoting the $B$ ($B^*$) meson mass),
\begin{equation}
p_{\rm typ}=\sqrt{m\, \delta} \simeq 500~\mbox{MeV},
\label{ptyp}
\end{equation}
 are treated as soft scales of the system, generically called $Q$ above.
\item In order to 
 remove the strong regulator dependence caused by the high-momentum contributions from the
$S$-wave-to-$D$-wave $B^{(*)}\bar B^{(*)} \to B^{(*)}\bar B^{(*)}$ transitions (in what follows, simply $S$-$D$ transitions) generated by the one-pion exchange (OPE)
a promotion of the ${\cal O}(Q^2)$ $S$-wave-to-$D$-wave counter term 
to leading order is required. 
Then, the fit to the data enforces that a large portion of the $S$-$D$ contribution from the OPE gets balanced by the $S$-$D$ contact interaction.
However, the residual effect from the OPE on the line shapes is visible and results in 
a quantitative improvement of the fits.
To check the convergence of the scheme the effect from the other contact interactions at the order ${\cal O}(Q^2)$, 
namely, from two $S$-wave-to-$S$-wave terms, was also studied. However, their 
effect on the line shapes was shown to be numerically small in line with the assumed power counting.
\item The effect of the inelastic channels $\Upsilon(nS)\pi$ ($n=1,2,3$) and $h_b(mP)\pi$ ($m=1,2$) 
is included by allowing them to couple to the $S$-wave $B^{(*)}$-meson
pairs. Following Refs.~\cite{Hanhart:2015cua,Guo:2016bjq}, transitions between inelastic channels
are omitted in the potential. As a result, the effective elastic potentials acquire imaginary parts 
driven by unitarity while the contributions to the real parts of the elastic potentials can be absorbed to the redefinition of the 
momentum-independent ${\cal O}(Q^0)$ contact interactions. The treatment of the inelastic channels used in this work is analogous to 
the construction of the annihilation potential in nucleon-antinucleon scattering --- see, for example, Ref.~\cite{Kang:2013uia}.
\item Extension of the approach to the SU(3) sector requires that all the other members of the lightest pseudoscalar Goldstone-boson octet are treated also as explicit degrees 
of freedom already at leading order. That is why the one-$\eta$ exchange (OEE) is also included as a part of the $B^{(*)}\bar B^{(*)} \to B^{(*)}\bar B^{(*)}$ effective potential. 
In the SU(2) sector, however, the effect from the explicit treatment of the $\eta$-meson appears to be negligible.
\item All low-energy constants (the two elastic couplings, the effective couplings to the inelastic channels, as well as the $S$-$S$ and $S$-$D$ contact interactions) 
are fixed from a combined fit to the experimental 
line shapes in the decays $\Upsilon(10860) \to B\bar B^*\pi$, $B^*\bar B^*\pi$, $h_b(1P)\pi\pi$, and $h_b(2P)\pi\pi$ which
proceed via the excitation of the $Z_b(10610)$ and $Z_b(10650)$ exotic states as well as from the total rates for the decays $\Upsilon(10860) \to \Upsilon(nS)\pi\pi$ $(n=1,2,3)$.
The line shapes in the $\Upsilon(10860) \to \Upsilon(nS)\pi\pi$ channels could not be included in the analysis so far since they require a proper treatment of the
two-pion final-state interaction.
\end{itemize}
In this work, we provide parameter-free predictions for the HQSS partner states of the $Z_b$ and $Z_b'$ molecules 
using the framework summarised above. 
In particular, we
exploit the fact 
that all the parameters of the elastic and inelastic potentials extracted from the experimental line shapes in the 
$J^{PC}=1^{+-}$ channel are the same in the partner channels up to spin symmetry violating corrections in the contact interactions that are expected to
be small. Thus we are able to
predict, for the first time, the line shapes and to extract the pole positions for the spin partner states $W_{bJ}$
with the quantum numbers $J^{++}$ ($J=0,1,2$). In particular, we discuss the impact of the one-pion exchange
on the observables.

It needs to be mentioned that the same formalism for the spin partner states of the $Z_b^{(\prime)}$'s was employed in Ref.~\cite{Baru:2017gwo}.
However, compared with that paper, this work marks progress in three important aspects: 
\begin{itemize}
\item[(i)] All parameters of the interaction are now fixed directly
 from a fit to the measured line shapes contrary to the earlier study
 where the masses of the $Z_b$ states obtained in different analyses were used as input.
 \item[(ii)] Relevant inelastic channels are included which makes it possible to predict the line shapes in the inelastic channels, in addition to the elastic ones.    
\item[(iii)] We investigate how the renormalisation programme works in the coupled-channel case. In particular, we are now in a position to study 
the effect of the $S$-$D$ counter term on the line shapes and the pole locations of the spin partner states. The need for this counter term is one of the conclusions of Ref.~\cite{Wang:2018jlv}.
\item[(iv)] The uncertainty of the EFT predictions for the pole positions, which comes from various sources, is estimated and discussed.
\end{itemize}

The paper is organised as follows. Section~\ref{sec:introduction} contains a brief introduction to the formalism employed. In Sect.~\ref{sec:Veff}, the effective potentials in the 
$Z_b$'s and $W_{bJ}$'s channels are discussed in detail. In Sect.~\ref{sec:rates}, the coupled-channel equations are provided and the expressions for the differential production rates in all elastic 
and inelastic channels are 
constructed. The resulting formulae are then employed in Sect.~\ref{sec:Zb_WbJ} to analyse the existing experimental data in the $Z_b$'s channel (in line with the results of 
Ref.~\cite{Wang:2018jlv}) and 
to predict the line shapes in the $W_{bJ}$'s channels with the quantum numbers $J^{++}$ ($J=0,1,2$). In addition, in Sect.~\ref{sec:poles},
the pole parameters (locations and residues) for the $W_{bJ}$ states are provided and an analysis of uncertainties is presented. We summarise in Sect.~\ref{sec:summary}. Appendix
\ref{app:Lag} contains the details of the NLO Lagrangian ${\cal O}(Q^2)$ used to build the suitable EFT, while Appendix \ref{app:projectors} provides the details of the partial wave projection operators applied to 
the effective potential.

\section{Effective potentials}
\label{sec:Veff}

\subsection{Some generalities and definitions}

The partial-wave-projected effective potential in the elastic channels used in our calculations reads
\bea
(V_{\rm eff})_{\alpha\beta} = (V^{\rm CT}_{\rm eff})_{\alpha\beta} +\left (V^{\pi}\right )_{\alpha\beta}+\left (V^{\eta}\right )_{\alpha\beta},
\label{pot2}
\eea
where $V^{\rm CT}_{\rm eff}$, $V^{\pi}$ and $V^{\eta}$ stand for the
effective contact interaction potential (composed of the elastic,
$V^{\rm CT}_{\rm NLO}$, and inelastic, $\delta V$, contributions, as discussed below), OPE, and OEE, 
respectively, and 
the indices $\alpha$ and $\beta$, which depend on the particle channel and quantum numbers ($J^{PC}$), are defined as
\begin{eqnarray}
&&1^{+-}:\hspace*{0.2cm}\alpha,\beta 
=\{ B\bar{B}^*({}^3S_1,-), B\bar{B}^*({}^3D_1,-),\nonumber\\ 
&&\hspace*{4.5cm}B^*\bar{B}^*({}^3S_1),B^*\bar{B}^*({}^3D_1) \} \nonumber\\
&&0^{++}:\hspace*{0.2cm}\alpha,\beta =\{B\bar{B}({}^1S_0),B^*\bar{B}^*({}^1S_0),B^*\bar{B}^*({}^5D_0) \}\nonumber\\[-3mm]
\label{basisvec}\\[-1mm]
&&1^{++}:\hspace*{0.2cm}\alpha,\beta =\{B\bar{B}^*({}^3S_1,+),B\bar{B}^*({}^3D_1,+),B^*\bar{B}^*({}^5D_1) \}\nonumber\\
&&2^{++}:\hspace*{0.2cm}\alpha,\beta=\{B\bar{B}({}^1D_2),B\bar{B}^*({}^3D_2),B^*\bar{B}^*({}^5S_2), \nonumber\\
&& \hspace*{2.5cm} B^*\bar{B}^*({}^1D_2),B^*\bar{B}^*({}^5D_2) ,
B^*\bar{B}^*({}^5G_2) \}.\nonumber
\end{eqnarray}
Here the individual partial waves are labelled as $^{2S+1}L_J$ with $S$, $L$, and $J$ denoting
the total spin, the angular momentum, and the total momentum of the two-meson system, respectively. Finally, the sign in the parentheses corresponds to 
the $B\bar{B}^*$ states with a given $C$-parity 
\be
\ket{B\bar B^*,\pm}=\frac{1}{\sqrt{2}}(\ket{B\bar B^*}\pm\ket{\bar B B^*}),
\label{Eq:Cpar}
\ee
with a universal definition of the $C$-parity transformation employed,
\begin{equation}
{\hat C}M=\bar{M},
\label{Cpardef}
\end{equation}
for any meson $M$.

\vspace{0.5cm}

\subsection{Contact interactions}

The ${\cal O}(Q^0)$ short-ranged elastic (open-bottom) interaction between the
states with given quantum numbers composed of the $B^{(*)}\bar{B}^{(*)}$ pairs is parameterised in terms 
of two contact terms $C_{10}$ and $C_{11}$ --- see Refs.~\cite{AlFiky:2005jd,Nieves:2012tt} and Lagrangian \eqref{Lag0} quoted in Appendix~\ref{app:Lag}. 
This appendix also contains the momentum-dependent order $\mathcal{O}(Q^2)$ contact interactions shown by Eq.~\eqref{Lag2} and originally derived in Ref.~\cite{Wang:2018jlv}.
Formally, the order $\mathcal{O}(Q^2)$ chiral Lagrangian contains also the contact terms which scale with the pion mass squared ($m_\pi^2$). However, as long as we work at a fixed light quark
mass those can be absorbed into the leading order counter terms.
Thus, to order $\mathcal{O}(Q^2)$ the contact potentials in the elastic channels for various quantum numbers relevant for this study read
\begin{widetext}
\begin{eqnarray}\label{vfull1^{+-}}
\hspace{-0.5cm}V_{\rm NLO}^{\rm CT}[1^{+-}]\left(p,p^\prime \right)=\left(\begin{array}{cccc}
\mathcal{C}_{d}+\mathcal{D}_d(p^2+p^{\prime 2}) \hspace{0.3cm} & \mathcal{D}_{SD}p^{\prime2} \hspace{0.3cm} & \mathcal{C}_{f}+\mathcal{D}_f(p^2+p^{\prime 2}) \hspace{0.3cm} & 
\mathcal{D}_{SD}p^{\prime2}\\
\mathcal{D}_{SD}p^{2} & 0 & \mathcal{D}_{SD}p^{2} & 0\\
\mathcal{C}_{f}+\mathcal{D}_f(p^2+p^{\prime 2}) \hspace{0.3cm} & \mathcal{D}_{SD}p^{\prime2} \hspace{0.3cm} & \mathcal{C}_{d}+\mathcal{D}_d(p^2+p^{\prime 2}) \hspace{0.2cm} & 
\mathcal{D}_{SD}p^{\prime2}\\
\mathcal{D}_{SD}p^{2} & 0 & \mathcal{D}_{SD}p^{2} & 0
\end{array}\right),
\end{eqnarray}
\bea\label{vfull0^{++}}
&&V_{\rm NLO}^{\rm CT}[0^{++}]\left(p,p^\prime \right)\nonumber\\
&&\hspace*{15mm}=\left(
\begin{array}{ccc}
\ds {\cal C}_d+ \frac{1}{2}{\cal C}_f+( {\cal D}_d+ \frac{1}{2}{\cal D}_f) \left(p^2+{p'}^2\right) & \ds\frac{1}{2} \sqrt{3} \left({\cal C}_f+{\cal D}_f \left(p^2+{p'}^2\right)\right) & -\sqrt{3} 
{\cal 
D}_{SD} {p'}^2 \\
\ds\frac{1}{2} \sqrt{3} \left({\cal C}_f+{\cal D}_f \left(p^2+{p'}^2\right)\right) &\ds {\cal C}_d+ \left(( {\cal D}_d-\frac{1}{2}{\cal D}_f) \left(p^2+{p'}^2\right)-\frac{1}{2}{\cal C}_f\right) & 
\ds -{\cal 
D}_{SD} {p'}^2 \\
-\sqrt{3} {\cal D}_{SD} p^2 & -{\cal D}_{SD} p^2 & 0 \\
\end{array}
\right),
\eea
\bea\label{vfull1^{++}}
V_{\rm NLO}^{\rm CT}[1^{++}]\left(p,p^\prime \right)=\left(
\begin{array}{ccc}
{\cal C}_d+{\cal C}_f+({\cal D}_d+{\cal D}_f) \left(p^2+{p'}^2\right) & -{\cal D}_{SD} {p'}^2 & -\sqrt{3} {\cal D}_{SD} {p'}^2 \\
-{\cal D}_{SD} p^2 & 0 & 0 \\
-\sqrt{3} {\cal D}_{SD} p^2 & 0 & 0 \\
\end{array}
\right),
\eea
\bea\label{vfull2^{++}}
&&V_{\rm NLO}^{\rm CT}[2^{++}]\left(p,p^\prime \right)\nonumber\\
&&\hspace*{15mm}=\left(
\begin{array}{cccccc}
0 & 0 & \ds-\sqrt{\frac{3}{5}} {\cal D}_{SD} p^2 & 0 & 0 & 0 \\
0 & 0 & \ds -\frac{3}{\sqrt{5}} {\cal D}_{SD} p^2 & 0 & 0 & 0 \\
\ds -\sqrt{\frac{3}{5}} {\cal D}_{SD} {p'}^2 & \ds -\frac{3}{\sqrt{5}}{\cal D}_{SD} {p'}^2 & {\cal C}_d+{\cal C}_f+({\cal D}_d+{\cal D}_f) \left(p^2+{p'}^2\right) & \ds -\frac{1}{\sqrt{5}}{\cal 
D}_{SD} {p'}^2 
& \ds\sqrt{\frac{7}{5}} {\cal D}_{SD} {p'}^2 & 0 \\
0 & 0 & \ds-\frac{1}{\sqrt{5}}{\cal D}_{SD} p^2 & 0 & 0 & 0 \\
0 & 0 & \ds\sqrt{\frac{7}{5}} {\cal D}_{SD} p^2 & 0 & 0 & 0 \\
0 & 0 & 0 & 0 & 0 & 0 \\
\end{array}
\right),
\eea
\end{widetext}
where each potential is given for the basis states defined by Eq.~\eqref{basisvec} and $p$ ($p'$) stands for the relative momentum of the initial (final) heavy meson pair.
 The potential given in Eq.~(\ref{vfull1^{+-}}) was derived and used already in Ref.~\cite{Wang:2018jlv}.

\subsection{Inelastic channels}
\label{sec:inelpot}

\begin{table*}[t!]
\caption{HQSS-constrained coefficients $\xi^\Upsilon_{i\alpha}$ and
 $\xi^\chi_{i\alpha}$ in front of the coupling constants in the
 multiplets $\Upsilon$ and $\chi$ calculated explicitly from the
 traces in Eq.~\eqref{Linel} --- see the vertices given in Eqs.~\eqref{via1} and \eqref{via2}. 
}
\label{tab:gYchi}
\begin{ruledtabular}
\begin{tabular}{cccccccc}
Multiplet& Channel & $B\bar{B}^*({}^3S_1,-)$ &$B^*\bar{B}^*({}^3S_1,-)$ &$B\bar{B}({}^1S_0)$ &$B^*\bar{B}^*({}^1S_0)$ &$B\bar{B}^*({}^3S_1,+)$ &$B^*\bar{B}^*({}^5S_2)$ \\
\hline
$\Upsilon$&$\pi \Upsilon_b$ & $1$ & $-1$ & --- & --- & --- & --- \\
$\Upsilon$&$\pi \eta_{b0}$ & --- & --- & $-1/\sqrt{2}$ & $\sqrt{3/2}$ & --- & --- \\
\hline
$\chi$&$\pi h_b$ & $1$ & $1$ & --- & --- & --- & --- \\
$\chi$&$\pi \chi_{b0}$ & --- & --- & --- & --- & $\sqrt{2/3}$ & --- \\
$\chi$&$\pi \chi_{b1}$ & --- & --- & $\sqrt{3/2}$ & $1/\sqrt{2}$ & $1/\sqrt{2}$ & $1/\sqrt{2}$ \\
$\chi$&$\pi \chi_{b2}$ & --- & --- & --- & --- & $\sqrt{5/6}$ & $\sqrt{3/2}$ \\
\end{tabular}
\end{ruledtabular}
\end{table*}

Since the interaction between the pion and heavy quarkonia is suppressed (see the discussion in Refs.~\cite{Hanhart:2015cua,Guo:2016bjq}), 
the direct transitions between the inelastic (hidden-bottom) channels can be safely neglected in the potential. 
Based on this assumption, in the approach 
employed in Refs.~\cite{Hanhart:2015cua,Guo:2016bjq,Wang:2018jlv}, the effect of the inelastic channels on observables is included through 
the transitions between the inelastic and elastic potentials only, while unitarity is preserved.
Furthermore, it is argued in Ref.~\cite{Wang:2018jlv} that all inelastic channels only couple to the $S$-wave elastic ones 
as their couplings to the $D$-wave elastic channels are suppressed by the factor $p_\text{typ}^2/m^2\ll 1$.
Transitions between the $S$-wave elastic and inelastic channels are described by the Lagrangian \cite{Mehen:2011yh}
\bea 
{\cal L}_{\rm HH}^{\rm inel}&=&\sum_{n=1,2,3}
\frac{1}{4}g_{\Upsilon(nS)}\tr[\Upsilon_n^\dagger H_a \bar H_b]u^0_{ab}\nonumber\\[-2mm]
\label{Linel}\\[-2mm]
&+&\sum_{m=1,2}
\frac14g_{\chi_b(mP)}\tr[{\chi^i_{m}}^{\!\!\! \dagger} H_a \sigma^j\bar{H}_b]\epsilon_{ijk}u^k_{ab}.\nonumber\\ 
\nonumber
\eea
Here the spin multiplets of the heavy-light mesons read
\be
H_a = P_a+ V^i_a \sigma^i,\quad\bar{H}_a=\bar{P}_a-\bar{V}^i_a\sigma^i,
\label{Hs}
\ee 
where $\sigma_i$ are the Pauli matrices, 
$P_a \,(\bar{P}_a)$ and $V^i_a \, (\bar{V}^i_a)$ are the pseudoscalar $B$ ($\bar{B}$) and vector $B^*$ ($\bar{B}^*$) mesons, respectively, with $a$ and $b$ for the isospin indices. 
The multiplets of the
heavy $(\bar{b}b)$ mesons are built as
\bea\label{eta}
&&\Upsilon_n=\sigma_i \Upsilon^i(nS) + \eta_b(nS),\label{Yn}\\
&&\chi^i_m=\sigma_\ell\Big(\chi_{b2}^{i\ell}(mP) + \frac{1}{\sqrt{2}}\epsilon^{i\ell n}\chi_{b1}^n(mP) \nonumber\\
&&\hspace*{20mm}+\frac{1}{\sqrt{3}}\delta^{i\ell}\chi_{b0}(mP)\Big)+h^i_b(mP),\label{chi}
\eea
and 
\be
u^\mu=-\frac{1}{f_\pi}\partial^\mu\Phi+\mathcal{O}(\Phi^3),
\label{Amu}
\ee
\be
\Phi=\left(\begin{array}{cc}
\pi^0+\sqrt{\frac 13}\eta & \sqrt2\pi^{+}\\
\sqrt2\pi^{-} & -\pi^0+\sqrt{\frac 13}\eta 
\end{array}\right),
\label{pieta}
\ee
with $f_{\pi}=92.4$ MeV being the pion decay constant \cite{Tanabashi:2018oca}.
Since the quarkonium states ($\Upsilon$,$\eta_b$) and ($h_b$,$\chi_b$)
form spin multiplets [see Eqs.~\eqref{eta} and \eqref{chi}], the coupling constants 
fixed in the analysis of the line shapes in the $1^{+-}$ channels~\cite{Wang:2018jlv} can be used to account for the inelastic transitions in the spin partner 
channels. 
As long as the direct inelastic transitions are neglected, the effect of the inelastic channels 
on the elastic ones can be included via an additional contribution, $\delta V$, to the effective contact elastic-to-elastic transition potential~\cite{Hanhart:2015cua},
\be
V^{\rm CT}_{\rm eff}=V^{\rm CT}_{\rm NLO}+\delta V, 
\label{eq:veffective} 
\ee
where
\begin{eqnarray} 
\delta V_{\alpha\beta}={\cal P}_{\alpha\beta}-\frac{i}{8\pi M}\sum_i 2m_{h_i} v_{i\alpha} v_{i\beta}\; p_i.
\label{Gindef}
\end{eqnarray}
Here ${\cal P}_{\alpha\beta}$ stands for the real part of $\delta V_{\alpha\beta}$, $m_{h_i}$ denotes the mass of the heavy $\bar{b}b$ meson in the $i$-th inelastic channel 
and $M$ is the total energy of the system. Further, $v_{i\alpha}(p_i,p)$ is the vertex function for the transitions between various heavy-meson states (\ref{basisvec})
(labelled by greek letters $\alpha$, $\beta$ and so on) and inelastic 
channels (labelled by latin letters $i$, $j$ and so on). The arguments $p_i$ and $p$ denote the on-shell momenta of the 
inelastic and elastic channels involved, respectively, measured in the rest frame of the system.
One finds for $i=\Upsilon(nS)$, $\eta_b(nS)$, with $n=1,2,3$, that
\begin{equation} 
 v^{\Upsilon}_{i\alpha}
=\xi^{\Upsilon}_{i\alpha}\, \frac{g_{\Upsilon(nS)}}{2\sqrt{2} \fpi} E_\pi(p_i) , 
\label{via1}
\end{equation} 
where $E_\pi(p_i)=\sqrt{m_\pi^2+p_i^2}$ denotes the pion energy for a
given inelastic momentum,
\begin{eqnarray} 
\label{eq:mominelastic}
p_i=\frac{1}{2M}\lambda^{1/2}(M^2,m_{h_i}^2,m_{\pi}^2),
\label{pi}
\end{eqnarray} 
with $\lambda(x,y,z)$ being the standard triangle function.

In contrast to Ref.~\cite{Wang:2018jlv}, we now keep the energy-dependence in the vertices $v^{\Upsilon}_{i\alpha}$ explicitly, as it comes out from the Lagrangian \eqref{Linel}. 
However, since the variation of the inelastic momenta with the energy is very minor near the elastic thresholds,
this correction does not affect the quality of the fits and merely results
in rescaling of the inelastic coupling constants compared with those used in Ref.~\cite{Wang:2018jlv}.
For $i=h_b(mP)$, $\chi_{bJ}(mP)$, with $m=1,2$ and $J=0,1,2$, the
expression for the vertex reads
\begin{equation} 
 v^{\chi}_{i\alpha}=\xi^{\chi}_{i\alpha} \frac{g_{\chi_b(mP)}}{2\sqrt{3}\fpi} p_i .
\label{via2}
\end{equation} 
The coefficients $\xi^{\Upsilon}_{i\alpha}$ and $\xi^{\chi}_{i\alpha}$, provided explicitly in Table~\ref{tab:gYchi}, are fixed by the HQSS and are straightforwardly calculated from the traces
appearing in Eq.~(\ref{Linel}). 
They do not depend on $n$ and $m$ although the individual coupling constants, in principle, do. We also note that the relative signs between various couplings are only 
relevant in 
the particle coupled channels (the channels with $J^{PC}= $ $1^{+-}$ and $0^{++}$ in Table~\ref{tab:gYchi} while for the channels $J^{PC}= $ $1^{++}$ and $2^{++}$ only the absolute 
values of $\xi$'s enter). 

The real parts induced by the inelastic channels, ${\cal P}_{\alpha\beta}$, are divergent and need to
be regularised. The scheme employed in Ref.~\cite{Wang:2018jlv} assumes that the whole real part of the inelastic contribution 
in the $J^{PC}=1^{+-}$ channel is absorbed into a redefinition of the LECs ${\cal C}_d$ and ${\cal C}_f$ --- see Eq.~(\ref{vfull1^{+-}}).
This is justified as the momentum dependence of ${\cal P}_{\alpha\beta}$ coming from remote inelastic channels is very weak
and, therefore, can be neglected. To proceed we need to ensure that, in the heavy quark limit, the same approach
works for the complete spin multiplet. Then we have
\bea
{\cal P}_{\alpha\beta}=\sum_n {\cal P}_{\alpha\beta}[\Upsilon_n]+\sum_m {\cal P}_{\alpha\beta}[\chi_m],
\label{PJPS}
\eea
with
\bea
 \label{PJPS_HQSS}
{\cal P}_{\alpha\beta}[\Upsilon_n]&=&
\frac{g_{\Upsilon(nS)}^2}{16 \fpi^2} I_{\Upsilon(nS)}\left[\sum_i\xi^{\Upsilon *}_{\alpha i}\xi^{\Upsilon}_{i\beta}\right],\nonumber \\[-2mm]
\\[-2mm]
{\cal P}_{\alpha\beta}[\chi_m]&=&\frac{g_{\chi_b(mP)}^2}{24 \fpi^2} I_{\chi_b(mP)}\left[\sum_i \xi^{\chi *}_{\alpha i}\xi^{\chi}_{i\beta}
\right],\nonumber
\eea
where
\bea
 I_{\Upsilon(nS)}=&&\Xint\frac{d^3 q}{(2\pi)^3}\frac{E_\pi^2(q)}{E_\pi(q) (M-E_\pi(q)-E_{\Upsilon_n})},\nonumber\\[-2mm]
\label{Is}\\[-2mm]
 I_{\chi_b(mP)}=&&\Xint\frac{d^3 q}{(2\pi)^3}\frac{q^2}{E_\pi(q) (M-E_\pi(q)-E_{\chi_m})} .\nonumber 
\eea
These principal value integrals are factored out of the brackets in Eq.~\eqref{PJPS_HQSS} since the masses of the members of the spin
multiplets coincide in the heavy quark limit. One finds by a direct evaluation using the coefficients from Table~\ref{tab:gYchi}
(no summation in $\alpha$ or $\beta$ is implied here):
\bea
&&\left[\sum_i \xi^\Upsilon_{i\alpha}\xi^{\Upsilon *}_{i\alpha}\right]_{1^{+-}}=\left[\sum_i \xi^\Upsilon_{i\beta}\xi^{\Upsilon *}_{i\beta}\right]_{1^{+-}}=1,\nonumber\\
&&\left[\sum_i \xi^\Upsilon_{i\alpha}\xi^{\Upsilon *}_{i\beta}\right]_{1^{+-}}=-\left[\sum_i \xi^\chi_{i\beta}\xi^{\chi *}_{i\alpha}\right]_{1^{+-}}=-1,\label{cz}\\
&&\left[\sum_i \xi^\chi_{i\alpha}\xi^{\chi *}_{i\alpha}\right]_{1^{+-}}=\left[\sum_i \xi^\chi_{i\beta}\xi^{\chi *}_{i\beta}\right]_{1^{+-}}=1,\nonumber
\eea
for $\alpha=B\bar{B}^*({}^3S_1,-)$ and
$\beta=B^*\bar{B}^*({}^3S_1,-)$; then
\bea
&&\left[\sum_i \xi^\Upsilon_{i\alpha}\xi^{\Upsilon *}_{i\alpha}\right]_{0^{++}}=\left[\sum_i \xi^\chi_{i\beta}\xi^{\chi *}_{i\beta}\right]_{0^{++}}=\frac12,\nonumber\\
&&\left[\sum_i \xi^\Upsilon_{i\alpha}\xi^{\Upsilon *}_{i\beta}\right]_{0^{++}}=-\left[\sum_i \xi^\chi_{i\beta}\xi^{\chi *}_{i\alpha}\right]_{0^{++}}=-\frac{\sqrt{3}}{2},\label{c0}\\
&&\left[\sum_i \xi^\Upsilon_{i\beta}\xi^{\Upsilon *}_{i\beta}\right]_{0^{++}}=\left[\sum_i \xi^\chi_{i\alpha}\xi^{\chi *}_{i\alpha}\right]_{0^{++}}=\frac32,\nonumber
\eea
for $\alpha=B\bar{B}({}^1S_0)$ and $\beta=B^*\bar{B}^*({}^1S_0)$;
and finally
\be
\left[\sum_i \xi^\chi_{i\alpha}\xi^{\chi *}_{i\alpha}\right]_{1^{++}}=\frac23+\frac12+\frac56=2, 
\label{c1}
\ee
where $\alpha=B\bar{B}^*({}^3S_1,+)$ and
\be
\left[\sum_i \xi^\chi_{i\alpha}\xi^{\chi *}_{i\alpha}\right]_{2^{++}} =0+\frac12+\frac32=2, 
\label{c2}
\ee
where $\alpha=B^*\bar{B}^*({}^5S_2)$. For the $1^{++}$ and $2^{++}$ channels, the individual contributions from the intermediate states $\pi \chi_{bJ}$ with $ J=0,1,2$, in order, are quoted 
explicitly on the right-hand side (r.h.s) of Eqs.~(\ref{c1}) and (\ref{c2}).
According to Eqs.~(\ref{eq:veffective}), (\ref{Gindef}), \eqref{PJPS_HQSS} and \eqref{cz} the real parts of the loops in the $1^{+-}$ channel can be absorbed into the bare LECs $\mathcal{C}_{d}$ and 
$\mathcal{C}_{f}$ 
entering the short-range interaction for the $Z_b$'s [see Eq.~(\ref{vfull1^{+-}})]
via 
\bea \nonumber
 \mathcal{C}_{d}&\to&\mathcal{C}_{d}+\\&&\nonumber
\frac{1}{16 \fpi^2} \left(\frac23\sum_m g_{\chi_b(mP)}^2I_{\chi_b(mP)}+
\sum_n g_{\Upsilon(nS)}^2I_{\Upsilon(nS)}\right),
\\ \nonumber
 \mathcal{C}_{f}&\to&\mathcal{C}_{f}+\\&&\nonumber
\frac{1}{16 \fpi^2}\left(\frac23\sum_m g_{\chi_b(mP)}^2I_{\chi_b(mP)}-
\sum_n g_{\Upsilon(nS)}^2I_{\Upsilon(nS)}\right).
\eea
It is also straightforward to see using Eqs.~\eqref{PJPS_HQSS} and
\eqref{c0}-\eqref{c2} that these redefinitions also hold for the
interactions in the spin partner channels $J^{++}$ ($J=0,1,2$). 
This procedure is correct up to the neglected terms that violate spin symmetry and contain the energy dependence of the integrals in Eq.~(\ref{Is}). 
Thus, in line with Ref.~\cite{Wang:2018jlv}, in what follows only the imaginary parts of the inelastic loops are retained in the effective contact interaction potential (\ref{eq:veffective}) for all 
spin partner states.

\subsection{Pion exchange}

\begin{figure}[t!]
\centerline{\includegraphics[width=0.2\textwidth]{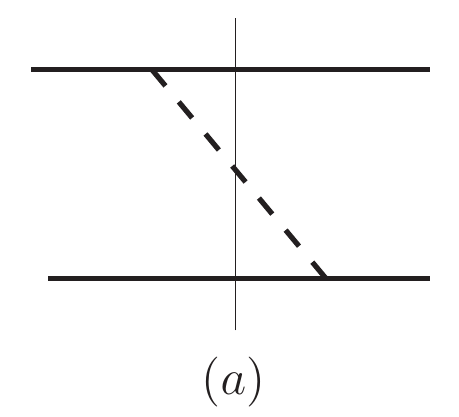}
\hspace*{5mm} 
\includegraphics[width=0.21\textwidth]{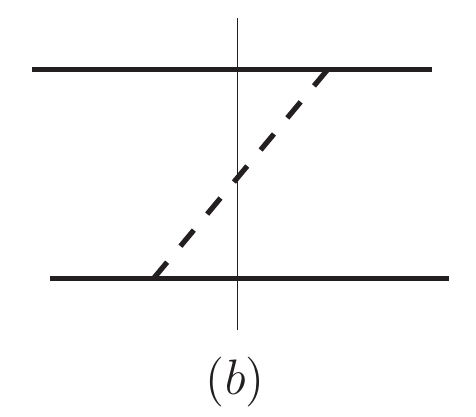}}
\caption{Diagrams in the time-ordered perturbation theory responsible for the two contributions to the OPE potential. The solid line is for the $B^{(*)}$ meson and the dashed line 
is for the pion. \label{fig:V1V2}}
\centerline{\epsfig{file=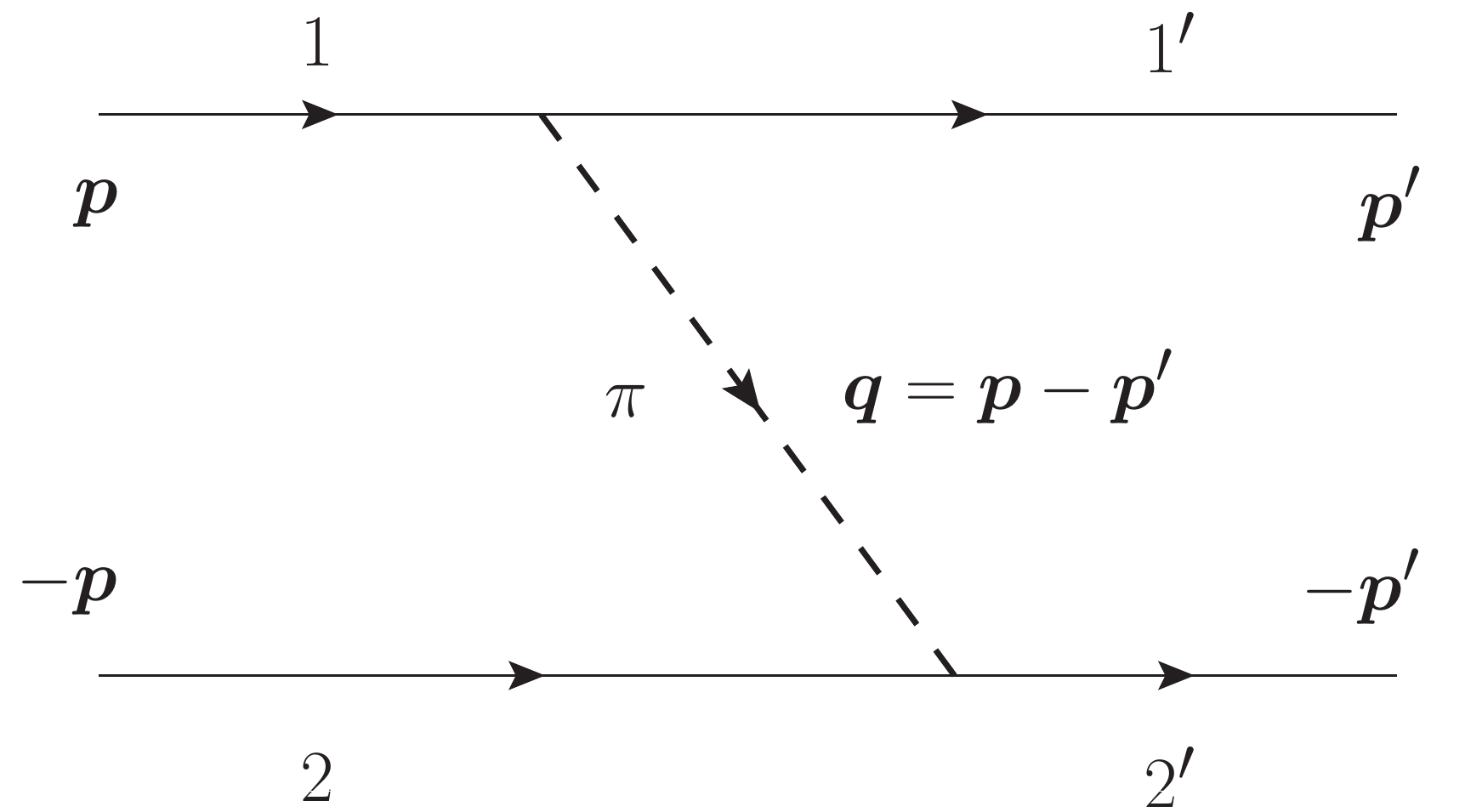, width=0.35\textwidth}}
\caption{Kinematics of the scattering due to the OPE as given by the first diagram in Fig.~\ref{fig:V1V2}.}\label{fig:diagram}
\end{figure}

The pion exchange in the $B^{(*)}B^*$ system is described by the Lagrangian \cite{Fleming:2007rp,Hu:2005gf}
\be
\mathcal{L}_{\Phi}=-\frac{g_{\scriptscriptstyle Q}}2\mbox{Tr}\left(\vec{\sigma}\cdot\vec{u}_{ab}H^\dagger_a H_b\right)+\mbox{h.c.},
\label{HHpi}
\ee
where { $u^\mu_{ab}$ is defined in Eq.~(\ref{Amu}) above.
The $B^{(*)}B^*\pi$ vertices extracted from the Lagrangian (\ref{HHpi}) take the form
\bea
v^a(B^*\to B\pi)&=&\frac{g_b}{2 f_\pi}\tau^a(\veep\cdot \veq),\label{vBstarBpi} \\
v^a(B^*\to B^*\pi)&=&-\frac{g_b}{\sqrt2 f_\pi}\tau^a(\veA \cdot\veq),
\label{vBstarBstarpi}
\eea
where $\veA=\frac{i}{\sqrt{2}}(\veep\times {\veep'}^*)$, then ${\veep}$, and ${\veep'}^*$ stand for the polarisation vectors of the initial and
final $B^*$ mesons, and $\veq$ is the pion momentum. These vertices 
agree with those used in Ref.~\cite{Nieves:2011vw}.

In order to determine the dimensionless coupling constant $g_b$ we rely on the heavy-quark flavour symmetry and set
\be
g_b=g_c\approx 0.57,
\label{gb}
\ee
where the numerical value of the $g_c$ is extracted from the most recent measurement of the $D^{*+}\to D^0\pi^+$ decay width,
\be
\varGamma(D^{*+}\to D^0\pi^+)=\frac{g_c^2m_{D^0} q^3}{12\pi f_\pi^2m_{D^{*c}}}.
\label{D*width2}
\ee
Here $m_{D^{*c}}$ and $m_{D^0}$ denote the masses of the $D^{*+}$ and $D^0$ mesons, 
respectively, and the final-state momentum $q=39$~MeV \cite{Tanabashi:2018oca}.
The value of $g_b$ quoted in Eq.~(\ref{gb}) agrees within 10\% with the result of a recent lattice QCD determination of the $B^*B\pi$ coupling constant \cite{Bernardoni:2014kla}. 

The isospin factor for the OPE potential is $\vetau_1\cdot\vetau_2^c=-\vetau_1\cdot\vetau_2=3-2I(I+1)$ that gives ($-1$) for the isotriplet states considered in this work. Here 
$\vetau^c=\tau_2\vetau^T\tau_2=-\vetau$ is the charge-conjugated Pauli matrix used for the antifundamental representation of the isospin group.

Finally, the overall sign of the OPE potential depends on the $C$-parity of the channel.
Using the definition of the $C$-parity given in Eq.~\eqref{Eq:Cpar} one finds
\be
V_\pm^{\rm OPE}=\braket{B\bar B^*,\pm|V^{\rm OPE}|B\bar B^*,\pm}=\pm V^{\rm OPE}, 
\ee
where
\be
V^{\rm OPE}\equiv \braket{B\bar B^*|V^{\rm OPE}|\bar B B^*}=\braket{\bar B B^*|V^{\rm OPE}|B\bar B^*}.
\ee
This additional sign from the $C$-parity for $C$-odd states is included in the integral equations explicitly, so that the potential is always defined without it.

We consider a coupled-channel system for the $B\bar{B}$, $B\bar{B}^*/\bar B{B}^*$ and $B^*\bar{B}^*$ channels.
Using the labels
\be
1\equiv B\bar{B},\quad 2\equiv B\bar{B}^*/\bar B{B}^*,\quad 3\equiv B^*\bar{B}^*,
\label{chan}
\ee
one can write for the OPE potentials 
\begin{widetext}
\bea
&&V_{11}^{\rm OPE}(\vep,\vep')=V_{12}^{\rm OPE}(\vep,\vep')=V_{21}^{\rm OPE}(\vep,\vep')=0,\label{V11}\\
&&V_{13}^{\rm OPE}(\vep,\vep')=\frac{2g_b^2}{(4\pi f_\pi)^2}({\veep_1'}^*\cdot \veq)({\veep_2'}^*\cdot \veq)\frac{2}{D_{B{B}^*\pi}(\vep,\vep')},\label{V13} \\
&&V_{22}^{\rm OPE}(\vep,\vep')=\frac{2g_b^2}{(4\pi f_\pi)^2}(\veep_1\cdot \veq)({\veep_2'}^*\cdot \veq) 
\left(\frac{1}{D_{BB\pi}(\vep,\vep')}+\frac{1}{D_{{B}^* {B}^*\pi}(\vep,\vep')}\right),\label{V22}\\
&&V_{23}^{\rm OPE}(\vep,\vep')=-\sqrt{2}\frac{2g_b^2}{(4\pi f_\pi)^2} \, (\veA_1 \cdot\veq)({\veep_2'}^* \cdot\veq) 
\left(\frac{1}{D_{B{B}^*\pi}(\vep,\vep')}+\frac{1}{D_{{B}^* {B}^*\pi}(\vep,\vep')}\right),\label{V23}\\
&&V_{33}^{\rm OPE}(\vep,\vep')=\frac{4g_b^2}{(4\pi f_\pi)^2} (\veA_1 \cdot\veq) (\veA_2 \cdot\veq)
\frac{2}{D_{{B}^* {B}^*\pi}(\vep,\vep')},\label{V33}
\eea
\end{widetext}
where the contributions from both time orderings as obtained in time-ordered perturbation theory (TOPT) are taken into account (see Figs.~\ref{fig:V1V2} and \ref{fig:diagram}). 
Further, $\veq=\vep-\vep'$, ${\veep_1}$ and ${\veep_2}$ (${\veep_1'}^*$ and 
${\veep_2'}^*$) stand for the polarisation vectors of the initial (final) $B^*$ mesons, $\veA_1=\frac{i}{\sqrt{2}}[\veep_1\times {\veep_1'}^*]$, $\veA_2=\frac{i}{\sqrt{2}}[\veep_2\times 
{\veep_2'}^*]$. The denominators $D_{B^{(*)}B^{(*)}\pi}(\vep,\vep')$ correspond to the $B^{(*)}B^{(*)}\pi$
propagators written in TOPT for the nonrelativistic $B$ and $B^*$ mesons,
\begin{widetext}
\bea
&&D_{B{B}^*\pi}(\vep,\vep')=D_{{B}^* B\pi}(\vep',\vep)=2E_{\pi}(\veq)\Bigl(m+m_{*}+\frac{\vep^2}{2m}+\frac{\vep'^2}{2m_{*}}+E_{\pi}(\veq)-M -i 0\Bigr),\label{V13D1NR1}\\
&&D_{BB\pi}(\vep,\vep')=2E_{\pi}(\veq)\Bigl(m+m+\frac{\vep^2}{2m}+\frac{\vep'^2}{2m}+E_{\pi}(\veq)-M-i 0\Bigr),\label{V22D1NR1}\\
&&D_{{B}^* {B}^*\pi}(\vep,\vep')=2E_{\pi}(\veq)\Bigl(m_{*}+m_{*}+\frac{\vep^2}{2m_{*}}+\frac{\vep'^2}{2m_{*}}+E_{\pi}(\veq)-M-i 0\Bigr) .
\label{V22D2NR1}
\eea
\end{widetext}
 The time-reversed transition potentials $V^{\rm OPE}_{31}(\vep,\vep')$ and $V^{\rm OPE}_{32}(\vep,\vep')$
are trivially obtained from Eqs.~\eqref{V13} and \eqref{V23} by interchanging the particle labels as $1\leftrightarrow 1', 2\leftrightarrow 2'$ 
(see also Fig.~\ref{fig:diagram}). 

Since our analysis covers the energy region between the $B\bar{B}$ and $B^*\bar{B}^*$ thresholds split by approximately $2\delta\approx 90$~MeV, 
which numerically appears to be of the order of the pion mass, a relativistic expression for the pion energy is used. 
It should be noted that, in the region of interest around the $B^{(*)}\bar{B}^{(*)}$ thresholds, $M<2m+m_\pi$, so that, unlike charmonium systems, one never hits the three-body cut in the OPE 
potentials defined above.

Since the pion is emitted by $B^{(*)}$ mesons in the $P$-wave --- see the Lagrangian (\ref{HHpi}) and the vertices (\ref{vBstarBpi}) and (\ref{vBstarBstarpi}) --- the OPE potential mixes $S$ and
$D$ waves. The 
partial wave projection of the OPE potentials (\ref{V13})-(\ref{V33}) can be done using the formalism of Refs.~\cite{Baru:2016iwj,Baru:2017gwo,Baru:2017pvh} that gives
\begin{widetext}
\bea
V^{\rm OPE}_{LL'}(p,p')=\frac1{2J+1}\int\frac{d \Omega_p}{4\pi}\frac{d \Omega_{p'}}{4\pi} {\rm Tr}\Bigl[P^\dagger(JLS;\bm n)V^{\rm OPE}(\bm p,\bm p') 
P(JL'S';\bm n')\Bigr],
\label{VPWA}
\eea
\end{widetext}
where $L=S$, $D$ and $G$, $\bm n=\bm p/p$ ($\bm n'=\bm p'/p'$), and a complete set of relevant properly normalised 
projection operators $P(JLS;\bm n)$ is given in Appendix~\ref{app:projectors}.

\section{Production rates of the $Z_b$'s and their spin partners $W_{bJ}$'s}
\label{sec:rates}

\subsection{Production vertex}
\label{sec:prodvert}

\begin{figure*}[t]
\epsfig{file=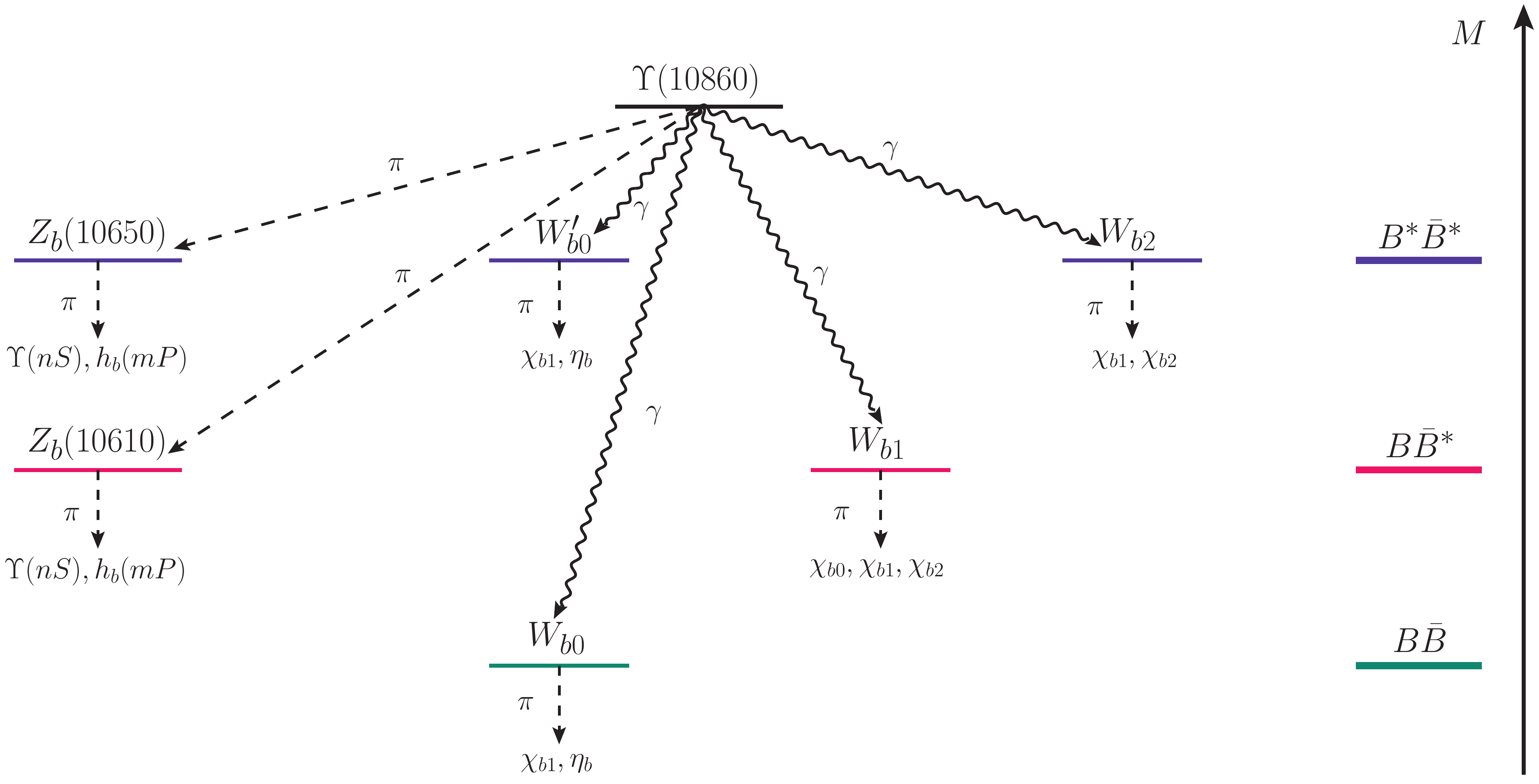, width=0.85\textwidth}
\caption{Summary of the production and decay channels for the 
$Z_b$'s and their spin partners $W_{bJ}$'s considered in this work. The states and thresholds
are arranged from bottom to top in accordance to the increasing energy.}
\label{fig:prod}
\end{figure*}

\begin{figure}[t!]
\hspace*{0.9cm}\centerline{\epsfig{file=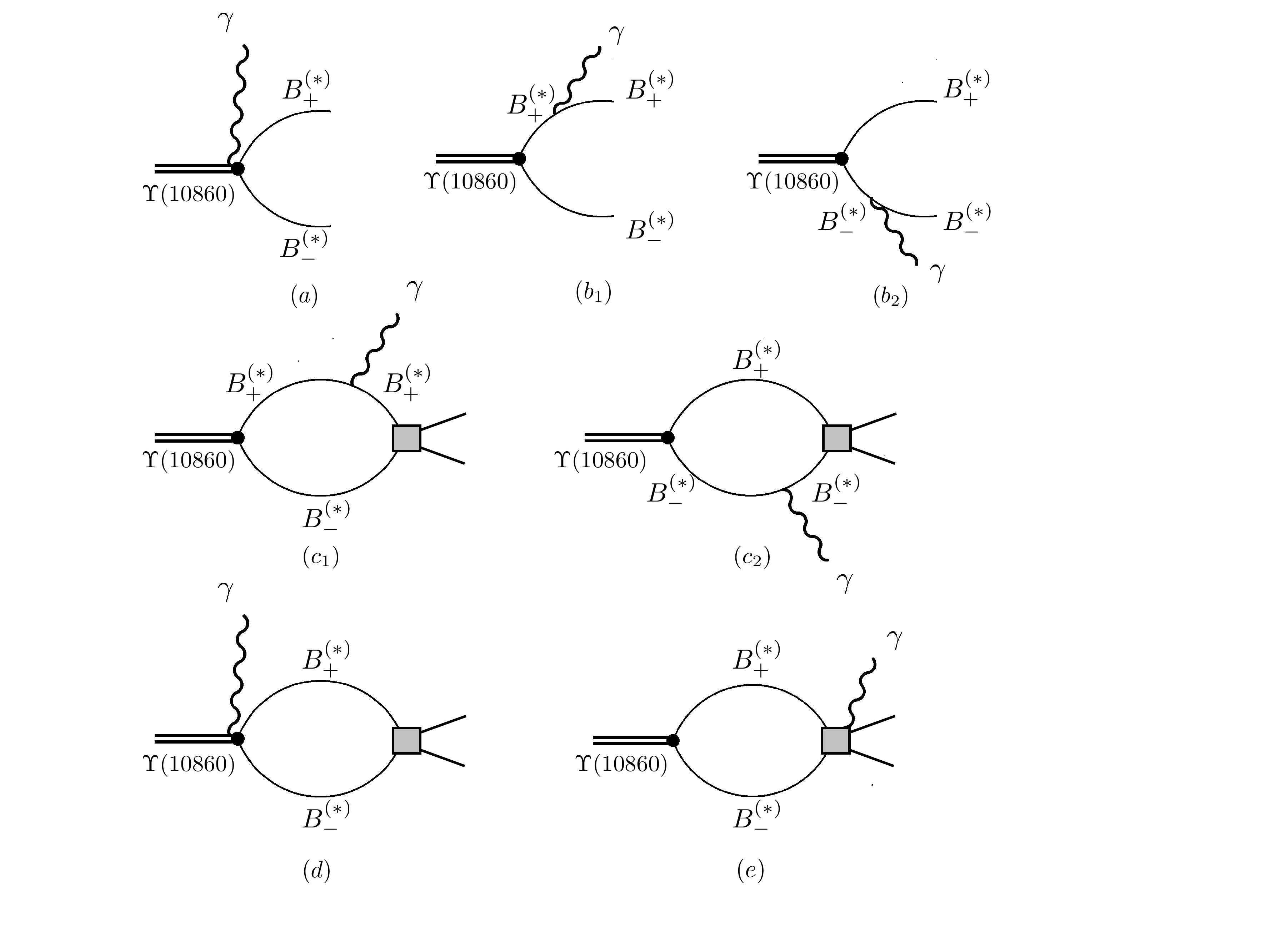, width=0.65\textwidth}}
\caption{Diagrams contributing to the $\Upsilon(10860)\to \gamma B^{(*)}\bar B^{(*)}$ decay amplitude: 
diagrams (a), (b1) and (b2) (in the first line) form a gauge invariant subset of tree level contributions, 
while diagrams (c1), (c2), (d) and (e) correspond to a gauge invariant subset of contributions at the one-loop level. 
The vertex in diagrams (a) and (d) comes from gauging the $\Upsilon(10860)\to B^{(*)}\bar B^{(*)}$ vertex; the photon vertices in (b1), (b2), (c1) and (c2)
are from gauging the kinetic terms of the heavy mesons. 
The diagram (d) is needed to account for gauging the regulator used in the loops and for a nonpointlike character of the amplitude in the final state. }
 \label{fig:source}
\end{figure}

The twin states $Z_b(10610)$ and $Z_b(10650)$ are produced in the one-pion decays of the $\Upsilon(10860)$ resonance, 
$$
\Upsilon(10860)\to\pi Z_b^{(\prime)}\to\mbox{final state}.
$$
Because of a different 
$C$-parity, the spin partner states $W_{bJ}$ should be produced in radiative decays of the $\Upsilon(10860)$, 
$$
\Upsilon(10860)\to\gamma W_{bJ}\to\mbox{final state}.
$$
Production and decay channels for the $Z_b$'s and $W_{bJ}$'s taken into account in our approach are summarised in Fig.~\ref{fig:prod}.
In line with the discussion in Sec.~\ref{sec:inelpot}, since the couplings of the $\Upsilon(10860)\gamma$ source term with the $D$-wave elastic channels are suppressed, 
we retain only the couplings of the $\Upsilon(10860)\gamma$ source term with the elastic 
channels in the $S$ wave.

A set of diagrams which contribute to the process $\Upsilon(10860)\to\gamma B^{(*)}B^{(*)}$ and provide a gauge invariant amplitude is shown in Fig.~\ref{fig:source}, where diagrams (a), (b1)
and (b2) contribute to the production operators at tree level while diagrams (c)-(e) represent contributions from the loops assuming that the intermediate particles are $B$ and $B^*$ mesons only. 
In this work, we do not aim at predicting the absolute rate of the decays $\Upsilon(10860)\to\gamma W_{bJ}$, which might involve some more sophisticated mechanisms (for example, as advocated in 
Ref.~\cite{Wu:2018xaa}),
but rather focus on the energy dependence of the line shapes with an arbitrary overall normalisation. To this end, we assume that like for the $Z_b$'s production
also for these processes the energy dependence from the production operator 
is rather smooth close to thresholds and can be merely neglected as compared with rapidly varying $B$-meson amplitudes in the final state. To advocate this approximation, below we discuss 
the diagrams shown in Fig.~\ref{fig:source}
in more detail.

To estimate the strength and the structure of the source term we start from 
the effective Lagrangian 
connecting the $\Upsilon(10860)$ bottomonium with the heavy meson fields at leading order
in the HQSS expansion~\cite{Mehen:2011yh},
\be 
{\cal L}_{\rm \Upsilon HH}=\frac12 g_{\Upsilon,5}\tr[\Upsilon_n^\dagger H_a \sigma^j i\lrbd_j \bar H_a]+\mbox{h.c.}
\label{HHprod}
\ee
Gauging this Lagrangian leads to a set of contact $\Upsilon(10860)\to\gamma B^{(*)}\bar{B}^{(*)}$ vertices which contribute to diagrams (a) and (d),
\be
v \left [ \Upsilon(10860)\to\gamma \left (B^{(*)}\bar{B}^{(*)}\right)^{(J^{++})} \right ]=\sum_{\alpha}v^{(J^{++})}_\alpha \hat{S}_\alpha^{(J^{++})},
\label{UpsWgam}
\ee
where, for a given $J$, index $\alpha$ runs over the relevant $S$-wave states, that is, $B\bar{B}({}^1S_0)$, $B^*\bar{B}^*({}^1S_0)$, $B\bar{B}^*({}^3S_1)$, $B^*\bar{B}^*({}^5S_2)$.
The spin operators, normalised according to $\sum_\lambda\left|\hat
 S^{(J^{++})}_\alpha\right|^2= 2 J_\Upsilon +1 = 3$ with $\lambda$
running over the polarisations of the spin-1 particles ($\gamma, 
\Upsilon$ and $B^*$), read
\bea\label{spin_B0++}
&&\hat{S}^{(0^{++})}_{B\bar{B}({}^1S_0)}=\sqrt{\frac32} (\vec\epsilon^{\gamma*}\cdot\vec \epsilon^{\Upsilon}) P^\dagger\Bigl(B\bar{B}({}^1S_0)\Bigr),\\ \label{spin_B*0++}
&&\hat{S}^{(0^{++})}_{B^*\bar{B}^*({}^1S_0)}=\sqrt{ \frac32}
 (\vec\epsilon^{\gamma*}\cdot\vec \epsilon^{\Upsilon}) P^\dagger\Bigl(B^*\bar{B}^*({}^1S_0)\Bigr), \\ \label{spin_1++}
&&\hat{S}^{(1^{++})}_{B\bar{B}^*({}^3S_1)}=-i\frac{\sqrt{3}}{2} \epsilon_{ijk} \epsilon^{\Upsilon}_i \epsilon^{\gamma*}_j \, P^\dagger\Bigl(B\bar{B}^*({}^3S_1)\Bigr)_k, \\ \label{spin_2++}
&&\hat{S}^{(2^{++})}_{B^*\bar{B}^*({}^5S_2)}=-\frac3{\sqrt{10}} \epsilon^{\Upsilon}_i \epsilon^{\gamma*}_j \, P^\dagger\Bigl(B^*\bar{B}^*({}^5S_2)\Bigr)_{ij}.
\eea
Here $\vec \epsilon^{\gamma*}$, $\vec \epsilon^{\Upsilon}$ and $\vec \epsilon_{1(2)}$ denote the polarisation vectors of the photon, $\Upsilon$ and $B^*$ mesons, respectively, 
and the explicit form of the projectors $P$ on relevant heavy meson states are given in Appendix~\ref{app:projectors}. 
Further, the partial-wave-projected vertices $v^{(J^{++})}_\alpha$ are defined as 
\be
\label{vert_vJ}
v^{(J^{++})}_\alpha=-\sqrt{\frac23}eg_{\Upsilon,5} \, \lambda^{(J^{++})}_\alpha,
\ee
where $e$ is the magnitude of the {electron charge} and the ratios of the coupling constants, $\lambda^{(J^{++})}_\alpha$, related by HQSS are quoted in Table~\ref{tab:ggamma}.
It is shown in Ref.~\cite{Mehen:2013mva} that experimental data might call for a significant amount of spin symmetry
violation in the transition $\Upsilon(10860) \to B^{(*)}\bar{B}^{(*)}$ (there is a tension of 2$\sigma$ between the
spin symmetric ratio of decay widths and the experimental data). Since the same couplings also contribute to the transitions $\Upsilon(10860) \to \gamma B^{(*)}\bar{B}^{(*)}$, HQSS violation is a 
potential additional source of uncertainty 
for our results --- we come back to this issue in the discussion below.

In addition to the contact diagram with the photon emission from the $\Upsilon(10860)\to\gamma B^{(*)}\bar{B}^{(*)}$ vertex, diagrams with the photon emission from the $B^{(*)}$-meson lines should be 
considered with 
the $B^{(*)}\to B^{(*)}\gamma$ vertices being of an electric or magnetic type. While the amplitudes with the magnetic photon emission vertices are gauge invariant by themselves, 
the additional amplitudes with the electric photon emission vertices are important to compensate for the gauge dependence of the contact 
$W_{bJ}\to\gamma B^{(*)}\bar{B}^{(*)}$ diagram and thus to provide an overall gauge invariance of the full amplitude. 
To estimate the electric contributions, we notice that in the nonrelativistic heavy meson formalism used here 
all momenta involved are 3-momenta and the photon momentum $\vec k$ fulfils the relation $\vec \epsilon^{\gamma} \cdot \vec k$=0. 
Then, one readily arrives at the following estimates for the tree-level diagrams (b1) and (b2) with the 
electric (thence superscript $e$) 
photon emission from the external $B^{(*)}$-meson lines relative to the contact $W_{bJ}\to\gamma B^{(*)}\bar{B}^{(*)}$ diagram,
\be
\frac{{\cal M}^{e}}{{\cal M}_{\rm cont}}\simeq p_\alpha\cdot\frac{1}{m\upomega}\cdot p_\alpha\simeq \frac{E_\alpha}{\upomega}\ll 1,
\label{estimate}
\ee
where one power of the relative $B^{(*)}$-meson momentum $ p_\alpha$ comes from the $\Upsilon(10860)\to B^{(*)}\bar{B}^{(*)}$ vertex extracted from Eq.~(\ref{HHprod}), the term 
$1/(m\upomega)$ is the $B^{(*)}$-meson propagator with $\upomega$ for the photon energy (here we do not distinguish between the $B$ and $B^*$ mass), and the second factor $ p_\alpha$ comes from the 
electric
photon emission vertex from the $B^{(*)}$ meson, which is derived by gauging the kinetic term in the Lagrangian \eqref{Lag0}. 
Further, to arrive at the very last relation in Eq.~(\ref{estimate}) 
we used that the energy of the $B$-meson pair relative to the threshold in the channel $\alpha$, $E_\alpha\approx p_\alpha^2/m$, 
does not exceed several dozen MeV while the photon energy $\upomega$ is an order of magnitude larger, $\upomega \approx M_{\Upsilon(10860)} - 2m \approx 200$-300 MeV.

One is, therefore, led to conclude that although the diagrams (b1) and (b2) with the electric photon emission from the $B^{(*)}$-meson lines are important to guarantee gauge invariance of 
the amplitude, in practical calculations they provide only small corrections 
and that the tree-level amplitude behaves basically as a constant in the energy region of relevance. 

The loop contributions from the diagrams (c)-(e) were already studied in the literature in the context of scalar mesons made of light quarks --- see, for example, 
Refs.~\cite{Close:1992ay,Kalashnikova:2004ta}. 
In particular, it is shown that for pseudoscalar mesons such loops form a gauge invariant subset of diagrams which yields a finite contribution to the amplitude. The arguments of 
Refs.~\cite{Close:1992ay,Kalashnikova:2004ta} can be generalised 
to find that, in the HQSS limit, these conclusions hold also for all members of the heavy meson spin multiplet and for all quantum numbers $J^{++}$.
Further, using the explicit results of Refs.~\cite{Close:1992ay,Kalashnikova:2004ta} for the loops with a point-like interaction between the mesons in the final state one concludes 
that, to a good approximation, also for diagrams (c)-(e) the production operator can be treated as a constant. 
 
To illustrate the argument, consider the resulting contribution from the diagrams shown in Fig.~\ref{fig:source} for the uncoupled case, 
\be
\hspace*{-0.1cm}{\cal M}_\alpha (p_\alpha) = v^{(J^{++})}_\alpha \hat{S}_\alpha^{(J^{++})} \biggl ( 1+ \bigl({\cal A}^\alpha(p_\alpha)+ i p_\alpha \bigr ) f_{\rm on}^\alpha(p_\alpha) \biggr ),
\ee
where the vertex and the spin structure in front of the parenthesis are from Eqs.~\eqref{spin_1++}-\eqref{vert_vJ}, ${\cal A}^\alpha(p_\alpha)$
and $ip_\alpha$ denote the real and imaginary parts of the pertinent loop, $ f_{\rm on}^\alpha(p_\alpha)$ is the on-shell $B$-meson
amplitude in the final state. Unitarity forces
$ f_{\rm on}^\alpha$ to have the form
\be
f_{\rm on}^\alpha(p_\alpha) = \frac{1}{{\cal B}^\alpha(p_\alpha)- i p_\alpha},
\ee
where $B^\alpha(p_\alpha)$ denotes the real part of the inverse scattering amplitude which is unconstrained by unitarity and is a real meromorphic function of $p_\alpha^2$ near the origin 
$p_\alpha$=0.
To leading order in a momentum expansion, ${\cal B}^\alpha(p_\alpha)$ is given by the inverse scattering length. Then one finds
\be
{\cal M}_\alpha (p_\alpha) = v^{(J^{++})}_\alpha \hat{S}_\alpha^{(J^{++})}\bigl ({\cal A}^\alpha(p_\alpha) +{\cal B}^\alpha(p_\alpha) \bigr)f_{\rm on}^\alpha(p_\alpha).
\ee
Thus, unitarity forces the production amplitude to be proportional to the scattering amplitude in the final state (a coupled-channel version of this
relation is provided in Ref.~\cite{Hanhart:2012wi}). In the heavy quark limit the functions ${\cal A}$ and ${\cal B}$ do not depend on the channel.
Moreover, since the momentum dependence of the functions ${\cal A}(p_\alpha)$ and ${\cal B}^\alpha(p_\alpha)$ is controlled by the
left-hand cuts of the production operator and the scattering amplitude, respectively, we expect that near thresholds both are well approximated by constants, which are also independent 
of the channel in the heavy quark limit. Based on this one can predict the ratios of the partial widths
for different decay channels of the $W_{bJ}$'s, up to spin symmetry violating corrections.

\begin{table}
\caption{Ratios of the coupling constants, $\lambda^{(J^{++})}_\alpha$, responsible for the production of the $W_{bJ}$ states in the radiative decays $\Upsilon(10860)\to\gamma W_{bJ}$.}
\label{tab:ggamma}
\begin{ruledtabular}
\begin{tabular}{cccc}
$B\bar{B}({}^1S_0)$ &$B^*\bar{B}^*({}^1S_0)$ &$B\bar{B}^*({}^3S_1,+)$ &$B^*\bar{B}^*({}^5S_2)$\\
\hline
1&$1/\sqrt{3}$&$2$&$\sqrt{20/3}$
\end{tabular}
\end{ruledtabular}
\end{table}

It is proposed in Ref.~\cite{Wu:2018xaa} that the most prominent production mechanism for the $Z_b$ states in the $\Upsilon(10860)$ and $\Upsilon(11020)$
decays involves $B_1'\bar B$ or $B_0\bar B$ intermediate states, with $B_0$ and $B_1'$ being the broad members of the quadruplet of the positive $P$-parity $B$ mesons. If this proposal is correct, 
the decay mechanism through the $B^{(*)}\bar B^{(*)}$ pairs considered above will give only a 
small contribution. However, it should be stressed
that the mechanism proposed in Ref.~\cite{Wu:2018xaa} should not change the line shapes but only the total rate of the
production cross sections, which is not a subject of the current study.

\subsection{Coupled-channel system}
\label{sec:LS}

The set of the allowed quantum numbers for the $B^{(*)}\bar{B}^{(*)}$ system is encoded in the basis vectors quoted in \eq{basisvec}. 
Inclusion of the OPE interaction enables transitions to the $D$ and even $G$ waves \cite{Baru:2016iwj}.

For a given set $J^{PC}$
the system of the partial-wave-decomposed coupled-channel Lippmann-Schwinger-type equations reads
\begin{eqnarray}\label{Eq:JPC}
T_{\alpha\beta}(M,p,p')&=&V^{\rm eff}_{\alpha\beta}(p,p')\\
&&\hspace*{-0.9cm}-\sum_\gamma \int \frac{d^3q}{(2\pi)^3} V^{\rm eff}_{\alpha\gamma}(p,q){G_\gamma(M,q)}T_{\gamma\beta}(M,q,p'),\nonumber
\end{eqnarray}
where $\alpha$, $\beta$, and $\gamma$ label the basis vectors defined in Eq.~(\ref{basisvec}), the effective potential is defined by \eq{pot2},
and the scattering amplitude $T_{\alpha\beta}$ is related with the invariant amplitude ${\cal M}_{\alpha\beta}$ as
\bea
T_{\alpha\beta}=-\frac{{\cal M_{\alpha\beta}}}{\sqrt{(2m_{1,\alpha})(2m_{2,\alpha})(2m_{1,\beta})(2m_{2,\beta})}},
\eea
with $m_{1,\alpha}$ and $m_{2,\alpha}$ ($m_{1,\beta}$ and $m_{2,\beta}$) being the masses of the $B^{(*)}$ mesons in the channel $\alpha$ ($\beta$).
The two-body propagator for the given set $J^{PC}$ takes the form 
\bea
G_\gamma=\left(q^2/(2\mu_\gamma)+m_{1,\gamma}+m_{2,\gamma}-M-i\epsilon\right)^{-1},
\eea
where the reduced mass is
\be
\mu_\gamma=\frac{m_{1,\gamma}m_{2,\gamma}}{m_{1,\gamma}+m_{2,\gamma}}.
\ee
It is convenient to define the energy $E_i$ relative to a particular threshold, namely,
\be\label{sqrts}
M=2m+E_1\equiv m+m_*+ E_2 \equiv 2m_*+E_3.
\ee

Finally, to render the loop integrals well defined we introduce a sharp ultraviolet cutoff $\Lambda$
which needs to be larger than all typical three-momenta related to
the coupled-channel dynamics. For the results presented below we choose $\Lambda=1$~GeV but we also address the problem of the renormalisability of the resulting EFT and estimate and discuss 
the theoretical uncertainty from the cutoff variation.

\subsection{Production rates}

\begin{table*}[t!]
\caption{\label{tab:par} The fitted values of the low-energy constants
and couplings to the $J^{PC}=1^{+-}$ data for the Contact and Pionfull fits as defined at the beginning of Sec.~\ref{sec:Zb_WbJ}. 
The $\mathcal{O}(Q^0)$ contact terms $\mathcal{C}_{d}$ and $\mathcal{C}_{f}$ are given in units of $\mathrm{GeV}^{-2}$, and the
${\cal O}(Q^2)$ contact terms $\mathcal{D}_D$ and $\mathcal{D}_{SD}$
are  in units of $\mathrm{GeV}^{-4}$. 
The couplings $g_{\Upsilon(nS)}$ ($n=1,2,3$) and $g_{h_b(mP)}$ ($m=1,2$) 
are given in units of $\gev^{-3/2}$. 
Only the absolute values of the coupling constants are presented 
since physical quantities are not sensitive to their signs. Uncertainties correspond to a $1\sigma$ 
deviation in the parameters.
The quality of each fit is assessed through the
reduced $\chi^2/\text{d.o.f.}$ quoted in the last column.} 
\begin{ruledtabular}
\begin{adjustbox}{width=2.01\columnwidth}
\begin{tabular}{lcccccccccc}
Fit & $\mathcal{C}_{d}$ & $\mathcal{C}_{f}$ & $\mathcal{D}_{D}$& $\mathcal{D}_{SD}$& $|g_{\Upsilon(1S)}|$ & $|g_{\Upsilon(2S)}|$ & $|g_{\Upsilon(3S)}|$ &
$|g_{h_b(1P)}|$ & $|g_{h_b(2P)}|$ & 
$\frac{\chi^{2}}{\text{d.o.f.}}$\\
\hline
Contact & $-3.30(11)$ & $-0.06(13)$ & $0$ & $0$& $0.04(1)$ & $0.23(4)$ & $0.61(15)$ & $0.55(4)$ & $1.91(15)$ & $1.29$ \\
Pionful fit 1 & $-0.10(36)$ & $-4.19(60)$ & $0$ & $-5.80(57)$ & $0.04(1)$ & $0.25(5)$ & $0.71(18)$ & $0.46(5)$ & $1.67(18)$ & $0.95$ \\
Pionful fit 2
& $1.33(40)$ & $-3.95(27)$ & $-3.36(54)$ & $-3.16(61)$ & $0.03(1)$ & $0.21(4)$ & $0.56(14)$ & $0.32(4)$ & $1.19(14)$ & $0.83$\\
\end{tabular}
\end{adjustbox}
\end{ruledtabular}
\end{table*}

 Since we are not interested in the absolute scale but only in the energy dependence of the line shapes, 
the production amplitude of the $\beta$-th elastic channel from a point-like source for some given quantum numbers $J^{PC}$ can be defined as 
\bea\label{eq:MelJ++}
{\cal M}_\beta(M,p)&=& N_\beta U_\beta, \\ 
U_\beta= \biggl(v_\beta &-& \sum_\alpha v_\alpha\int \frac{d^3q}{(2\pi)^3} 
G_\alpha(M,q)T_{\alpha\beta}(M,q,p) \biggr),\nonumber
\eea
where the relativistic normalisation factor is
\bea
N_{\beta}&=&\sqrt{(2m_{1,\beta})(2m_{2,\beta})(2 m_\Upsilon)},
\eea
and the nonvanishing partial-wave-projected production vertices $v_\alpha$ are from \eq{vert_vJ}.

Since the direct interactions between the inelastic channels are neglected in the formalism applied
here, the $i$-th inelastic channel in the final state 
can only be reached via a transitions through the intermediate elastic channels. 
In particular, for a given set $J^{PC}$, the inelastic amplitude ${\cal M}_i$ is obtained by convolving the relevant elastic amplitude ${\cal M}_\beta(M,p)$ 
from \eq{eq:MelJ++} with the corresponding elastic-to-inelastic transition vertex from Eqs.~(\ref{via1}) and (\ref{via2}), that is,
\bea
&&{\cal M}_i(M,p_i)= \, N_i U_i, \\ 
\label{eq:MinJ++}
&&U_{\chi_{bJ}}{=}-\sum_\beta{\int}\frac{d^3q}{(2\pi)^3}
U_\beta(M,q)G_\beta(q) 
\, { v^{\chi}_{\beta,\chi_{bJ}}} ,\nonumber \\
&&U_{\eta_{b}}{=}-\sum_\beta{\int}\frac{d^3q}{(2\pi)^3} U_\beta(M,q)G_\beta(q) 
\, { v^{\Upsilon}_{\beta,\eta_{b}}}
,\nonumber
\eea
where $N_i=\sqrt{ (2 m_{\Upsilon(10860)})(2 m_{h_i})}$. 

Finally, the differential widths in the elastic and inelastic channels
read 
\be
\frac{d\Gamma_\beta}{dM}=\frac{|{\cal M}_\beta|^2\, k\, p_\beta}{32\pi^3 m_{\Upsilon(10860)}^2},\quad
\frac{d\Gamma_i}{dM}=\frac{|{\cal M}_i|^2\, k\, p_i}{32\pi^3 m_{\Upsilon(10860)}^2},
\label{eq:gamma}
\ee
respectively, where $k$ is the three-momentum of the photon in the rest frame of the $\Upsilon(10860)$ and $p_{\beta}$ 
is the three-momentum in the $\beta$-th elastic channel in the rest frame of the $B^*\bar{B}^{(*)}$ 
system, namely, 
\bea
&&p_\beta=\frac{1}{2M}\lambda^{1/2}(M^2,m_{1,\beta}^2, m_{2,\beta}^2),\\
&&k=\frac{1}{2m_{\Upsilon(10860)}}\lambda^{1/2}(m_{\Upsilon(10860)}^2,M^2, 0),
\eea
and the momentum in the inelastic channel $p_i$ is defined in Eq.~(\ref{pi}) above.

\section{Data analysis for the $Z_b$'s and prediction of the line shapes for the $W_{bJ}$ states}
\label{sec:Zb_WbJ}

\subsection{Line shapes in the $1^{+-}$ channels}

\begin{figure*}[t]
\begin{center}
\epsfig{file=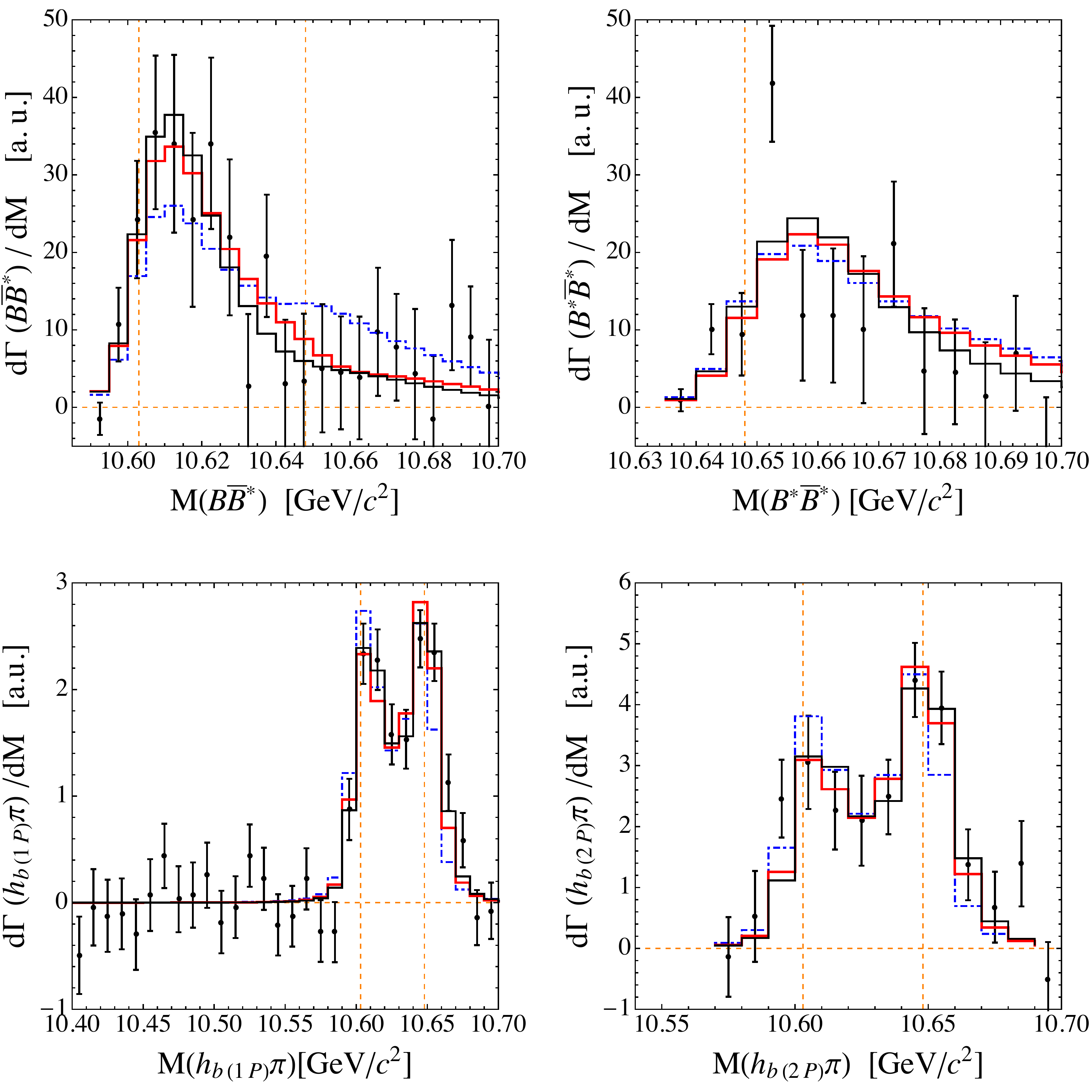, width=0.9\textwidth}\\
\caption{The fitted line shapes in the $1^{+-}$ channel.
Upper row: elastic channels $B\bar{B}^*$ and $B^*\bar{B}^*$. Lower row: inelastic channels $h_b(1P)\pi$ and $h_b(2P)\pi$. The line shapes which correspond to the Contact and Pionful fits 1 and 2 are 
shown by
the blue dashed, red thick solid and black solid curves, respectively. The vertical dashed lines indicate the position of the $B\bar{B}^*$ and $B^*\bar{B}^*$ thresholds. The experimental data given 
by black points with 
error bars are from Refs.~\cite{Belle:2011aa,Garmash:2015rfd}.}\label{fig:1pm}
\end{center}
\end{figure*}

In Ref.~\cite{Wang:2018jlv} an analysis of the experimental line shapes corresponding to the decays of $\Upsilon(10860) \to B^{(*)}\bar{B}^*\pi$ and $h_b(mP)\pi\pi$ ($m=1,2$) 
channels was carried out.

In what follows, we consider three fitting strategies introduced in Ref.~\cite{Wang:2018jlv}:
\begin{enumerate}
\item {\bf Contact fit}: purely $S$-wave momentum-independent contact interactions (analogous to fit A in Ref.~\cite{Wang:2018jlv});
\item {\bf Pionful fit 1}: 
complete leading-order potential that involves $S$-wave contact terms plus the OPE, 
plus the ${\cal O}(Q^2)$ $S$-wave-to-$D$-wave counter term promoted to LO (analogous to fit E in Ref.~\cite{Wang:2018jlv}).
\item {\bf Pionful fit 2}: 
Pionful fit 1 supplemented by the ${\cal O}(Q^2)$ $S$-wave-to-$S$-wave contact terms at NLO and the 
$\eta$-meson exchange (analogous to fit G in Ref.~\cite{Wang:2018jlv}).
\end{enumerate}
The line shapes in the $1^{+-}$ channel, where the $Z_b$'s states reside, corresponding to the best fits for the three schemes quoted above are compared with
the experimental data in Fig.~\ref{fig:1pm}. The parameters extracted from these fits are collected in Table~\ref{tab:par}.
One can see that the quality of the line shape description by the pionful fits is better than that by the Contact fit that is reflected in the change of the $\chi^{2}/\text{d.o.f.}$ 
from $1.29$ for the Contact fit to $0.95$ for the Pionful fit 1 and $0.83$ for the Pionful fit 2. 

\subsection{Renormalisability of the heavy hadron EFT with pions}

\begin{figure*}[t]
\begin{center}
\epsfig{file=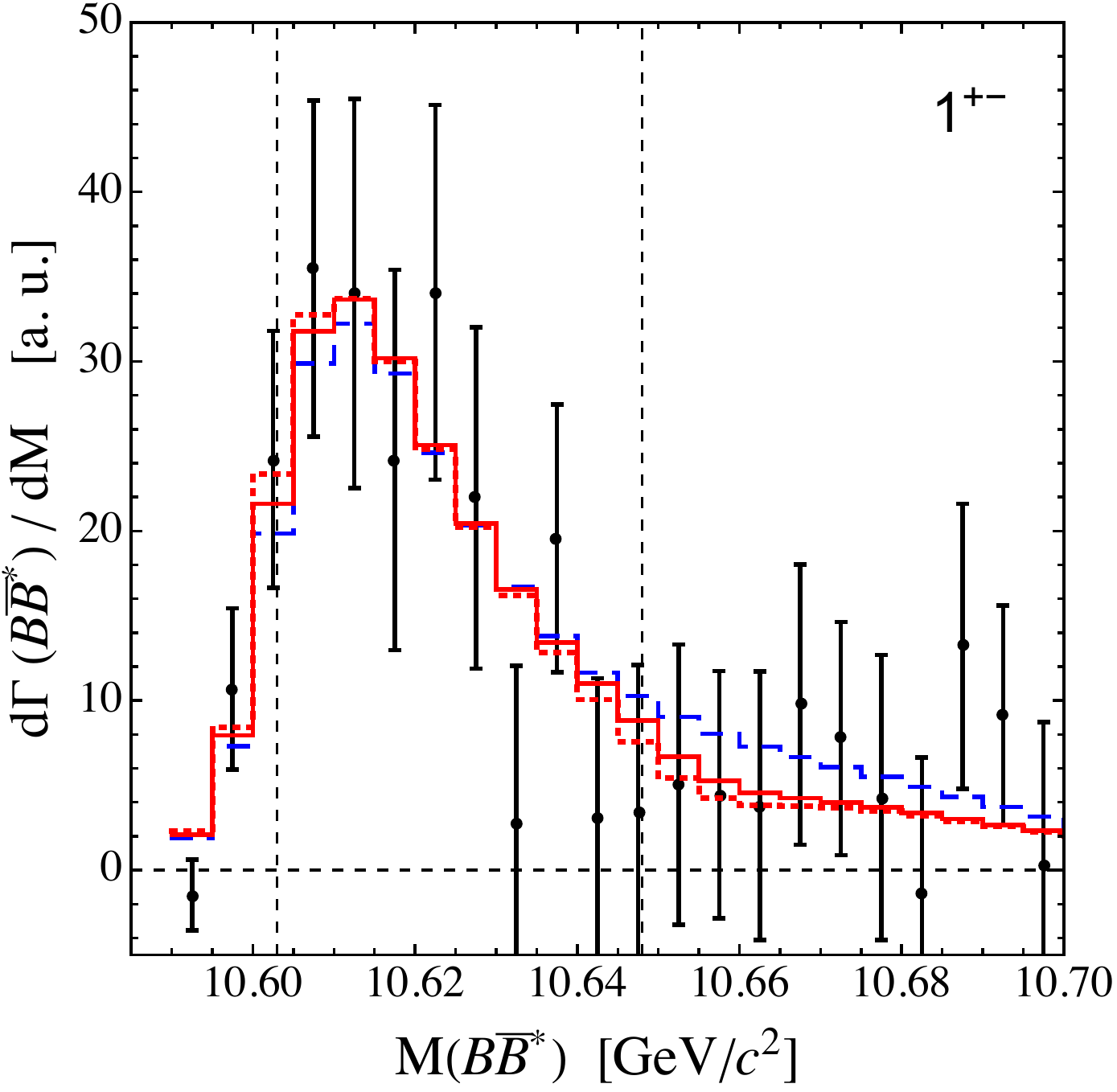, width=0.305\textwidth}\hspace{0.3cm}
\epsfig{file=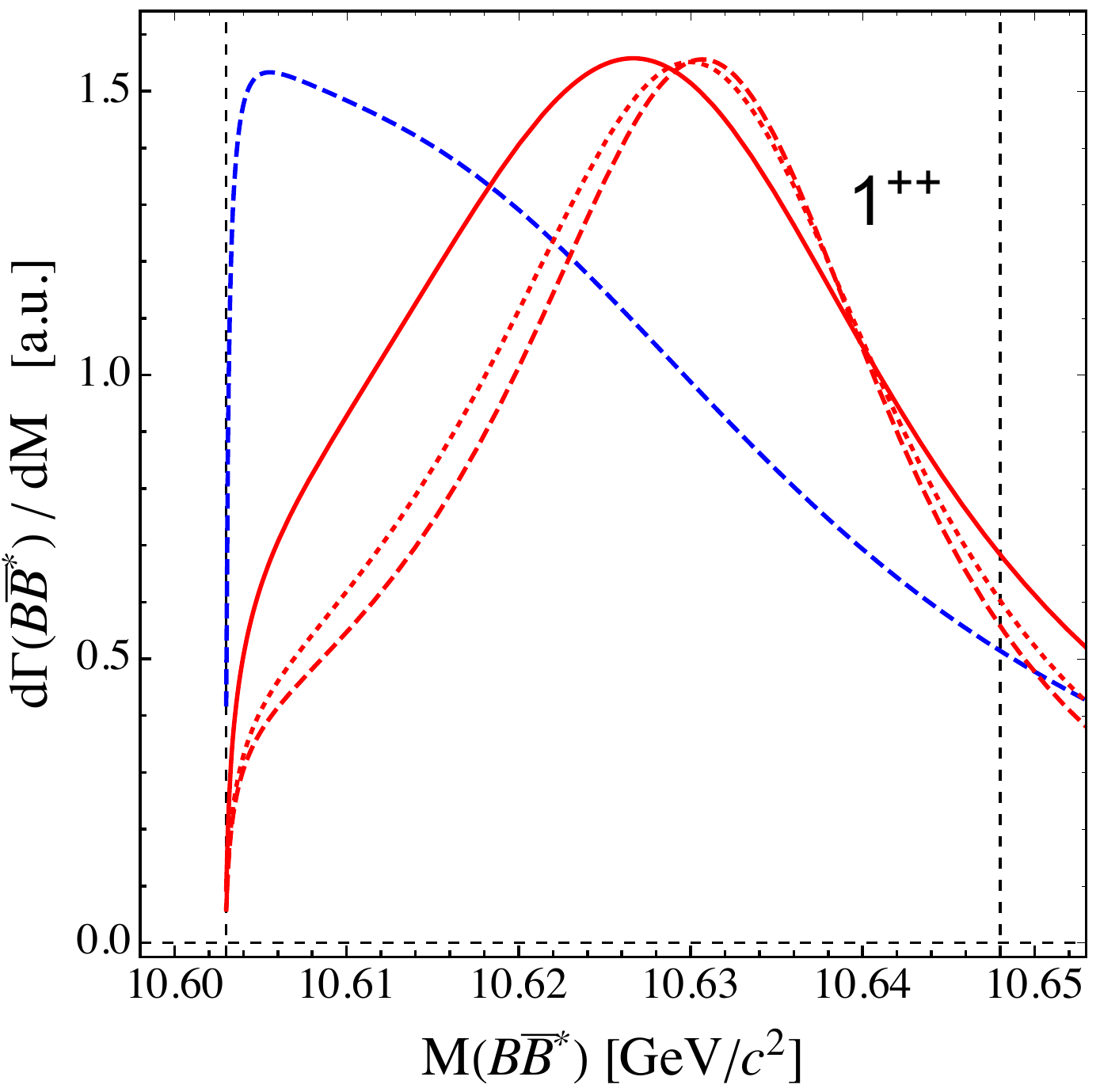, width=0.3\textwidth}\hspace{0.5cm}
\epsfig{file=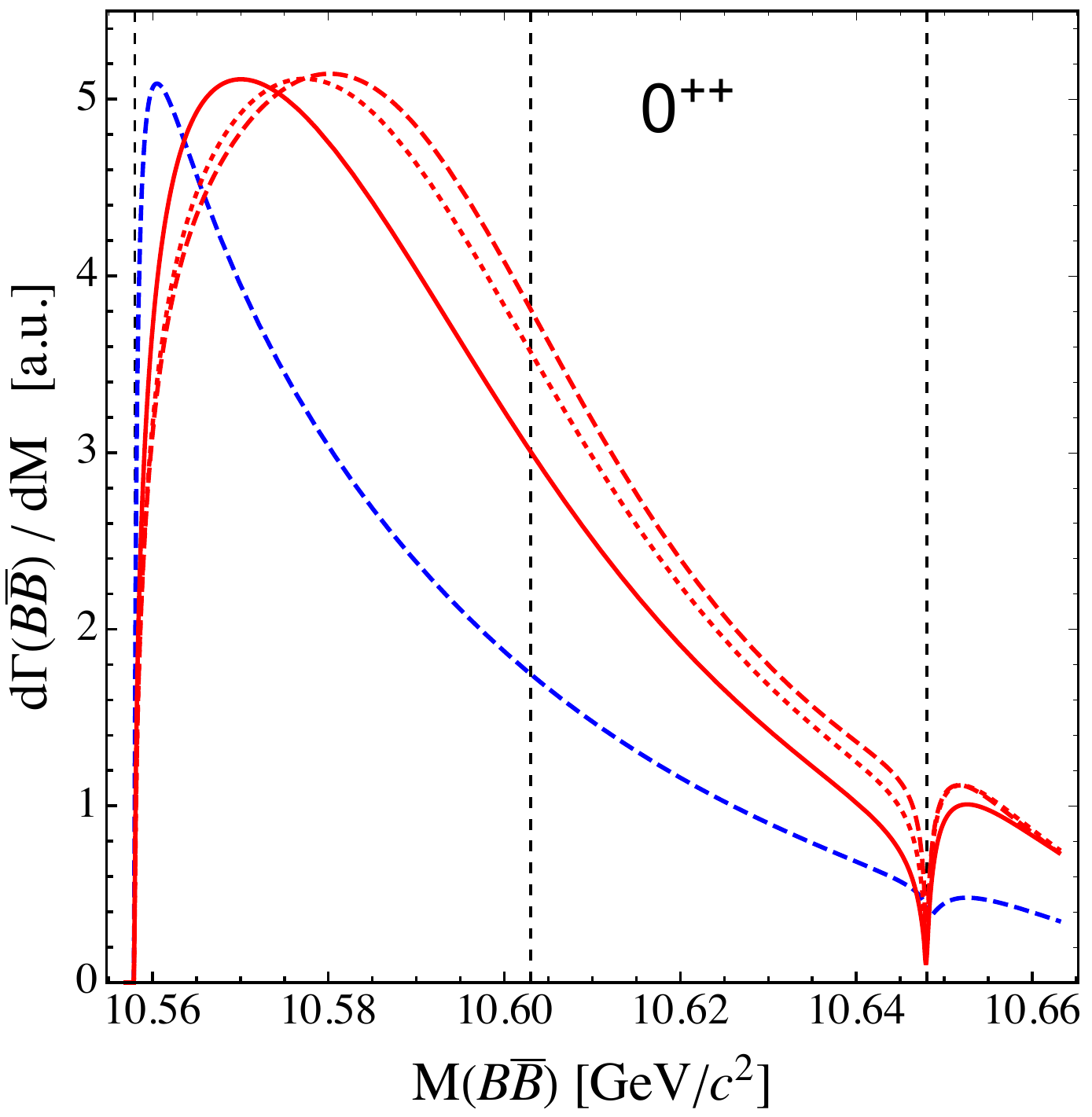, width=0.29\textwidth}
\\
\epsfig{file=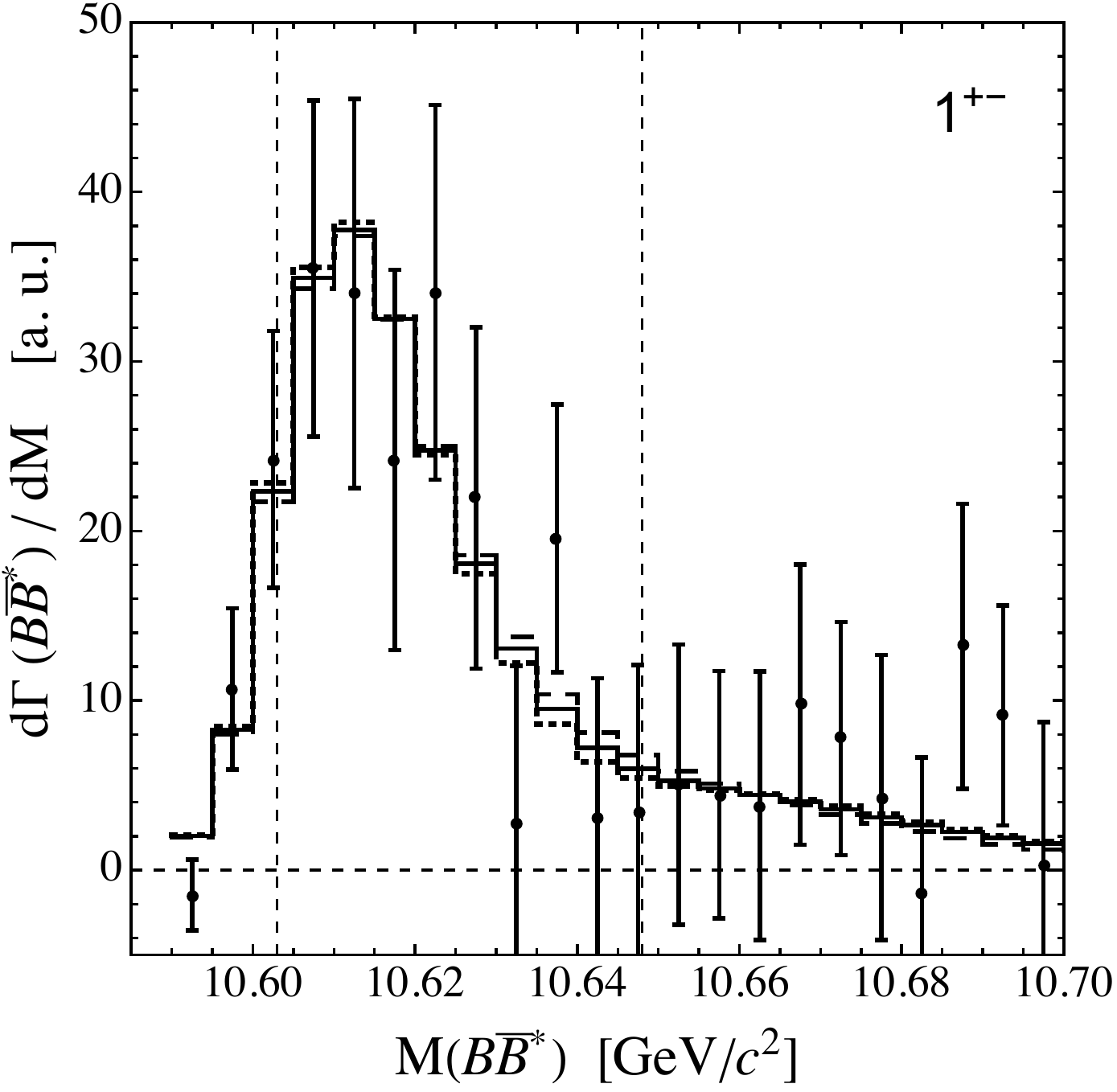, width=0.305\textwidth}\hspace{0.3cm}
\epsfig{file=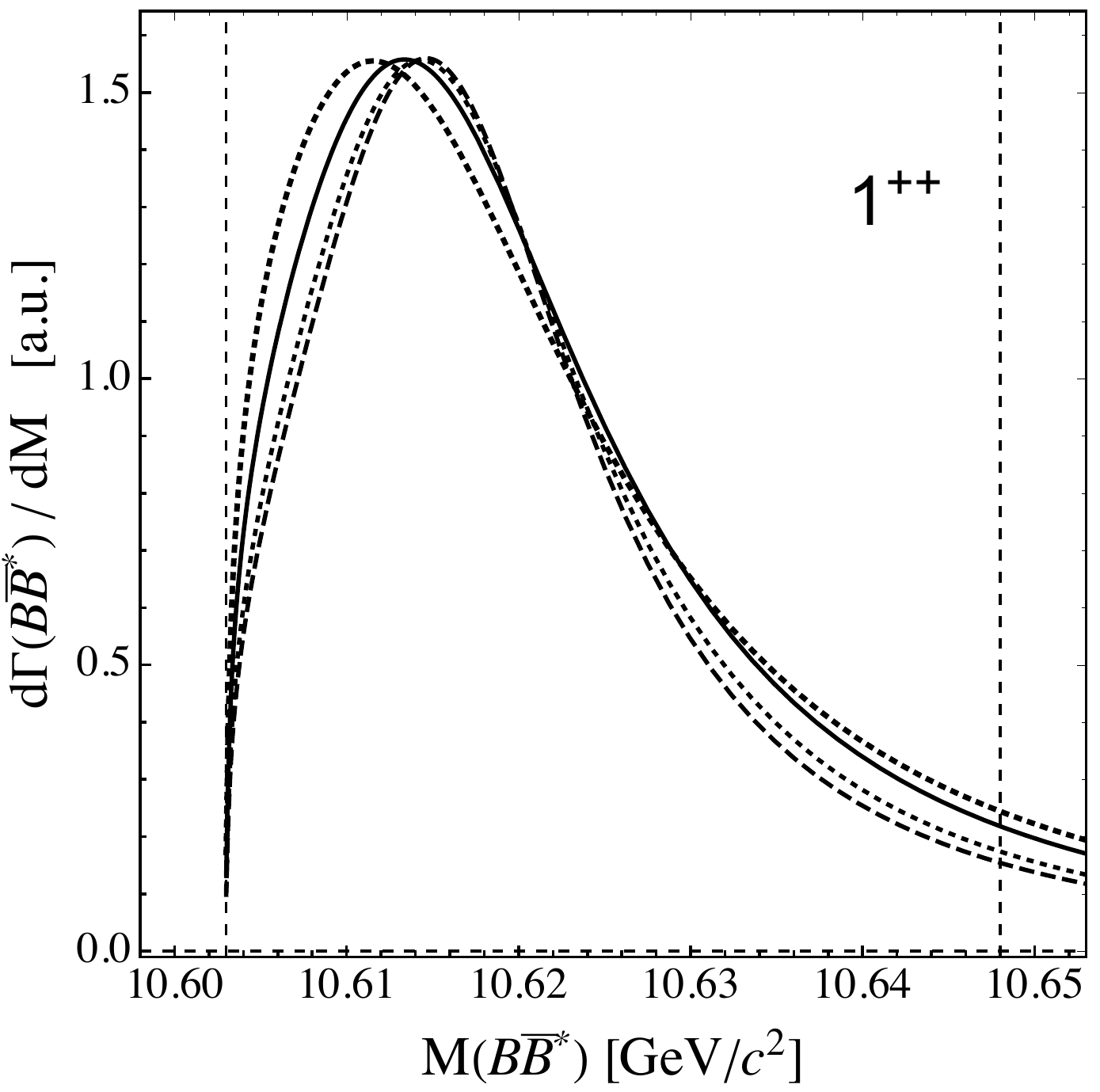, width=0.3\textwidth}\hspace{0.5cm}
\epsfig{file=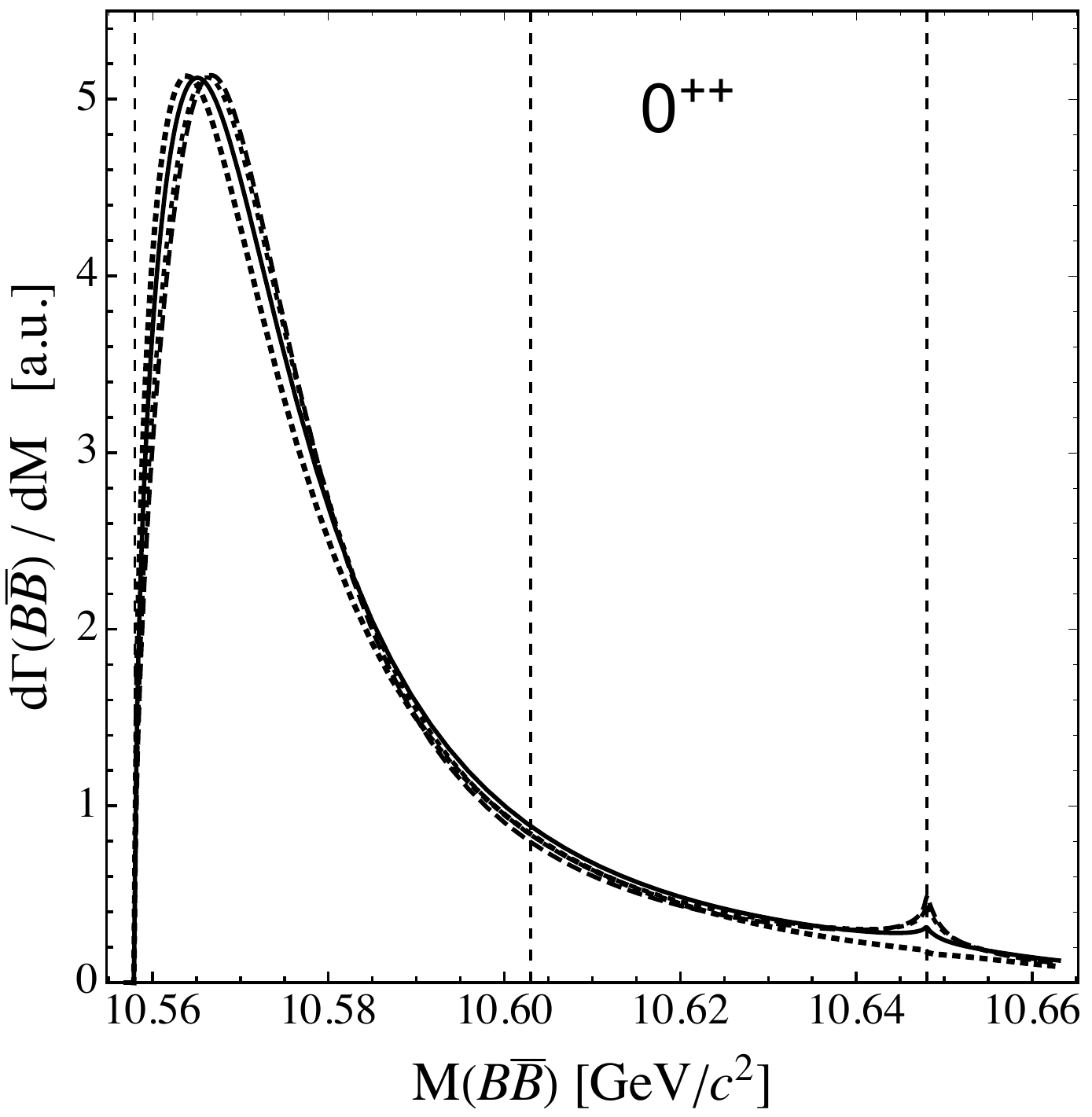, width=0.29\textwidth}
\caption{Propagation of the cutoff dependence of the theoretical fits
 from the $1^{+-}$ channel used as input to the spin partner channels
 $J^{++}$ ($J=0,1,2$) which are parameter-free predictions. 
The upper panel shows the elastic line shapes for the Pionful fit 1 (fit E in Ref.~\cite{Wang:2018jlv}) and the lower panel corresponds to the Pionful fit 2 (fit G in Ref.~\cite{Wang:2018jlv}). 
Notation of the curves in the upper panel: 
blue dashed --- $\Lambda=0.8$ GeV, red solid --- $\Lambda=1$ GeV, red dotted --- $\Lambda=1.2$ GeV, red dashed --- $\Lambda=1.3$ GeV. 
Notation of the curves in the lower panel: thick black dotted --- $\Lambda=0.8$ GeV, black solid --- $\Lambda=1$ GeV, black dotted --- $\Lambda=1.2$ GeV, black dashed --- $\Lambda=1.3$ 
GeV.}\label{fig:cutoff_dep}
\end{center}
\end{figure*}

The use of the standard nonrelativistic approach to heavy mesons, as employed in Ref.~\cite{Wang:2018jlv} and also used here,
leads to coupled-channel integral equations for the scattering amplitudes which, at leading order in the EFT
expansion, are linearly divergent. Therefore, when the potential truncated at a given order is iterated within the integral equations 
an infinite number of ultraviolet (UV) divergent higher-order contributions is generated. 
The problem is well known in the context of nuclear chiral EFT --- see, for example,
Refs.~\cite{Nogga:2005hy,Epelbaum:2006pt} and references therein. The standard way to cure this problem in practical calculations is to employ a 
finite UV cutoff of the order of a natural hard scale in the problem,
so that the unwanted higher-order contributions turn out to be suppressed \cite{Lepage:1997cs}; see also a recent discussion in 
Ref.~\cite{Epelbaum:2018zli}. For an alternative approach
with relativised integral equations of the Kadyshevsky type 
in the context of an NN and heavy-meson EFTs see Refs.~\cite{Epelbaum:2012ua} and \cite{Baru:2015tfa}, respectively. 
 
The logic explained above is the basis for the renormalisation programme used in Ref.~\cite{Wang:2018jlv} and, hence, is also employed here.
It should be stressed that the formulation of an EFT for the $Z_b$'s and their spin partners is much more challenging than that for the NN problem
because of larger soft scales involved here. Indeed,
since the $B$ mesons are, roughly, by a factor of five heavier than nucleons, an EFT for the $Z_b(10610) $ and $Z_b(10650)$ states, separated by 
$\delta =45$ MeV, unavoidably involves the momenta of the order of $\sqrt{ m \delta} \approx 500$ MeV treated as soft. 
Moreover, in the course of practical fits of the experimental line shapes the momenta as large as 800 MeV are included from the high-energy tail of the experimental distributions. 
Clearly, the influence of such high momenta on the
dynamics close to the relevant thresholds is minor 
whereas the renormalisation of the theory (and the possible residual cutoff dependence) is severely affected by this high-momentum range. 

In Ref.~\cite{Wang:2018jlv} it was found that the cutoff dependence
generated via the iterations of the OPE in the $S$ waves can be almost
completely absorbed into the momentum-independent contact terms 
${\cal C}_d$ and ${\cal C}_f$ at LO. On the other hand, the cutoff dependence of the line shapes in the $1^{+-}$ channel from iterations of the $S$-wave-to-$D$-wave OPE turns out to be sizeable 
which calls for 
the promotion to leading order of the contact term ${\cal D}_{SD}$ providing the $S$-wave-to-$D$-wave transitions between heavy mesons, which na{\"i}vely would appear only at NLO. 
In Fig.~\ref{fig:cutoff_dep}, we illustrate the cutoff dependence for the elastic line shapes corresponding to the quantum numbers $1^{+-}$, $1^{++}$ and $0^{++}$ for the cutoff variation from 0.8 to 
1.3 GeV.
We start the discussion with the results for the quantum numbers
$1^{+-}$, where the fits to the experimental data were performed (see the left plots in Fig.~\ref{fig:cutoff_dep}).
As a general trend, the results demonstrate a mild cutoff dependence and a saturation for larger cutoffs. 
Meanwhile, as expected, the result for the smallest cutoff $\Lambda =0.8$ GeV for the Pionful fit 1 deviates from the other curves 
(\emph{cf.} the blue dashed curve for $\Lambda =0.8$ GeV with the red solid and dotted curves corresponding to $\Lambda =1.0$ GeV and $\Lambda =1.2$ GeV, respectively).
Indeed, in order to maintain approximate $\Lambda$-independence in the pionful calculations
for smaller cutoffs, we found empirically that the magnitude of the contact term ${\cal D}_{SD}$
must be increased such that it generates an increasing
$S$-wave-to-$S$-wave higher-order contribution through iterations. The
latter induces a strong $\Lambda$-dependence 
unless an additional ${\cal O}(Q^2)$ 
$S$-$S$ contact 
term is included in the potential. 
As a consequence, the results for the Pionful fit 1 still show some cutoff dependence for the observables in the $1^{+-}$ channel while the cutoff dependence for the Pionful fit 2, where the order 
${\cal O}(Q^2)$ $S$-$S$ contact term ${\cal D}_d$ is included, 
is diminished significantly. Exactly the same pattern, though somewhat enhanced, can also be seen in Fig.~\ref{fig:cutoff_dep} for the spin partners. 
It is obvious that, for the smallest cutoff, the results for the Pionful fit 1 (blue dashed curve) possess an unwanted sizeable $S$-wave-to-$S$-wave higher-order contribution which, for the Pionful 
fit 2, is largely absorbed by the 
$S$-$S$ contact term ${\cal D}_d$. 
Still, the results for the Pionful fit 1 for the cutoffs from 1 GeV onward quickly saturate with the cutoff increase and may be regarded as reasonable predictions at leading order. 
In what follows, we will discuss the line shapes and extract the poles of the amplitude for both Pionful fits 1 and 2. 
Still, we regard the predictions obtained for the Pionful fit 2 as our main results since in this case the cutoff-related artefacts induced by the iteration of the 
truncated potential are significantly reduced. 
It should be clear that the results for the Pionful fit 2 correspond to an incomplete NLO calculation and that, in addition, there are long-range contributions from the two-pion exchange (TPE) 
not included in the present study. 
It remains to be seen whether or not their inclusion affects the predictions for the spin partner states. However, given that, numerically, the long-range part of the OPE plays a role 
of a correction as compared with the short-range mechanisms, the effect from the long-range TPE is expected to be small.

\begin{table*}[t!]
\caption{The ratios of the individual widths for the elastic and inelastic channels
in the $\Upsilon(10860)$ radiative decays via the $W_{bJ}$ states
relative to the sum of all individual partial widths for a given $J$ (all ratios in each line add up to unity) obtained for the Pionful fit 2 (fit G in Ref.~\cite{Wang:2018jlv}).}
\label{tab:fractions} 
\begin{ruledtabular}
\begin{tabular}{c|cccccccccccc}
$J^{PC}$ & $B\bar{B}$ &$B\bar{B}^{*}$& $B^{*}\bar{B}^{*}$ & $\chi_{b0}(1P)\pi$ & $\chi_{b0}(2P)\pi$ & $\chi_{b1}(1P)\pi$ & $\chi_{b1}(2P)\pi$ & $\chi_{b2}(1P)\pi$& 
$\chi_{b2}(2P)\pi$ & $\eta_{b0}(1S)\pi$& $\eta_{b0}(2S)\pi$\\
\hline
$2^{++}$ & $0.06 $ & $0.07$ & $0.54$ & --- & --- & $0.03$ & 0.06 & 0.09 & 0.16 & --- & --- \\
$1^{++}$ & --- & $0.76$ & --- &0.03 & 0.06& $0.02$ & 0.04 & 0.04 & 0.05 & --- & --- \\
$0^{++}$ & $0.73$ & --- & 0.14 & --- & --- &0.05 & 0.06& --- & --- & 0.002 & 0.01 
\end{tabular}
\end{ruledtabular}
\end{table*}

\subsection{Line shapes in the spin partner channels}

\begin{figure*}[ht]
\begin{center}
\epsfig{file=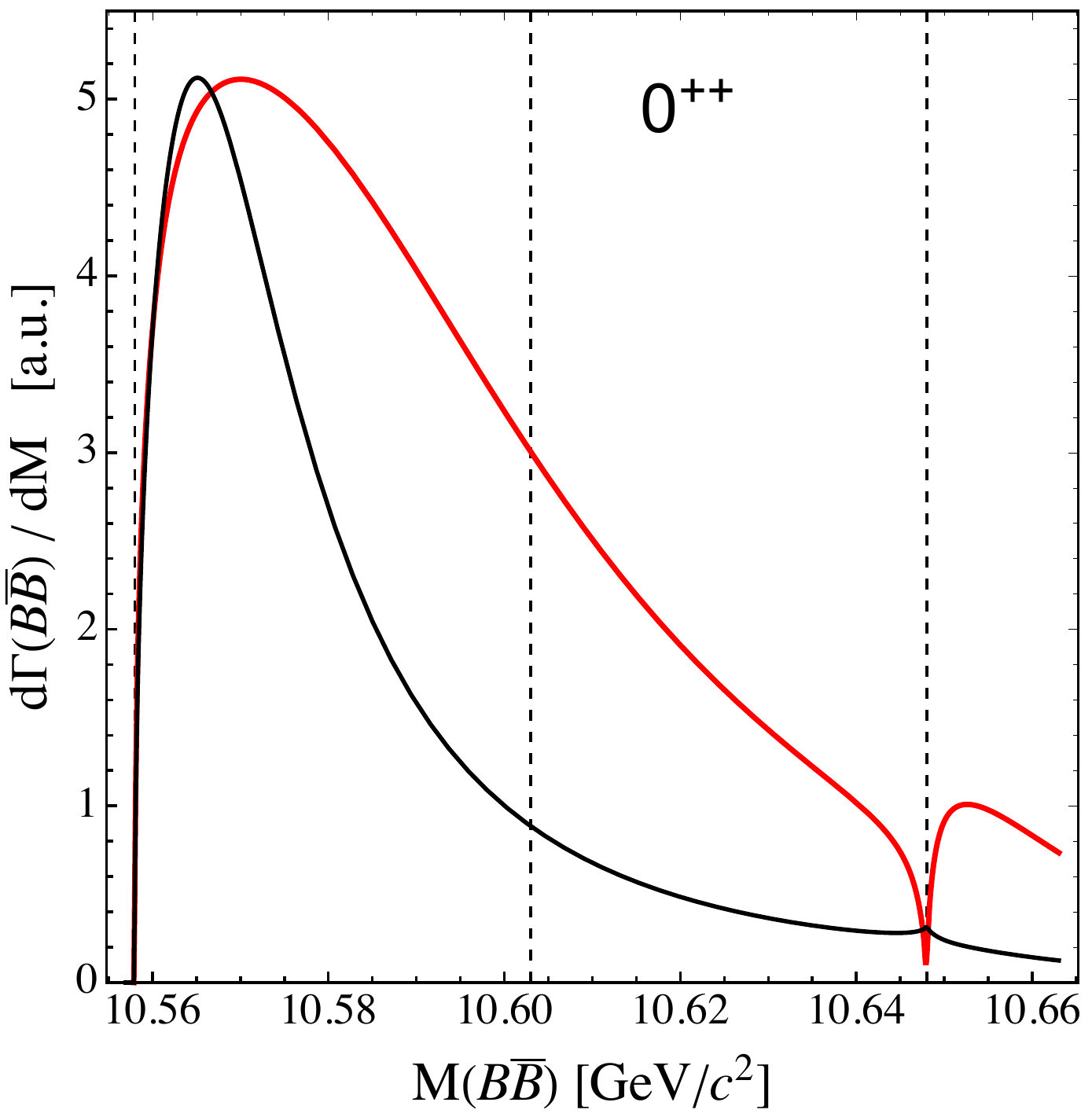, width=0.42\textwidth}\hspace{0.75 cm}
\epsfig{file=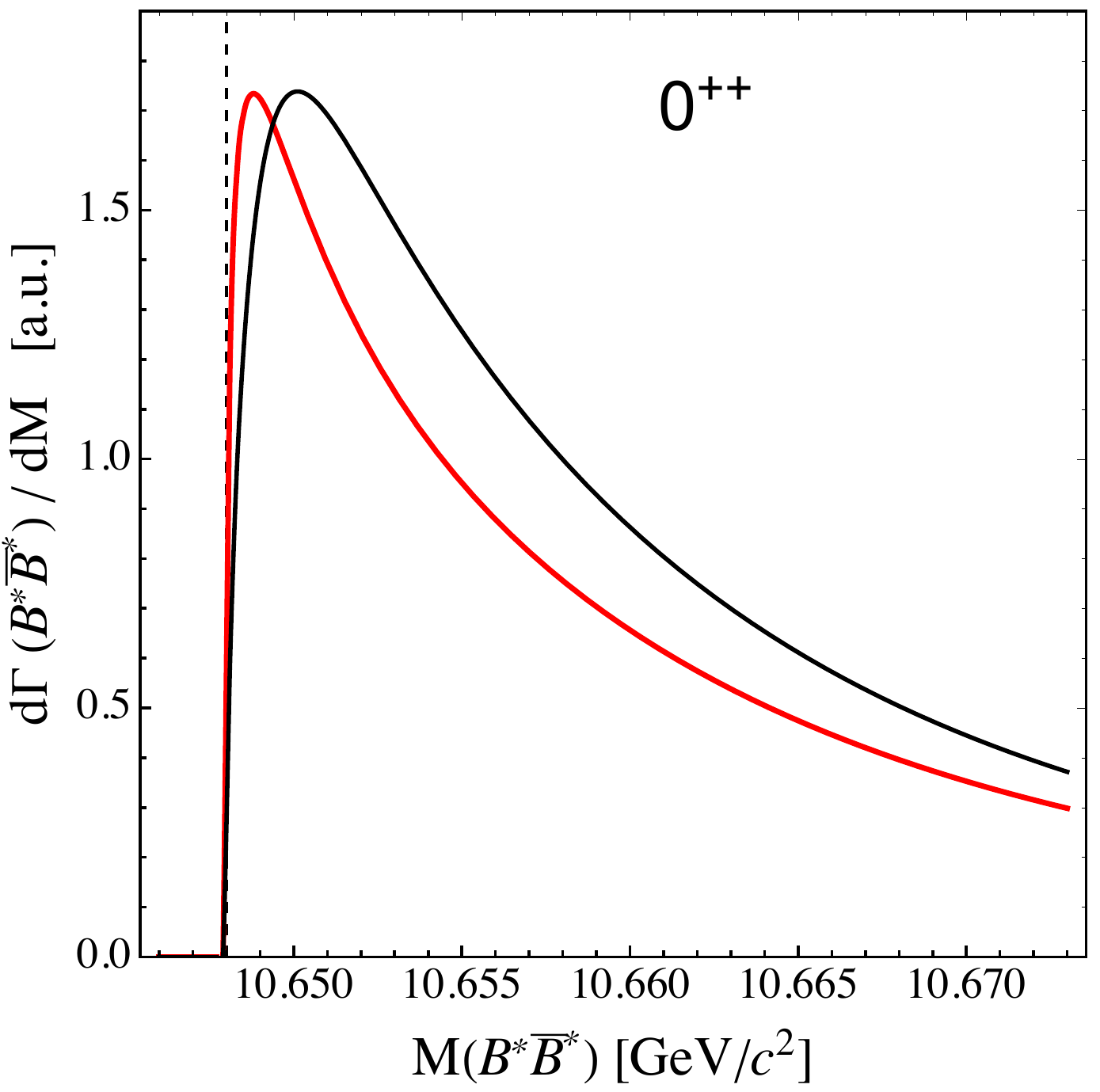, width=0.435\textwidth}\vspace*{5mm}
\epsfig{file=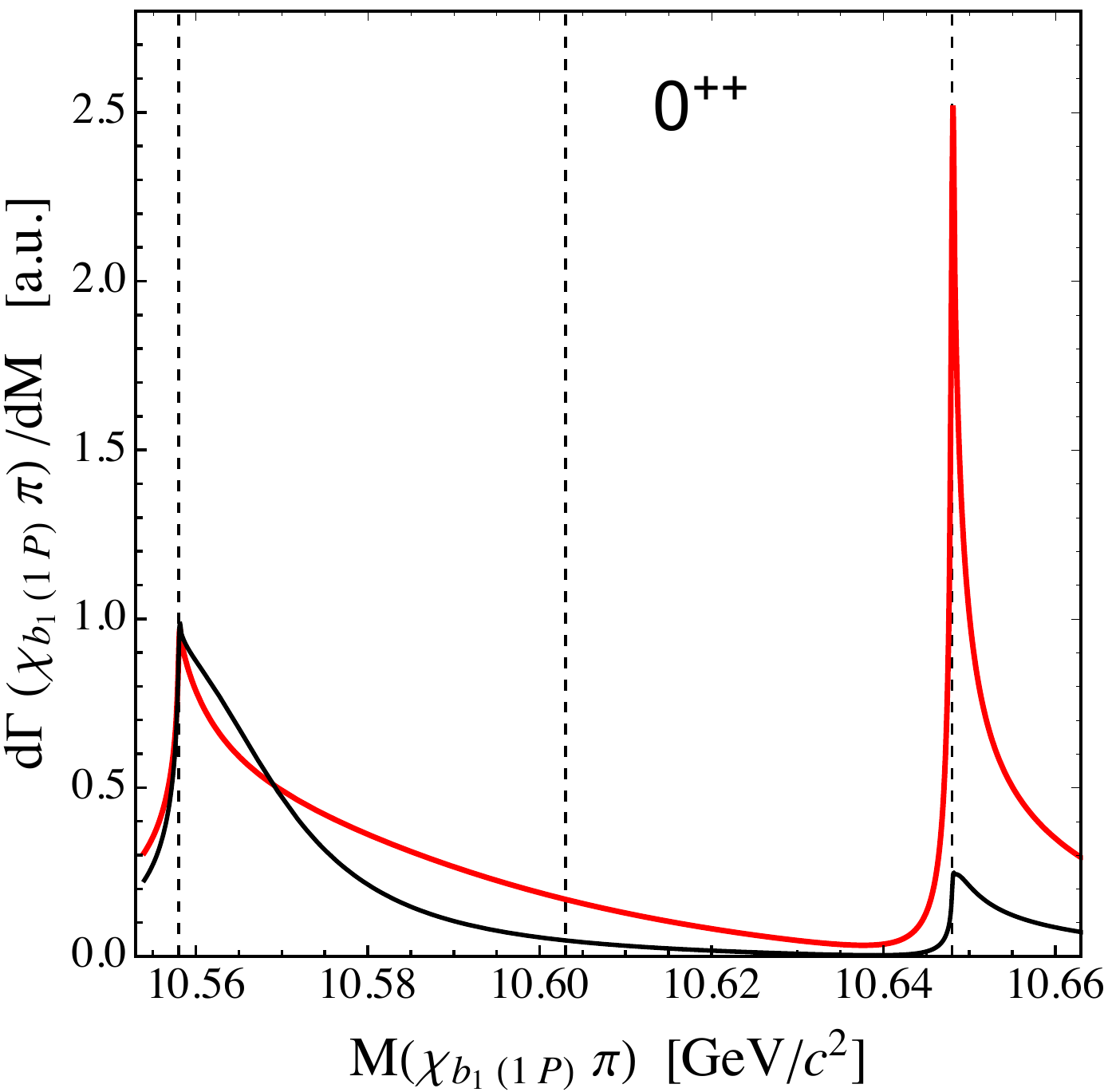, width=0.44\textwidth}\hspace{0.75 cm}
\epsfig{file=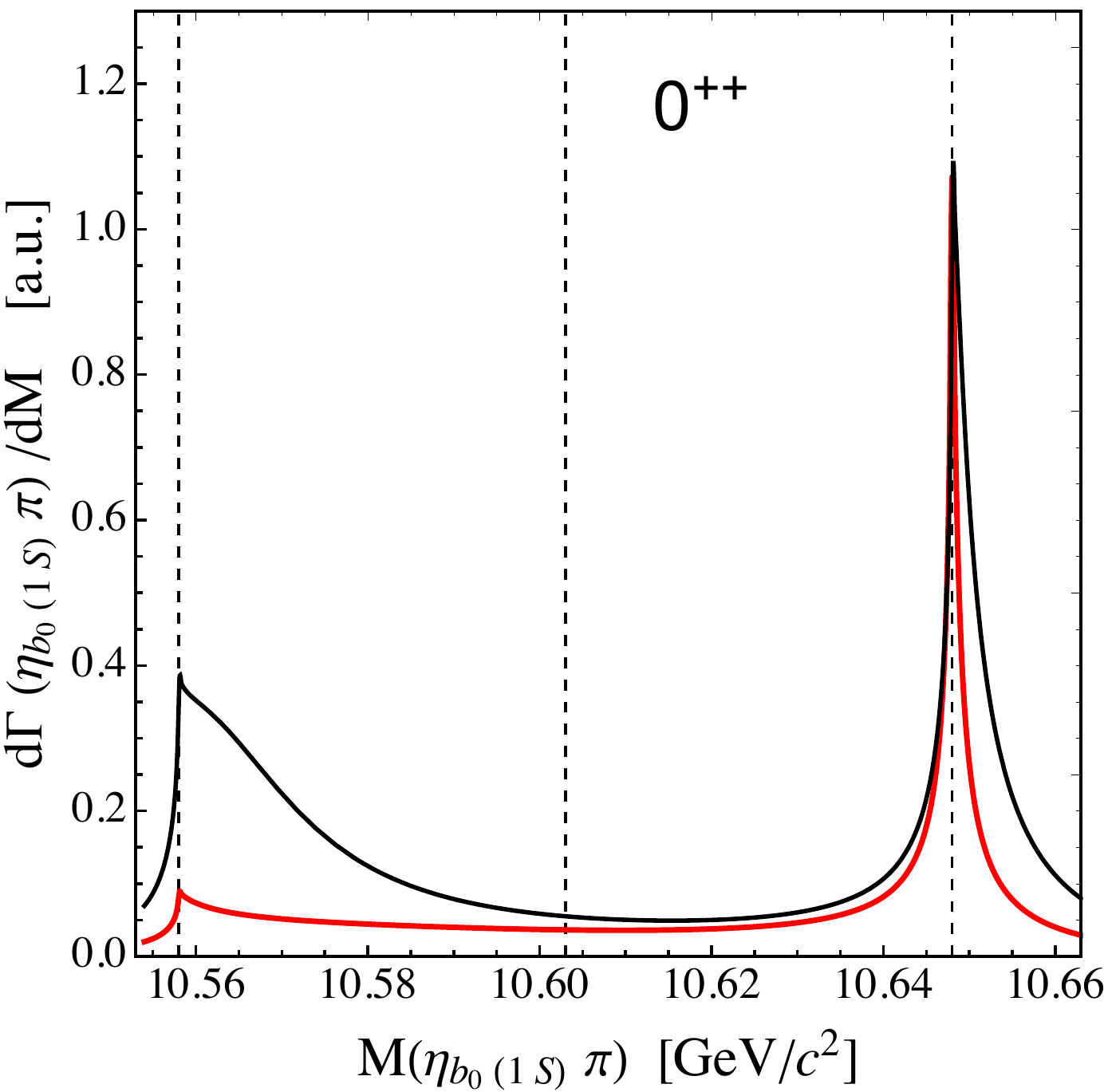, width=0.44\textwidth}
\caption{Predicted line shapes in the $0^{++}$ channel. Upper panel:
the line shapes in the $B\bar B$ and $B^*\bar B^*$ channels. Lower panel: the line shapes in the $\chi_{b1}(1P)\pi$
and $\eta_{b0}(1S)\pi$ channels. The red and black lines show the results for the Pionful fits 1 and 2, respectively, and the vertical dashed lines indicate the position of the $B\bar{B}, B\bar{B}^*$ 
and $B^*\bar{B}^*$ thresholds.}\label{fig:0pp}
\end{center}
\end{figure*}

\begin{figure*}[ht]
\begin{center}
\epsfig{file=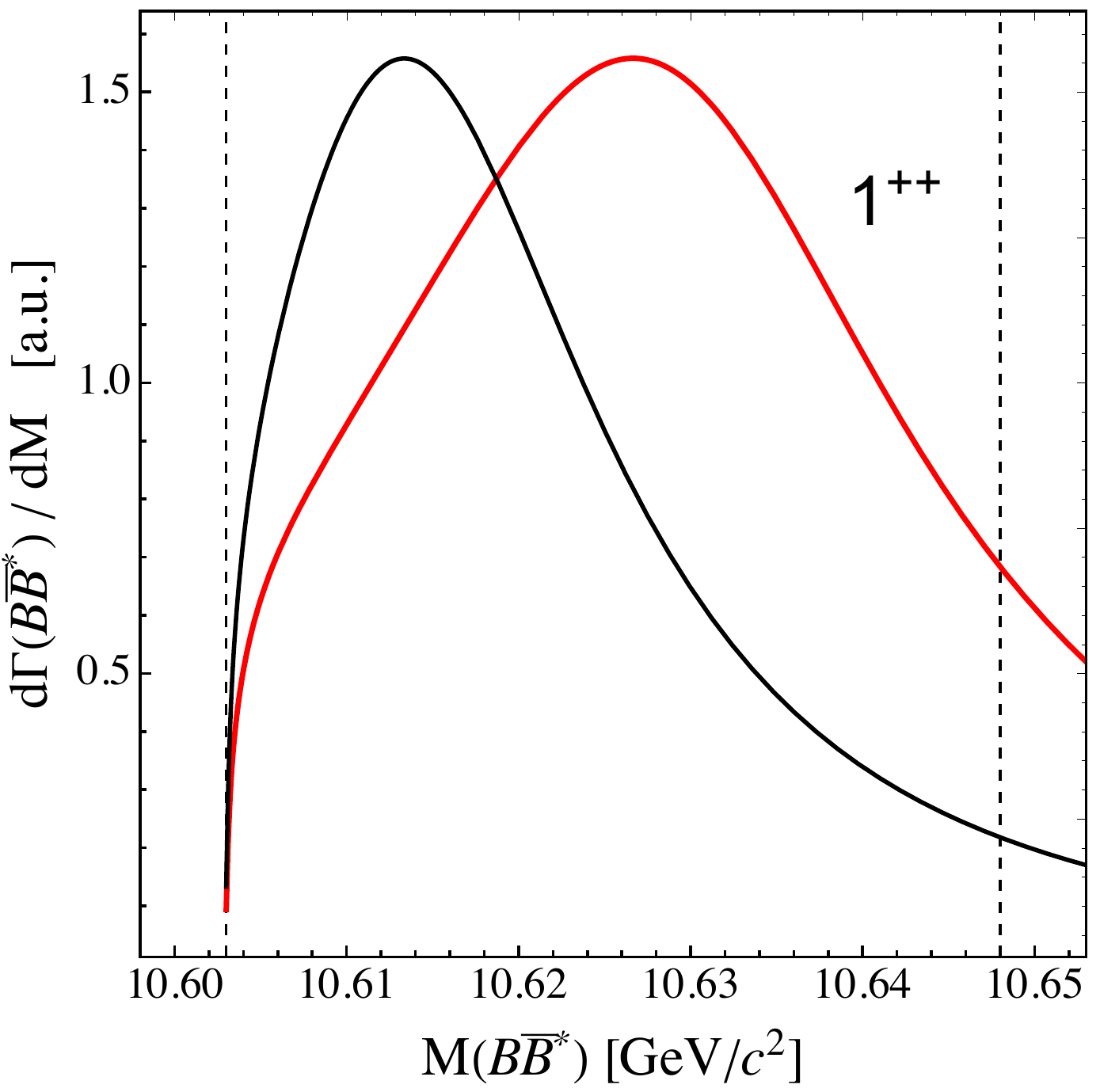, width=0.42\textwidth}\hspace{0.75 cm}
\epsfig{file=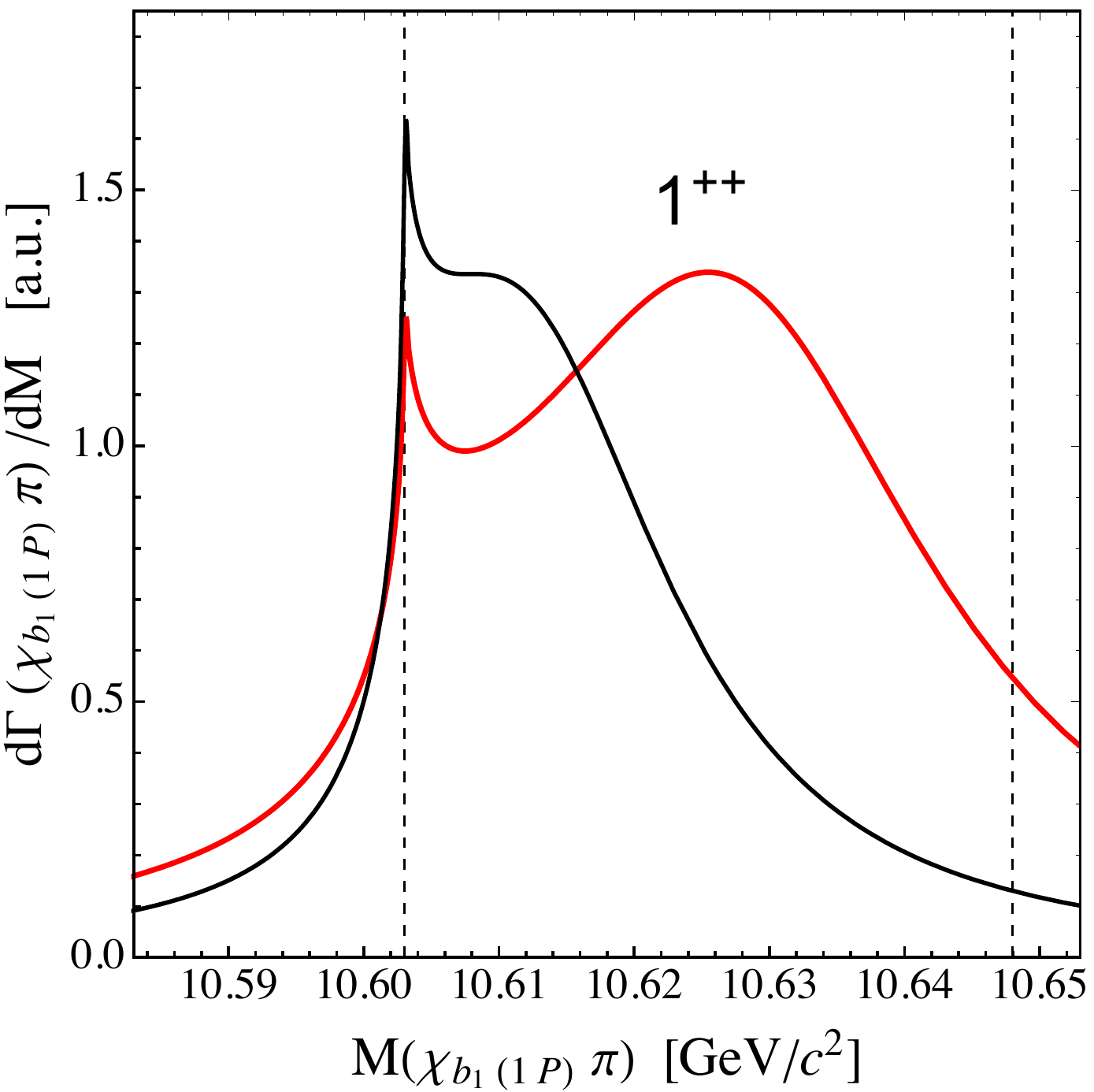, width=0.42\textwidth}
\caption{Predicted line shapes in the $1^{++}$ channel. Left panel: the line shape in the $B\bar B^*$ channel; Right panel: the line shape in the $\chi_{b1}(1P)\pi$ channel. 
The vertical dashed lines indicate the position of the $ B\bar{B}^*$ and $B^*\bar{B}^*$ thresholds. For notation see Fig.~\ref{fig:0pp}}\label{fig:1pp}
\end{center}
\end{figure*}

\begin{figure*}[ht]
\begin{center}
\epsfig{file=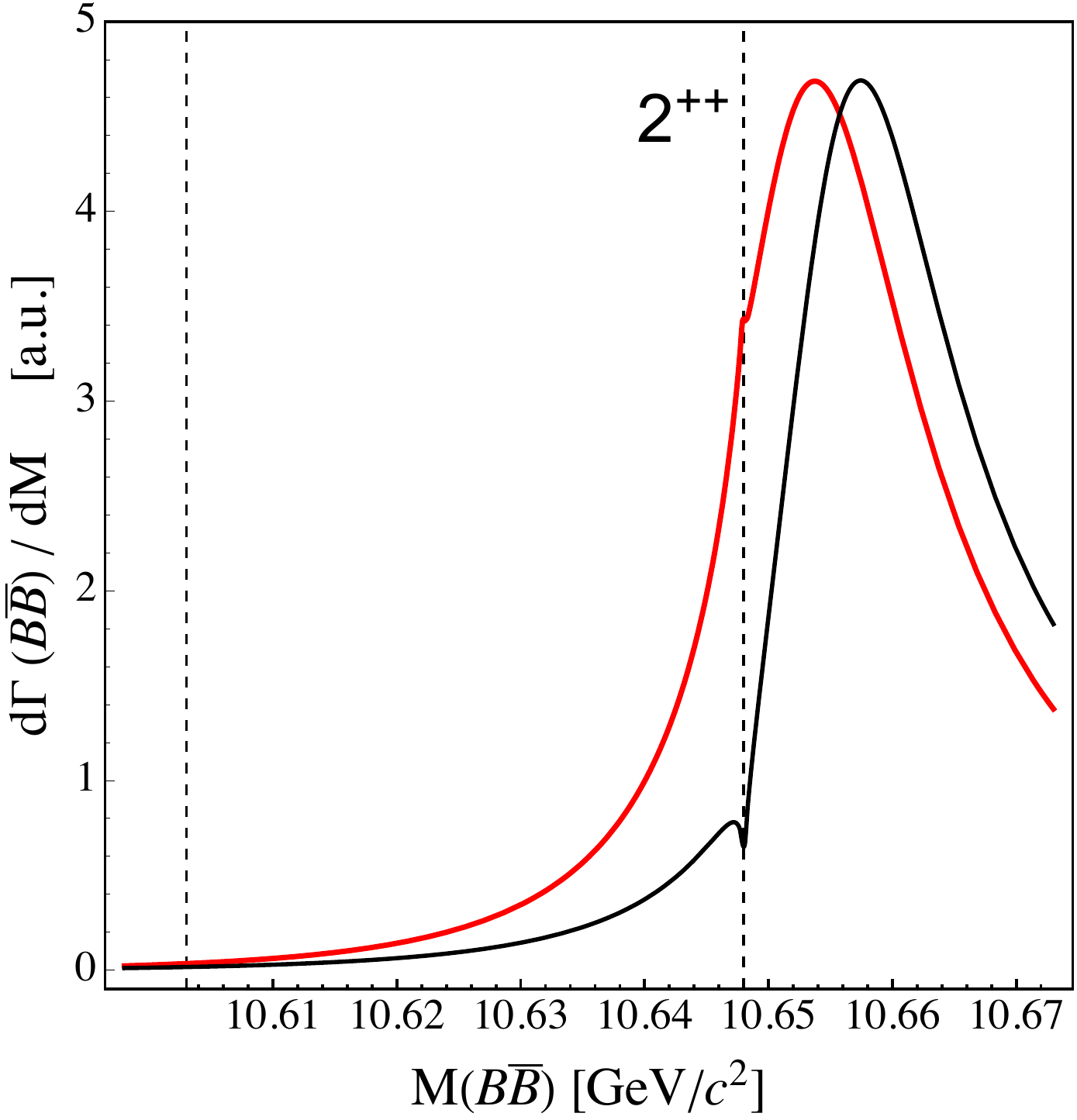, width=0.42\textwidth}\hspace{0.75 cm}\vspace{0.75 cm}
\epsfig{file=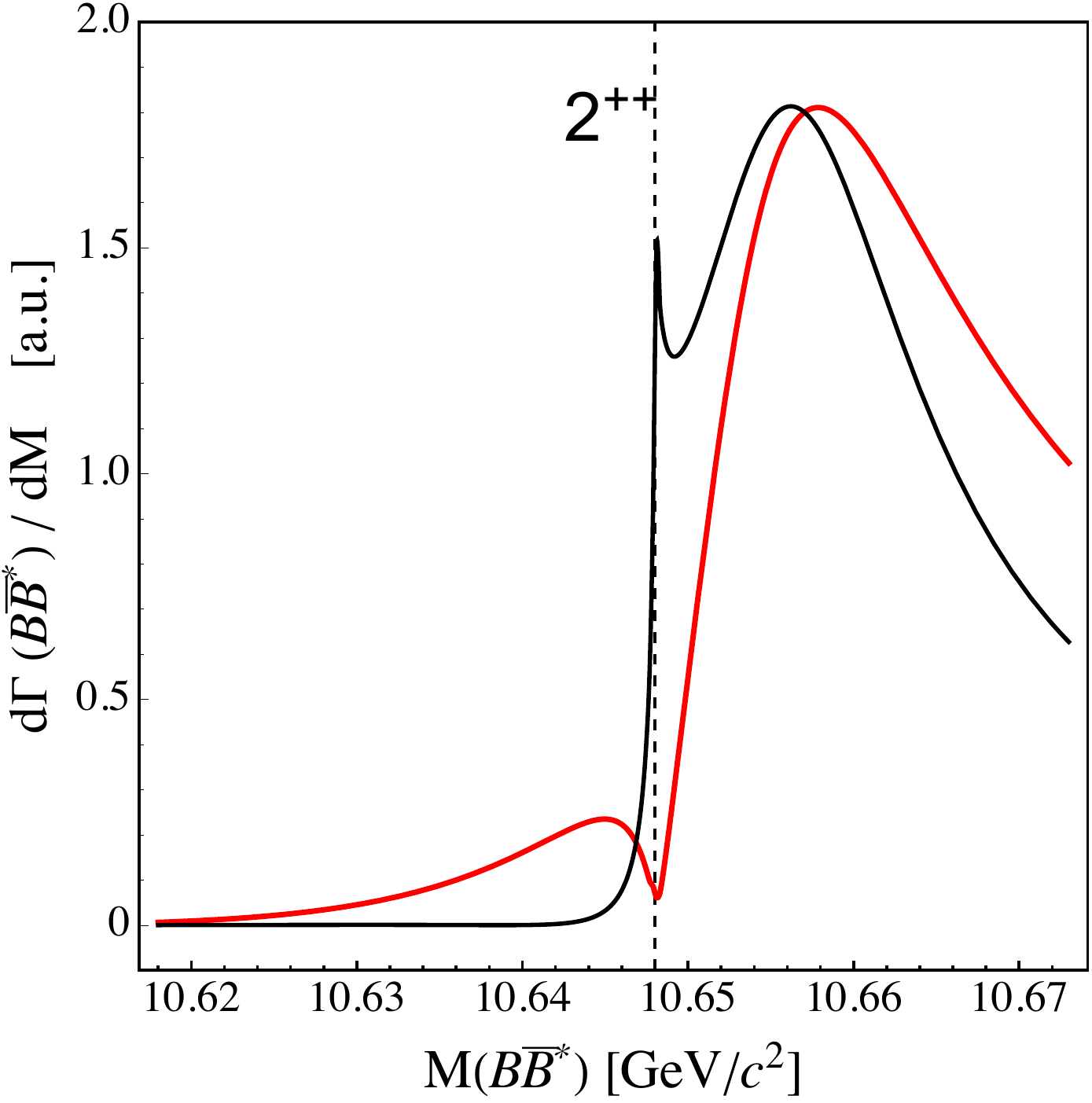, width=0.435\textwidth}\vspace*{5mm}
\epsfig{file=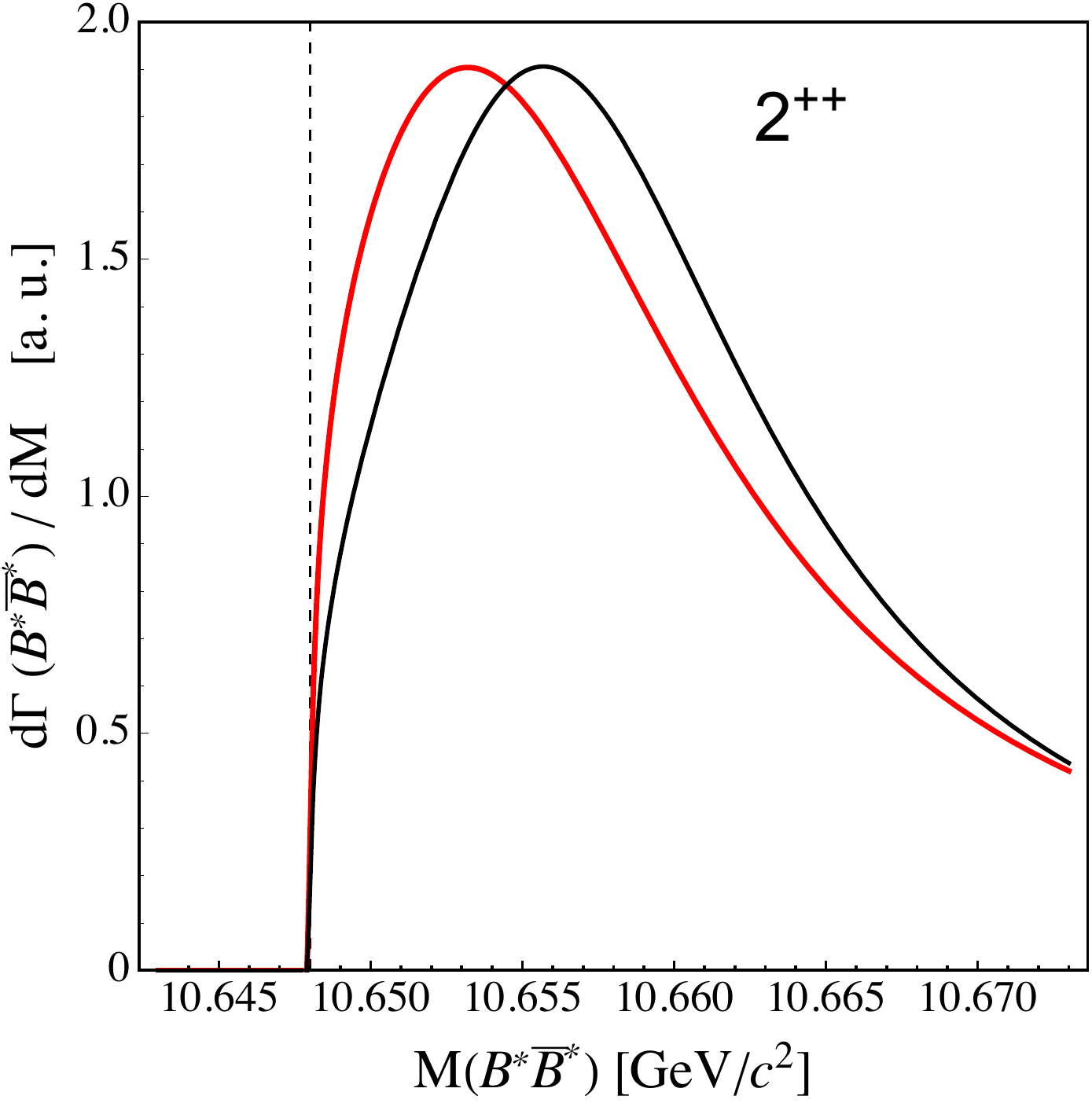, width=0.44\textwidth}\hspace{0.75 cm}
\epsfig{file=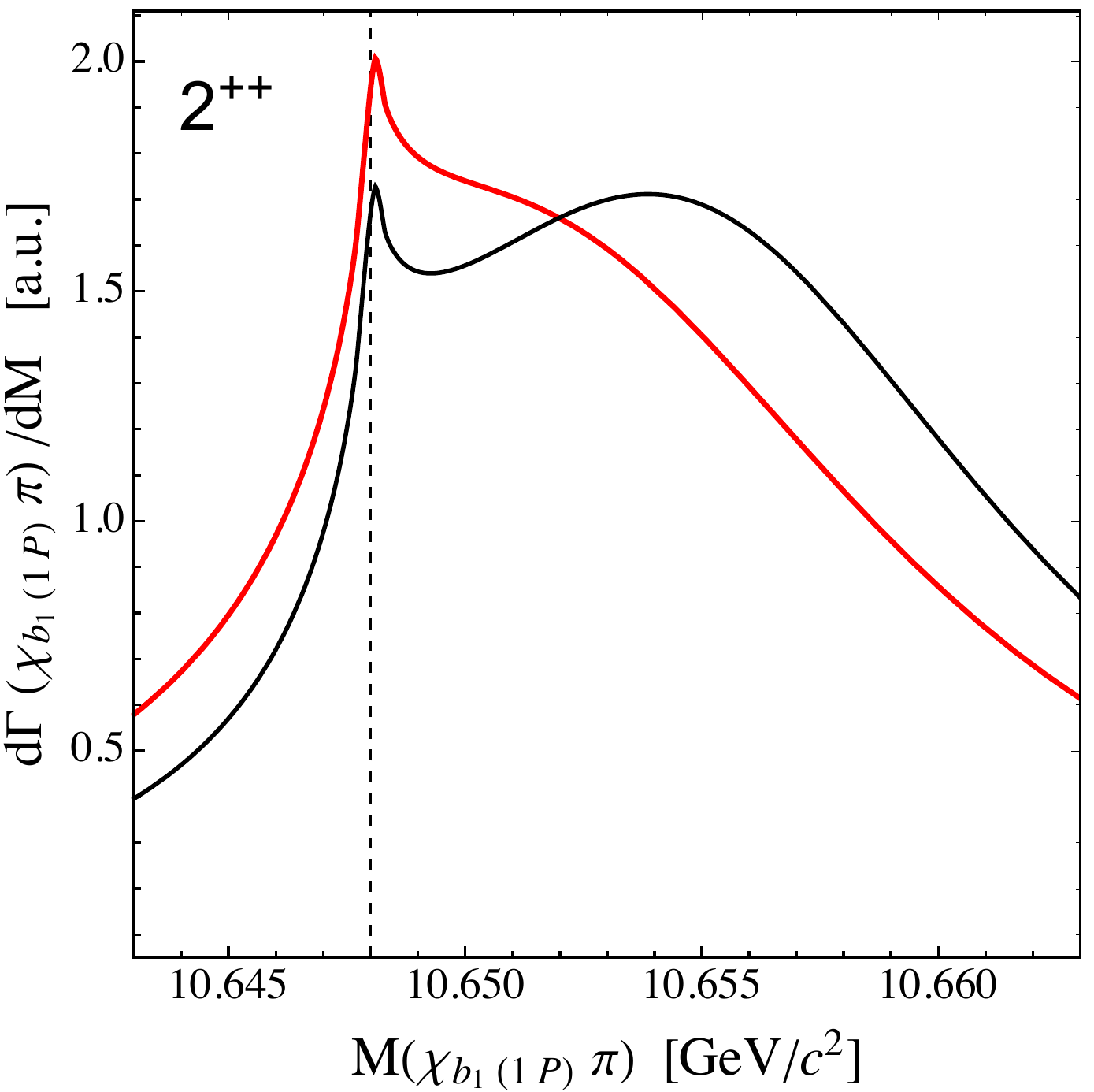, width=0.44\textwidth}
\caption{Predicted line shapes in the $2^{++}$ channel. Upper panel: the line shape in the $B\bar B$ and $B\bar B^*$ channels generated from the $S$-wave-to-$D$-wave transitions
in the OPE; Lower panel: the line shape in the $B^*\bar B^*$ and $\chi_{b1}(1P)\pi$ channel. 
The vertical dashed lines indicate the position of the $B^*\bar{B}^*$ threshold. For notation see Fig.~\ref{fig:0pp}.}\label{fig:2pp}
\end{center}
\end{figure*}

\begin{figure*}[t]
\begin{center}
\epsfig{file=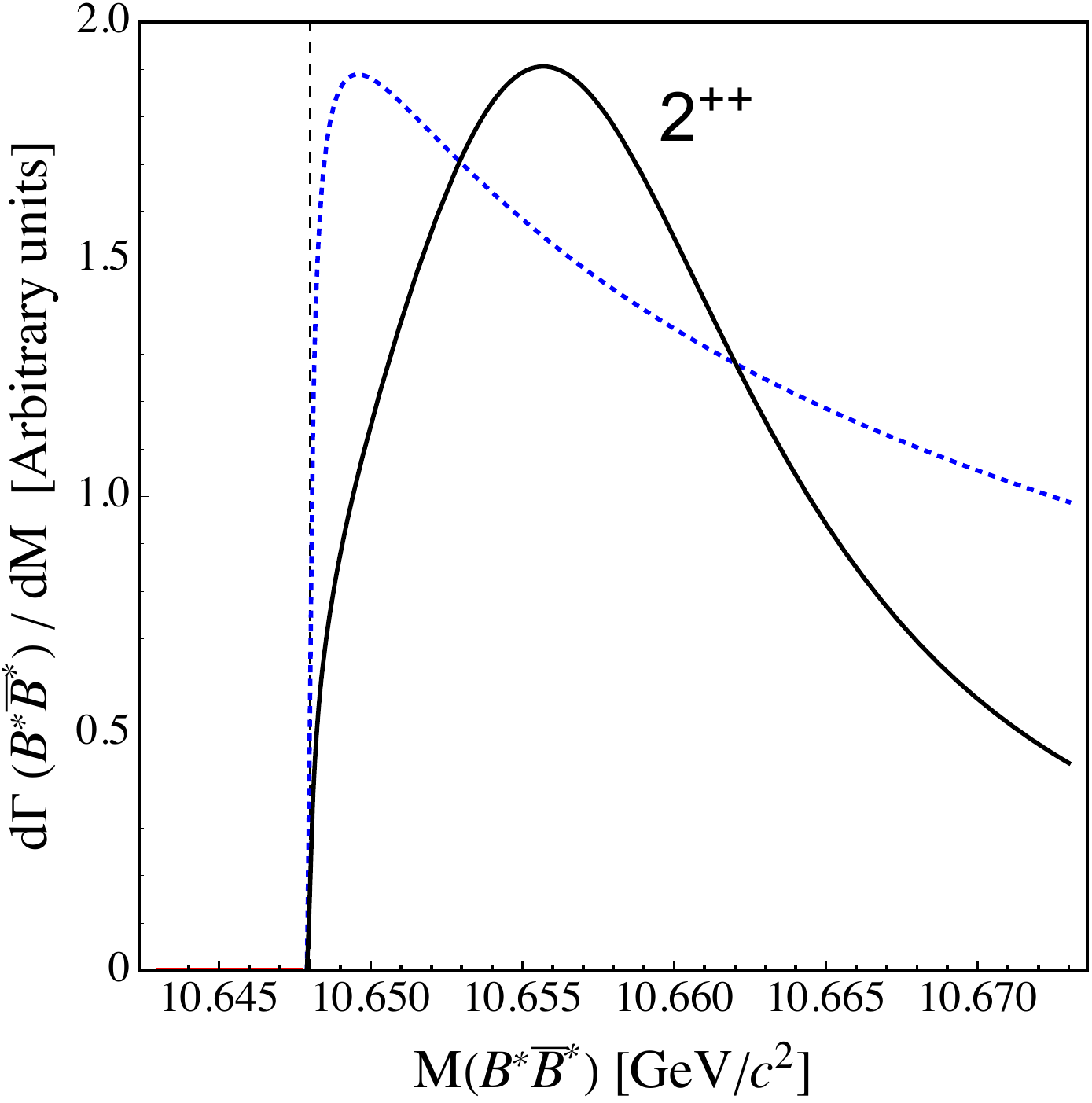, width=0.4\textwidth}\hspace{1.4 cm}
\epsfig{file=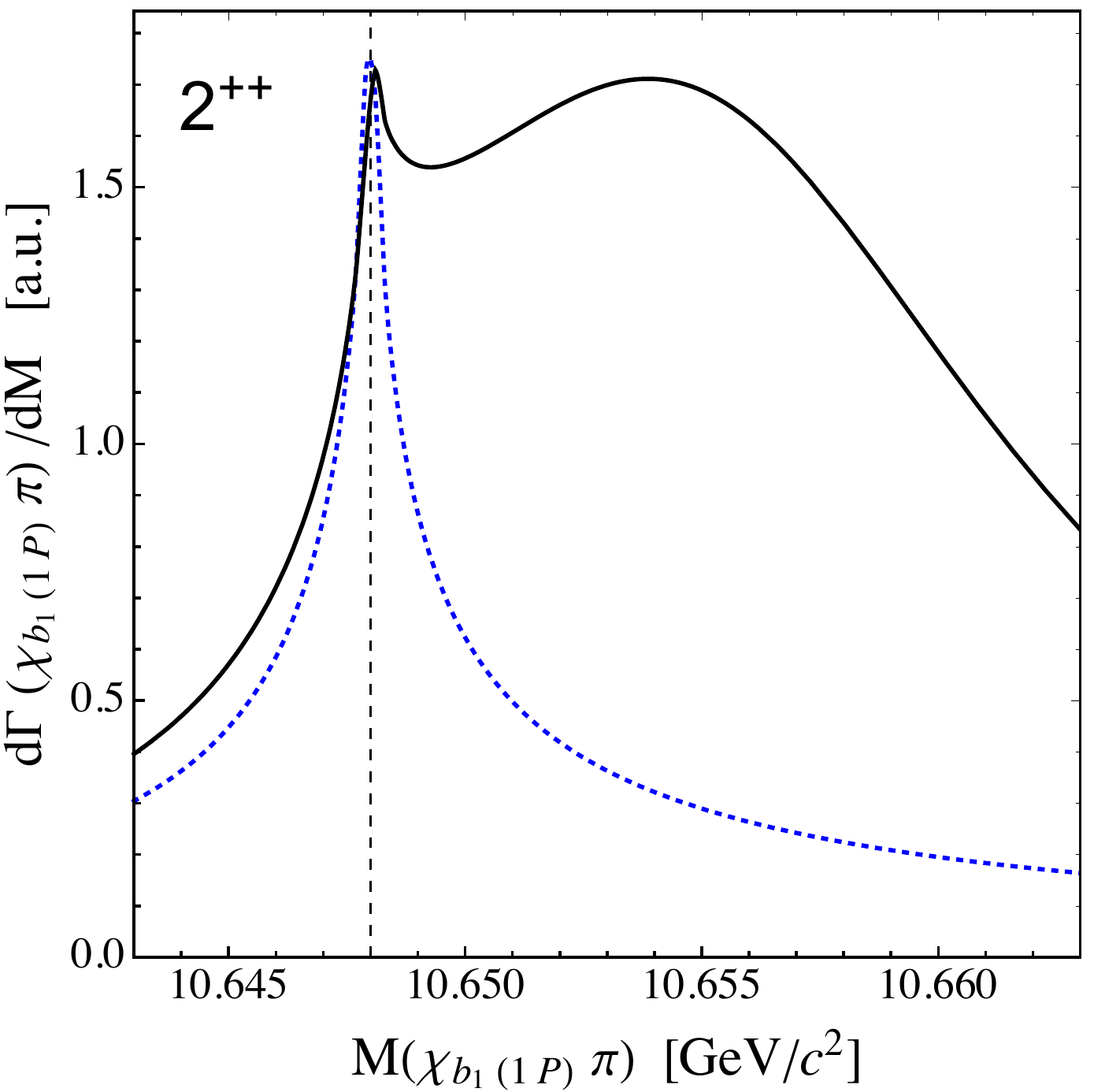, width=0.4\textwidth}
\caption{Comparison of the line shapes in the $2^{++}$  $B^*\bar B^*$
  (left) and $\chi_{b1}(1P)\pi$ (right) channels for the Contact fit (blue dotted curve) and Pionful fit 2 (black solid curve). 
The line shapes for the Contact fit for any $J$ are alike and reveal a near-threshold behaviour typical for a virtual state. 
}\label{FitG_vs_A}
\end{center}
\end{figure*}

In Figs.~\ref{fig:0pp}-\ref{fig:2pp} the line shapes in the spin partner channels with $J=0,1,2$ are shown for the two pionful fit schemes introduced above. 
For each scheme the line shapes are calculated employing the best-fit parameters extracted from the analysis of the data in the $1^{+-}$ channel for the cutoff 1 GeV.
Specifically, in each plot we present the relevant elastic $B^{(*)}\bar B^{(*)}$ and inelastic differential widths (in arbitrary units) defined by \eq{eq:gamma}. The relative normalisation of the 
curves 
for the Pionful fits 1 and 2 is chosen such that the two curves have the same magnitude at the resonance peak. In the case of the $0^{++}$ channel, where two states are present, the 
curves are normalised to have the same strength in one of the peaks. While the overall scale of the line shapes is not a subject of the current investigation, 
as discussed in Sec.~\ref{sec:prodvert}, the branching fractions defined relative to the total width for each $J$ are predicted here --- see Table ~\ref{tab:fractions} and  
discussions below.
For the inelastic channels, as examples, we show the distributions in the $\chi_{b1}(1P)\pi$ and $\eta_{b0}(1S)\pi$ final states.
The energy behaviour of the line shapes in the other channels not shown here ($\chi_{bJ}(mP)\pi$ with $m=1,2$ and $\eta_{b0}(2S)\pi$) is completely analogous 
to that for the $\chi_{b1}(1P)\pi$ and $\eta_{b0}(1S)\pi$ channels, respectively, while their relative scales can be read off from Table~\ref{tab:fractions}.
The differential rates for the pionful fits exhibit either 
a well-pronounced hump above the relevant threshold, as seen in 
Figs.~\ref{fig:1pp} and \ref{fig:2pp} for the $1^{++}$ and $2^{++}$ channels, 
or a sizeable near-threshold distortion, as in the $0^{++}$ case --- see Fig.~\ref{fig:0pp}. This picture is typical for a resonance which is supported by the position of the 
poles of the amplitude in the energy complex plane --- see the discussion in Sec.~\ref{sec:poles} below. 

In Fig.~\ref{FitG_vs_A}, for illustrative purposes, we compare the line shapes in the $2^{++}$ channel for the Contact and Pionful fit 2 schemes. For all $J$'s,
the inelastic line shapes corresponding to the Contact fit reveal only a cusp-like structure at the relevant elastic threshold enhanced by the presence of a near-threshold pole --- the 
behaviour typical for a virtual state scenario. On the contrary, 
when the pions supplemented by the $O(p^2)$ contact terms are included, the poles move to the complex plane, as 
will be discussed in the next section, resulting in
above-threshold resonance-type structures in the line shapes. 

The partial widths $\Gamma$ in all considered elastic and inelastic channels can be obtained as integrals over the entire relevant energy interval. 
The ratios of the individual partial widths to the sum of all contributions for a given $J$ are shown in Table~\ref{tab:fractions}.
Such ratios do not depend on the overall scale and, therefore, can be regarded as a parameter-free prediction of our approach. 
In particular, for the elastic widths one finds the relations
\bea
&\Gamma^{1^{++}}_{B\bar{B^*}({}^3 S_1)}:\Gamma^{2^{++}}_{B^*\bar{B^*}({}^5 S_2)}:\Gamma^{0^{++}}_{B\bar{B}({}^1 S_0)}:\Gamma^{0^{++}}_{B^*\bar{B^*}({}^1 S_0)}&\nonumber\\[-2mm]
\label{GammaX1}\\[-2mm]
&\approx 15:12:5:1,&\nonumber
\eea
\bea\label{GammaXD}
&\Gamma^{2^{++}}_{B\bar{B}({}^1 D_2)}:\Gamma^{2^{++}}_{B\bar{B}^*({}^3 D_2)}:\Gamma^{0^{++}}_{B^*\bar{B^*}({}^1 S_0)}\approx 3:3:2.&
\eea
Although the potential spin symmetry violation in the elastic source terms discussed in Ref.~\cite{Mehen:2013mva} can somewhat distort these results, the general pattern should persist.

Let us summarise the findings we arrived at.
\begin{itemize}
\item As expected in the molecular scenario, for each $W_{bJ}$ state, the decay rate to the corresponding elastic channel with the nearest threshold is the largest while
the inelastic channels are strongly suppressed compared with it. The
decay rates to remote elastic channels are also suppressed. For
example, the contribution of the $W_{b0}'$ state to the $B \bar B$ rate
is quite marginal, as can be seen in Fig.~\ref{fig:0pp}.
This is a direct consequence of the properties demonstrated by the data 
in the $1^{+-}$ channel --- see Fig.~\ref{fig:1pm}: although the coupled-channel dynamics allows for such transitions, the data do not favour them.
\item The largest rates correspond to the $\Upsilon(10860)$ decays to the $\gamma B\bar B^*$ and $\gamma B^*\bar B^*$ channels via the $W_{b1}$ and $W_{b2}$ partners, respectively --- see 
Eq.~(\ref{GammaX1}).
\item The ratios predicted in Eq.~(\ref{GammaX1}) from the measured line shapes of the $Z_b$ states are consistent with the estimates presented in a recent study 
\cite{Voloshin:2018pqn}. 
\end{itemize}

As to the absolute scale of the rates $\Upsilon(10860)\to\gamma W_{bJ}\to \gamma B^{(*)}\bar{B}^{(*)}$, a two order of magnitude suppression is trivially expected as compared with the rates 
$\Upsilon(10860)\to\pi 
Z_b^{(\prime)}\to \pi B^{(*)}\bar{B}^{*}$ because of the standard fine structure penalty for electromagnetic processes, that also agrees with the estimates made in 
Ref.~\cite{Voloshin:2018pqn}. Meanwhile, this suppression is expected to be overcome by the Belle-II experiment due to its large luminosity and, as a result, an almost 
two-order-of-magnitude 
increase 
of the statistics as compared with the previous-generation experiment Belle.

\section{The pole positions of the $Z_b$, $Z_b'$ and $W_{bJ}$ states}
\label{sec:poles}

\subsection{Extracting the poles in a multichannel scattering problem}
\label{sec:omega}

In this section we employ the approach developed above to predict in a parameter-free way the pole positions for the spin partners $W_{bJ}$ ($J=0,1,2$) with the quantum numbers $J^{++}$. 
For the pole search in the complex energy plane we follow the approach of Refs.~\cite{Guo:2016bjq,Wang:2018jlv} and stick to the four-sheet Riemann surface corresponding to two elastic 
channels --- either the $B\bar B$ and $B^* \bar B^*$ channels in case of $J^{PC}=0^{++}$ or the $B\bar B^*$ and $B^* \bar B^*$ ones for all other quantum numbers. All inelastic 
thresholds are remote and their impact on the poles of interest, which are located near the elastic thresholds, is minor --- see the discussion in Ref.~\cite{Wang:2018jlv}. 
Then, for two coupled channels with the thresholds split by the mass difference $\Delta$,
the four-sheeted Riemann surface can be mapped onto a single-sheeted plane of a new variable, which is traditionally denoted as
$\omega$ \cite{kato,Badalian:1981xj}, via the relations\footnote{The situation in the $2^{++}$ channel is more complicated since all three elastic channels are coupled.
However, also in this case, it appears to be convenient to employ the mapping onto the $\omega$-plane for 
the channels $B\bar B^*$ and $B^* \bar B^*$ while the $B\bar B$ channel is treated explicitly.}
\be
\ds k_1=\sqrt{\frac{\mu_1\Delta}2}\left(\omega+\frac1{\omega}\right),\quad
k_2=\sqrt{\frac{\mu_2\Delta}2}\left(\omega-\frac1{\omega}\right). 
\label{omegak1k2}
\ee
Then, the energy defined relative to the lowest threshold of the two, $E=M-m^{(1)}_{1}-m^{(1)}_{2}=M-m^{(2)}_{1}-m^{(2)}_{2}+\Delta$, reads 
\bea
&\ds E=\frac{k_1^2}{2\mu_1}=\frac{k_2^2}{2\mu_2}+\Delta=\frac{\Delta}4\left(\omega^2+\frac1{\omega^2}+2\right),&\nonumber
\eea
where $\mu_1$ and $\mu_2$ are the reduced masses in the first and second elastic channels labelled as (1) and (2), respectively. Specifically, in the $0^{++}$ channel, 
\be
\Delta=2\delta, \quad m^{(1)}_{1}=m^{(1)}_{2}=m, \quad m^{(2)}_{1}=m^{(2)}_{2}=m_* ,
\ee
while in the channels $1^{++}$  and  $2^{++}$ one has
\be
\Delta=\delta, \quad m^{(1)}_{1}=m,\quad m^{(1)}_{2}=m^{(2)}_{1}=m^{(2)}_{2}=m_*.
\label{Deqd}
\ee
Then the one-to-one correspondence between the four Riemann sheets in the $E$-plane 
(denoted as RS-N, where N=I, II, III, IV) and various regions in the $\omega$-plane reads
\begin{eqnarray*}\label{sheets_two_ch}
\mbox{RS-I} (++):&&\quad{\rm Im}~k_1>0,\quad{\rm Im}~k_2>0,\\
\mbox{RS-II}(-+):&&\quad{\rm Im}~k_1<0,\quad{\rm Im}~k_2>0,\\
\mbox{RS-III}(--):&&\quad{\rm Im}~k_1<0,\quad{\rm Im}~k_2<0,\\
\mbox{RS-IV}(+-):&&\quad {\rm Im}~k_1>0,\quad{\rm Im}~k_2<0,
\end{eqnarray*}
where the signs in the parentheses correspond to the signs of the imaginary parts
of the momenta $k_1$ and $k_2$.
These regions in the $\omega$-plane are depicted 
in Fig.~\ref{fig:omega}.
The thick solid line corresponds to the real values of the energy lying on (physical) RS-I.
It is easy to see that the physical region between the two thresholds 
corresponds to
$|\omega| = 1$, with both Re$(\omega)$ and Im$(\omega)$ positive,
and the thresholds at $E=0$ and $E=\Delta$ are mapped to the points $\omega=\pm i$ and 
$\omega=\pm 1$, respectively. 

\begin{figure}
\epsfig{file=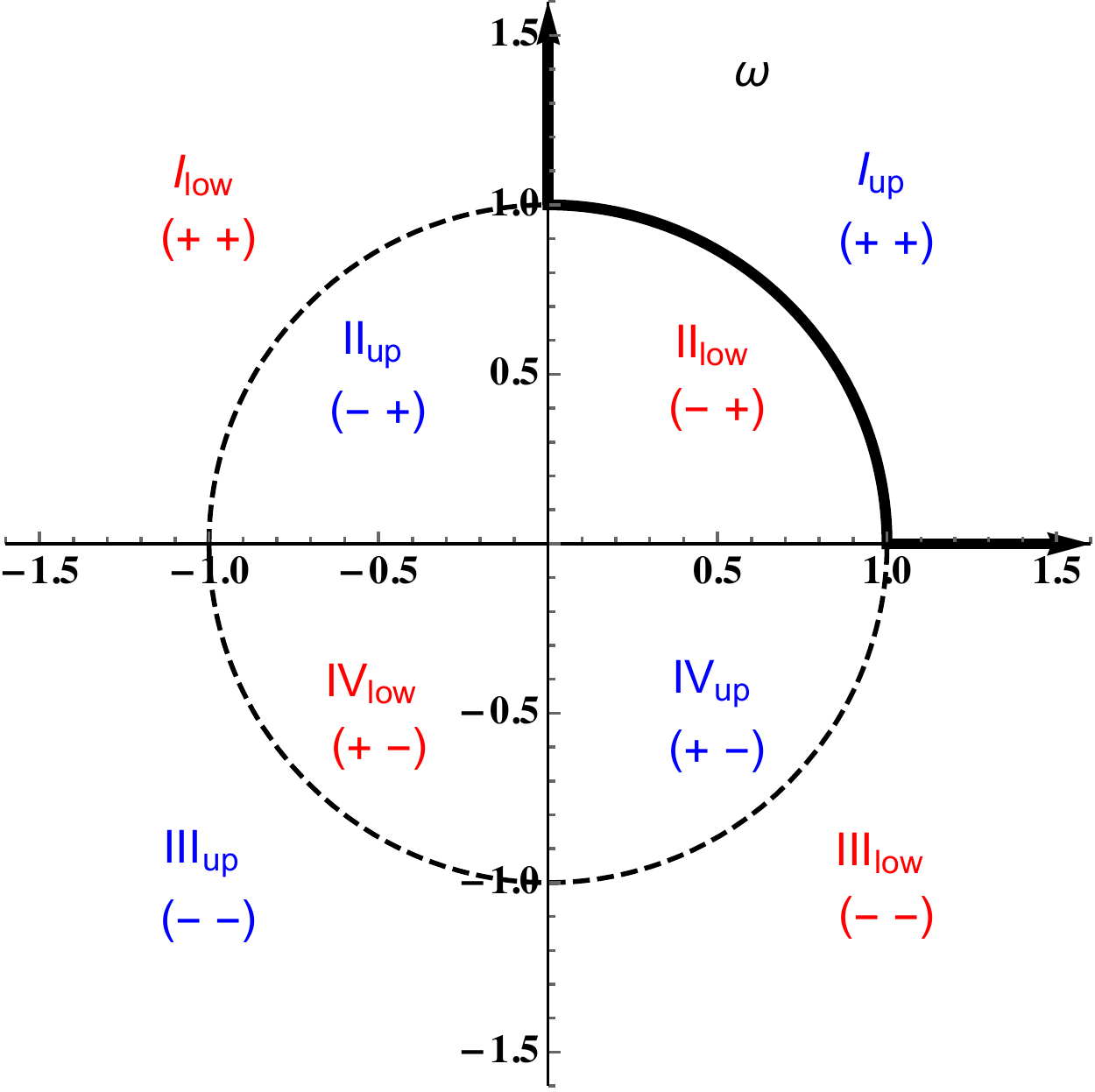, width=0.35\textwidth}
\caption{The unitary-cut-free complex $\omega$-plane for the two elastic channels
obtained from the four-Riemann-sheeted complex energy plane by the conformal transformation (\ref{omegak1k2}). 
The eight regions separated by the unit circle and by the two axes correspond to the upper and lower half-planes (see the subscripts 'up' and 'low') 
in the four Riemann sheets of the energy plane denoted as RS-N with N=I,II,III,IV \cite{kato,Dudek:2016cru}. The bold line indicates the physical 
region of a real energy $E$ \cite{kato}. \label{fig:omega}}
\end{figure}

\begin{figure*}[t]
\begin{center}
\epsfig{file=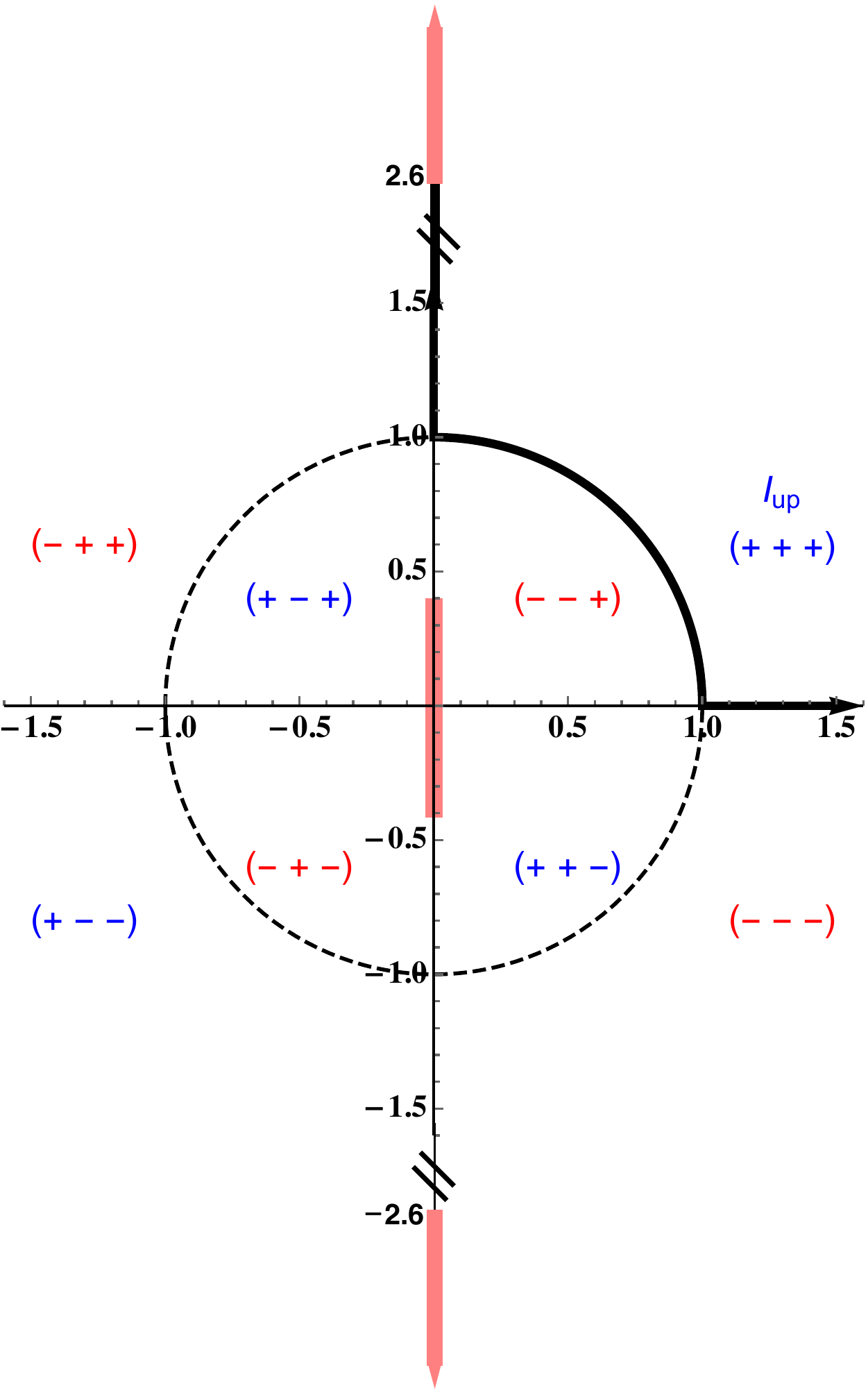, width=0.35\textwidth}\hspace{1cm}
\epsfig{file=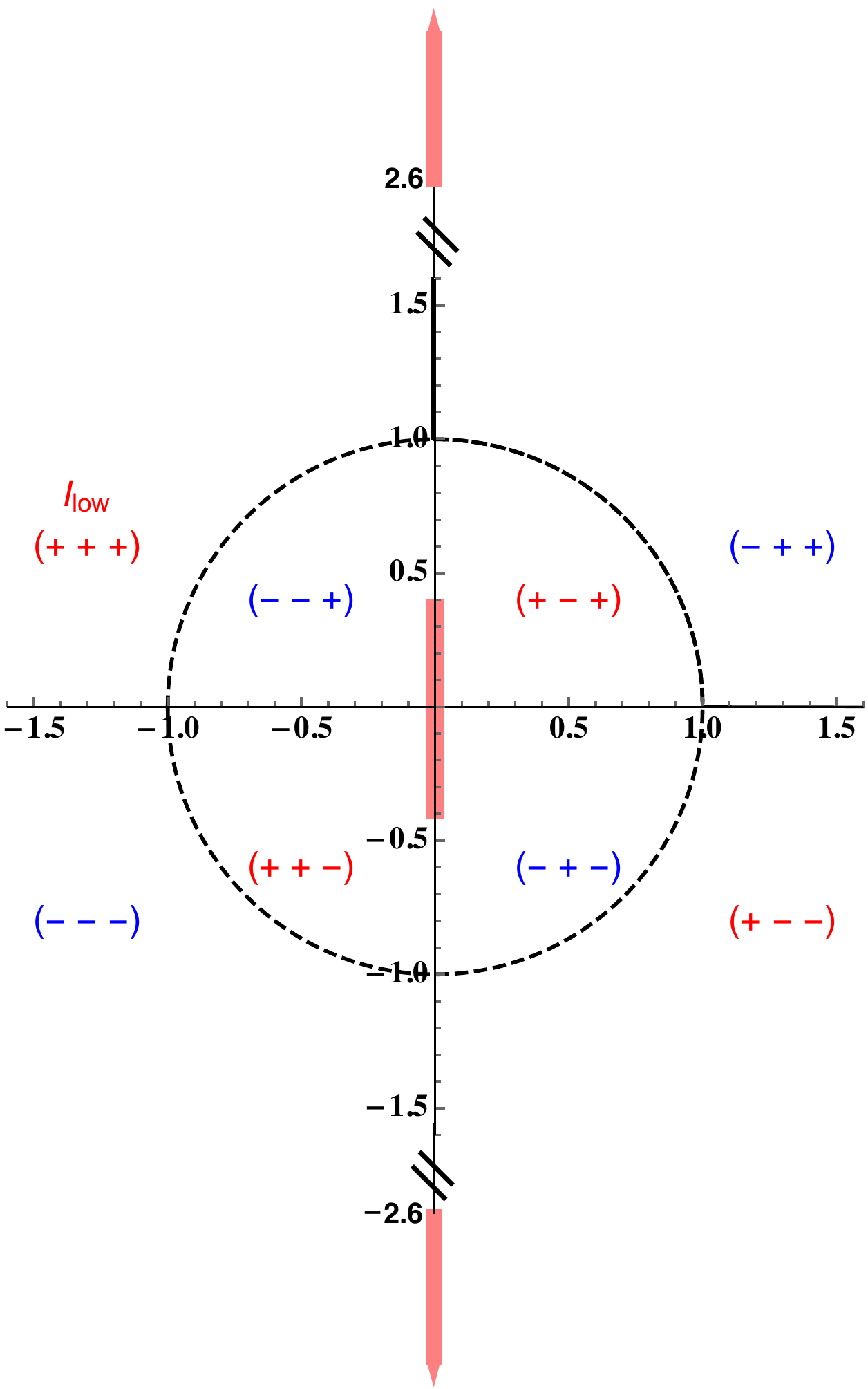, width=0.35\textwidth}
\caption{Generalisation of the two-channel complex $\omega$-plane to the 
three-channel case with two elastic and one inelastic channel. 
Left panel: the $\omega$-plane sheet closest to the physical region of a real energy indicated by the black bold line. In this $\omega$-plane,
the inelastic momentum for the particles with the masses $m_{\rm in1}$ and $m_{\rm in2}$
is related to the energy as $p_{\rm in} =\frac{1}{2M}\lambda^{1/2}(M^2,m_{\rm in1}^2,m_{\rm in2}^2) $.
Right panel: the $\omega$-plane sheet distant from the physical region. Here the inelastic momentum 
is related to the energy as $p_{\rm in} =-\frac{1}{2M}\lambda^{1/2}(M^2,m_{\rm in1}^2,m_{\rm in2}^2) $.
Transitions from one omega plane to the other are possible through the right-hand cut from the inelastic channels denoted by
the pink fat line in both panels.
\label{fig:two_omega}}
\end{center}
\end{figure*}

The nomenclature of the Riemann sheets as described above is relevant for a two-channel situation while in the presence of additional channels it needs to be generalised.
As an illustration, 
consider a three-channel case with two elastic and one inelastic channel. In this case, the three-channel complex omega plane can be schematically viewed as a 
two-sheeted $\omega$-plane with the sheets connected by analyticity through the inelastic cut --- see Fig.~\ref{fig:two_omega} where, in line with the two-channel case above, the sheets are labelled 
by the signs of 
the imaginary parts of 
the momenta in each channel. 

Clearly, 
not all poles found on all Riemann sheets are of a physical significance. Specifically, only those poles which have a short (compared with the thresholds splitting $\Delta$)
path to the real energies on the physical sheet RS-I (labelled as $I_{\rm up}$ $(+++)$ 
in the left $\omega$-plane --- see Fig.~\ref{fig:two_omega}) 
can leave their imprint on observables.
It is easy to see that the poles residing in the right
$\omega$-plane do not meet this 
criterion and, therefore, can be safely disregarded. 
For example, in order to reach
$I_{\rm up}$ $(+++)$ starting from the lower domain of the RS-I 
($I_{\rm low}$ $(+++)$ in the right
$\omega$-plane --- see Fig.~\ref{fig:two_omega})
one would need to travel a long way around the inelastic branch point.
Among the poles residing in the left
$\omega$-plane those on the sheets $(--+)$ and $( --- )$ are the most important ones since they are the closest to the physical region of the real energy 
(the fat black line in Fig.~\ref{fig:two_omega}). A pole on the sheet $(--+)$ located in the vicinity of the lower elastic threshold (the point $\omega=i$) will result in a significant 
near-threshold distortion of the line shapes while if the pole is located deeper in the complex plane (but still on the same sheet) it will show up as a clear resonance peak. 
The poles residing on the sheet $( --- )$ will manifest themselves in the observables in exactly the same manner but with respect to the upper threshold. 
Next in importance are the poles residing on the sheets $(+-+)$ and $(++-)$. 
They cannot generate resonance humps in the 
line shapes above threshold but can significantly enhance the cusp-like structure 
at threshold provided these poles reside not too deep in the complex plane. 

It is also important to notice that imposing constraints from unitarity and analyticity 
on the scattering amplitude $T$ requires that
\bea
T (k_1, k_2, k_3) = T^*(-k_1^*, -k_2^*, -k_3^*).
\eea
This implies that if there is a pole at $\omega = \omega_0$ in the 
left (main) omega plane there must be also a pole at $- \omega_0^*$ in the 
right omega plane --- see Fig.~\ref{fig:two_omega}. 
This mirror pole, however, has no physical significance, as already explained. 

Generalisation of the logic discussed above to a larger number of channels is straightforward if one bears in mind that, 
as before, only one $\omega$-plane sheet containing the domain of the physical energies on RS-I is relevant and that the signs of the imaginary parts for all remote inelastic channels coincide.

For the case at hand, we arrive effectively at a 3-channel (two elastic plus an effective inelastic) problem for the quantum numbers $1^{+-}, 1^{++}$ and $0^{++}$
while for $2^{++}$, where all three elastic channels are present, the effective problem contains 4 channels.

\subsection{Poles of the $Z_b$'s and their spin partners}
\label{sec:poles_Zb_Wb}

\begin{table*}[t!]
\caption{The pole positions and the residues $g^2$ (see the definition in Eq.~(\ref{resdef})) in various $S$-wave $B^{(*)}\bar B^{(*)}$ channels for the Contact fit. 
The energy $E_{\rm pole}$ is given relative to the nearest open-bottom threshold quoted in the third column, so that it is one of the energies $E_n$ ($n=1,2,3$) defined in Eq.~(\ref{sqrts}). 
The Riemann Sheet (RS) is defined by the signs of the imaginary parts of the corresponding momenta (quoted in the columns 4-7); 
a missing sign indicates that this channel is uncoupled. Uncertainties correspond to a $1\sigma$ 
deviation in the parameters allowed by the fit to the data in the channels with $J^{PC}=1^{+-}$ where the $Z_b^{(\prime)}$ states reside \cite{Wang:2018jlv}. For the estimate of the theoretical 
uncertainties see chapter \ref{theoruncert}. The poles are 
calculated for the cutoff $\Lambda=1$ GeV. 
}\label{tab:Contact}
\begin{ruledtabular}
\begin{tabular}{lllcccccc}
$J^{PC}$& State & Threshold &Im~$p_{\rm in}$ & Im~$p_{B\bar B}$ & Im~$p_{B\bar{B^*}}$ & Im~$p_{B^*\bar{B}^*}$ &$E_{\rm pole}$ w.r.t. threshold [MeV] & Residue at $E_{\rm pole}$ \\
\hline
$1^{+-}$& $Z_b$ & $B\bar{B}^*$ &$+$ & & $-$ & $+$ &$(-0.9\pm 0.4)+i(1.0\pm 0.3)$ & $ (-1.4\pm 0.2)+i(0.5\pm 0.1)$\\
$1^{+-}$& $Z_b'$ & $B^*\bar{B}^*$ &$+$ & & $+$ & $-$ &$(-0.8\pm 0.5)+i(1.3\pm 0.4)$ & $(-1.4\pm 0.3)+i(0.7\pm0.1)$\\
$0^{++}$& $W_{b0}$ & $B\bar{B}$ &$+$ & $-$ & & $+$ &$(-1.0\pm 0.6)+i(1.0\pm 0.3)$ & $(-1.4\pm 0.3)+i(0.5\pm 0.1)$\\
$0^{++}$& $W_{b0}'$& $B^*\bar{B}^*$ &$+$ & $+$ & & $-$ &$(-1.2\pm 0.6)+i(0.9\pm 0.3)$ & $(-1.4\pm 0.3)+i(0.4\pm 0.1)$\\
$1^{++}$& $W_{b1}$ & $B\bar{B}^*$ &$+$ & & $-$ & &$(-0.3\pm 0.6)+i(1.6\pm 0.8)$ & $(-1.3\pm 0.4)+i(0.9\pm 0.1)$\\
$2^{++}$& $W_{b2}$ & $B^*\bar{B}^*$ &$+$ & & &$-$ &$(0.4\pm 0.6)+i(1.9\pm 0.9 )$ & $(-1.2\pm 0.4)+i(1.3\pm 0.2 )$
\end{tabular}
\end{ruledtabular}
\caption{The same as in Table~\ref{tab:Contact} but for the Pionful fit 1.
}\label{tab:pf1}
\begin{ruledtabular}
\begin{tabular}{lllcccccc}
$J^{PC}$& State & Threshold &Im~$p_{\rm in}$ & Im~$p_{B\bar B}$ & Im~$p_{B\bar{B^*}}$ & Im~$p_{B^*\bar{B}^*}$ &$E_{\rm pole}$ w.r.t. threshold [MeV] & Residue at $E_{\rm pole}$ \\
\hline
$1^{+-}$& $Z_b$ & $B\bar{B}^*$ &$-$ & & $-$ & $+$ &$(-1.3\pm 0.2)-i(0.6\pm 0.1)$ & $(-0.6\pm 0.1)-i(0.1\pm 0.1) $ \\
$1^{+-}$& $Z_b'$ & $B^*\bar{B}^*$ &$-$ & & $-$ & $-$ &$(2.1\pm 2.2)-i(12.9\pm 2.4)$ & $(0.8\pm 0.1)-i(0.4 \pm 0.2)$ \\
$0^{++}$& $W_{b0}$ & $B\bar{B}$ &$+$ & $-$ & & $+$ & $(-8.5 \pm 2.8) + i (1.5 \pm 0.2)$ & $(-2.0 \pm 0.7) - i (0.1 \pm 0.3)$ \\
$0^{++}$& $W_{b0}'$& $B^*\bar{B}^*$ &$-$ & $-$ & & $-$ & $(-1.2\pm 0.1) - i (0.7 \pm 0.3)$ & $(-0.4 \pm 0.1) - i (0.2 \pm 0.1)$ \\ 
$1^{++}$& $W_{b1}$ & $B\bar{B}^*$ &$-$ & & $-$ & $+$ & $(25.0 \pm 2.6) - i (20.5 \pm 3.3)$ &$(0.9 \pm 0.1) - i (0.4 \pm 0.2)$\\ 
$2^{++}$& $W_{b2}$ & $B^*\bar{B}^*$ &$-$ & $-$ & $-$ &$-$ & $(4.0 \pm 2.1) - i (10.4 \pm 1.5)$ & $(0.4 \pm 0.1) - i (0.2 \pm 0.1)$ 
\end{tabular}
\end{ruledtabular}
\caption{The same as in Table~\ref{tab:Contact} but for the Pionful fit 2.
}\label{tab:pf2}
\begin{ruledtabular}
\begin{tabular}{lllcccccc}
$J^{PC}$& State & Threshold &Im~$p_{\rm in}$ & Im~$p_{B\bar B}$ & Im~$p_{B\bar{B^*}}$ & Im~$p_{B^*\bar{B}^*}$ &$E_{\rm pole}$ w.r.t. threshold [MeV] & Residue at $E_{\rm pole}$ \\
\hline
$1^{+-}$& $Z_b$ & $B\bar{B}^*$ &$-$ & & $-$ & $+$ & $(-2.3\pm 0.5)-i(1.1\pm 0.1)$ & $ (-1.2\pm 0.2)+i(0.3\pm 0.2) $\\ 
$1^{+-}$& $Z_b'$ & $B^*\bar{B}^*$ &$-$ & & $-$ & $-$ &$(1.8 \pm 2.0) - i (13.6 \pm 3.1)$ & $(1.5 \pm 0.2) - i (0.6 \pm 0.3)$\\ 
$0^{++}$& $W_{b0}$ & $B\bar{B}$ &$-$ & $-$ & & $+$ & $(2.3 \pm 4.2) - i (16.0 \pm 2.6)$ & $(1.7 \pm 0.6) - i (1.7 \pm 0.5)$ \\
$0^{++}$& $W_{b0}'$& $B^*\bar{B}^*$ &$-$ & $-$ & & $-$ & $(-1.3 \pm 0.4) - i (1.7 \pm 0.5)$ & $(-0.9 \pm 0.3) - i (0.3 \pm 0.2)$\\ 
$1^{++}$& $W_{b1}$ & $B\bar{B}^*$ &$-$ & & $-$ & $+$ & $(10.2 \pm 2.5) - i (15.3 \pm 3.2)$ & $(1.3 \pm 0.2) - i (0.4 \pm 0.2)$\\
$2^{++}$& $W_{b2}$ & $B^*\bar{B}^*$ &$-$ & $-$ & $-$ &$-$ & $(7.4 \pm 2.8) - i (9.9 \pm 2.2)$ & $(0.7 \pm 0.1) - i (0.3 \pm 0.1)$
\end{tabular}
\end{ruledtabular}
\end{table*}

In the vicinity of a pole located at $M=M_{R_\alpha}$
the elastic scattering amplitude $T_{\alpha\alpha}(M,p,p')$ given in \eq{Eq:JPC} takes the form
\bea
T_{\alpha\alpha}= \frac{g_\alpha^2}{M^2-M_{R_\alpha}^2}\approx \frac{g_\alpha^2}{2M_{R_\alpha}}\frac{1}{M-M_{R_\alpha}},
\label{resdef}
\eea
where the energy $M$ is defined in \eq{sqrts} and $g_\alpha^2$ and $M_{R_\alpha}$ stand for the residue and the pole position in the channel $\alpha$, respectively. 

The most relevant poles for the case of the quantum numbers $J^{PC}=1^{+-}$ are collected in Tables~\ref{tab:Contact}-\ref{tab:pf2}, together with the residues at these poles. 
As was explained in detail above, we regard a pole as relevant if it has a short (compared with the splitting between the nearest elastic thresholds $\delta$) path to the physical RS-I and as such 
affects the form of the line shapes.
In the pionful fits, the poles representing the $Z_b(10610)$ and $Z_b(10650)$ states inhabit the sheets $(--+)$ and $(---)$, respectively, and, according to the logic discussed above, 
reveal themselves in the line shapes as peaks above thresholds.

Also, in Tables~\ref{tab:Contact}-\ref{tab:pf2}, we present the relevant poles (counted relative to the nearby elastic thresholds) 
and the corresponding residues predicted for the spin partners in the $0^{++}$, $1^{++}$ and $2^{++}$ channels. 
For the Pionful fit 2 regarded here as the most reliable calculation, the poles in all channels reside on the sheets closest to the upper domain of the physical RS-I. 
The poles for the $Z_b$ and $W_{b0}'$ are located just in the vicinity of the $B\bar B^*$ and $B^*\bar B^*$ 
threshold, respectively, and show up as near-threshold distortions in the line shapes (see the black solid lines in Figs.~\ref{fig:1pm} and \ref{fig:0pp}).
Meanwhile, the poles representing the other states are shifted from the respective thresholds to the complex plane by about 10--15 MeV 
and manifest themselves as humps ($Z_b'$ and $W_{b0}$) or pronounced above-threshold peaks ($W_{b1}$ and $W_{b2}$) (see the black solid lines in Figs.~\ref{fig:0pp}--\ref{fig:2pp}).
The shift of the pole positions in the Pionful fit 2 as compared with the Contact fit, where all poles correspond to virtual states, 
appears mainly due to the pion dynamics --- this effect is fully in line with the findings of Ref.~\cite{Baru:2017gwo}.

\subsection{Uncertainty estimate}
\label{theoruncert}

Uncertainties of the poles and residues given in Tables~\ref{tab:Contact}-\ref{tab:pf2} correspond to a 
$1\sigma$ deviation in the parameters of fit from the central values shown in 
Table~\ref{tab:par}. The source of this uncertainty is the experimental errors in the data.

As for the theoretical uncertainty, it can be estimated as the maximum of the two errors 
from the truncation of the EFT expansion at a given order (for the discussion of the truncation errors in the NN sector see, for example, Ref.~\cite{Epelbaum:2014efa}) and from the cutoff 
variation. 
To explain the truncation error method, let us introduce an observable quantity $X^{(\nu)} (Q)$ calculated to a given order $\nu$ in the EFT expansion in the momentum $Q$,
\bea\label{EFTexp}
X^{(\nu)} (Q) = \sum_{n=0}^\nu \alpha_n \, \chi^n, \quad \chi = \frac{Q}{\Lambda_h},
\eea
where $Q \sim p_{\rm typ} = 0.5$ GeV and $\Lambda_h \sim 4 \pi f_{\pi} \simeq 1$ GeV with $f_{\pi}$ denoting the pion decay constant; $\{\alpha_n\}$ are the expansion coefficients with $\alpha_1=0$ 
since there are no operators at the order $Q$. 

Then, assuming that the expansion \eqref{EFTexp} converges, the error at the given order $\nu$ is expected to come from the first neglected chiral order, that is, it should
scale as $\chi^{\nu +1}$ unless the coefficient $ \alpha^{\nu +1}$
vanishes. In the latter case, the uncertainty is estimated based on
the nonvanishing result at the order $\chi^{\nu +2}$, and so on. 
For example, the observable at LO ($\nu=0$) and its truncation error read
\bea\nonumber
X^{(0)} (Q) &=& \alpha_0, \\
\Delta X^{(0)} (Q) &=&X^{(2)} (Q) - X^{(0)} (Q) = \alpha_2 \, \chi^2, 
\eea
where we used that $\alpha_1=0$.

Although the poles are not observed directly, they manifest themselves in observable quantities such as line shapes, 
and thus the truncation error method is expected to work for them too. 
To estimate the truncation error at LO ($\nu=0$), we compare the results calculated explicitly at the orders $\nu=0$ (Pionful fit 1) and $\nu=2$ (Pionful fit 2) to find that the truncation error 
for the $W_{b2}$ does not exceed 5 MeV while it is only about 1 MeV for the near-threshold state $W_{b0}'$ as well as for the $Z_b$ and $Z_b'$ states. 
On the other hand, the truncation error at LO for the states $W_{b1}$ and $W_{b0}$ is of the order of 15 MeV which does not look unnatural either given the large 
expansion parameter of the pionful EFT. 
It also needs to be emphasised that the chiral expansion for the 
$W_{b0}$ state might converge slower than expected in this work since this state resides near the $B\bar B$ threshold that lies by $\delta =45$ MeV lower than the energy region used in the fits for 
the $Z_b$'s. 
It remains to be seen how the pole position for this state is affected by the inclusion of higher-order interactions. 

The truncation error method becomes particularly useful when the results at least at several chiral orders are calculated explicitly, which is not yet feasible. 
Still, from the results presented above one can conclude that the partner states $W_{b2}$ and $W_{b0}'$, both residing near the $B^*\bar B^*$ threshold, indicate a very good stability with respect to 
the inclusion 
of higher-order interactions. Since the truncation error estimate for the NLO results (Pionful fit 2) is not possible at present (it would require a complete N$^2$LO calculation) we rely on 
naturalness to 
provide a rough estimate of this uncertainty which might be especially useful for the $W_{b1}$ and $W_{b0}$ states. 
Specifically, to be more conservative we pick the maximal value from the poles given in Tables~\ref{tab:Contact}--\ref{tab:pf2} at NLO and multiply it by the expansion 
parameter to get
\bea
15\ {\rm MeV} \cdot \chi \simeq 7.5{\rm MeV}
\eea
This estimate gives roughly a twice larger uncertainty for the Pionful fit 2 than the cutoff variation. 

\section{Summary}
\label{sec:summary}
In this paper we address the properties of the spin partners $W_{bJ}$ with the quantum numbers $J^{++}$ ($J=0,1,2$) of the bottomonium-like states $Z_b(10610)$ and $Z_b(10650)$. We employ the EFT 
approach developed previously in Ref.~\cite{Wang:2018jlv} and fix all unknown low-energy constants and couplings from the data on the line shapes in the elastic and inelastic channels for the 
negative $C$-parity 
states $Z_b$ and $Z_b'$. After that, the same EFT approach consistent with requirements from unitarity, analyticity and HQSS is employed to predict in a parameter-free way the line shapes of the 
positive $C$-parity spin partner states $W_{bJ}$ in the corresponding elastic ($B^{(*)}\bar{B}^{(*)}$) and inelastic ($\eta_b(nS)\pi$ and $\chi_{bJ}(mP)\pi$) channels. 

Because of the 
positive $C$-parity the $W_{bJ}$'s should be produced in the radiative
decays of the vector bottomonium $\Upsilon(10860)$. It is argued that
the production operator which involves 
the tree-level and one-loop contributions 
behaves as a smooth function of the energy in the near-threshold
region of interest here. Therefore, in agreement with the Watson's theorem,
the energy dependence of the 
line shapes can be predicted based on the strong interaction 
amplitudes between the heavy mesons in the final state. These amplitudes contain the poles in the vicinity of the thresholds which are associated with the excitation of the $W_{bJ}$ states.
In addition, the ratios of the partial branchings to all aforementioned elastic and inelastic decay channels of the $W_{bJ}$ partners come as predictions 
of our approach, since the overall normalisation constant from the production operator drops out in these ratios in the HQSS limit. 

With a multi-channel amplitude at hand which possesses the correct analytic structure we extract the poles of the amplitude in the complex energy plane and its residues at these poles
for all four partner states and find that our most advanced pionful
analysis of the data on the $Z_b$'s (the Pionful fit 2) is consistent
with all $W_{bJ}$'s being above-threshold resonances. 
In contrast to this, in the pionless approach, all $W_{bJ}$'s appear
as virtual below-threshold states. Since these two scenarios reveal themselves differently in the line shapes (\emph{c.f.} threshold cusp versus above threshold hump in the
inelastic channels in Fig.~\ref{FitG_vs_A}), the experimental data in
various channels relevant for the $W_{bJ}$'s should provide key information on the role of the pion dynamics for
the system at hand.

The uncertainties in the pole positions are estimated and discussed in detail. The errors accounted for in this work come from (i) a statistical $1\sigma$ 
deviation in the parameters allowed by the fit to the data, (ii) truncation of the EFT expansion at a given order and (iii) the cutoff variation. Although the evaluated uncertainties appear to 
be of a natural size a better estimate of the truncation error would be very desirable. That would call for the inclusion of the two-pion exchange contributions to the elastic potentials at 
next-to-leading-order
which does not involve any new parameters.

We conclude by stating that, although the electro-magnetic fine structure penalty suppresses the probability of the $W_{bJ}$'s production in radiative decays of the $\Upsilon(10860)$ by two orders of 
magnitude compared with the 
$Z_b^{(\prime)}$ production in the
one-pion decays of the $\Upsilon(10860)$, this suppression is to be overcome by the large statistics anticipated for the Belle-II $B$-factory. We, therefore, expect the spin partners of the $Z_b$ 
states to be copiously produced in this experiment.

\begin{acknowledgments}
This work was supported in part by the DFG (Grant No. TRR110) and the NSFC (Grant No. 11621131001) through the
funds provided to the Sino-German CRC 110 ``Symmetries and the Emergence of Structure
in QCD''. Work of V.B. and A.N. was supported by the Russian Science Foundation (Grant No. 18-12-00226).
\end{acknowledgments}

\appendix

\section{Effective Lagrangians}
\label{app:Lag}

The low-energy $B^{(*)} \bar{B}^{(*)}$ scattering at leading order ${\cal O}(Q^0)$ is described by the Lagrangian \cite{Mehen:2011yh}\footnote{Because of a different convention 
for the $C$-parity transformation adopted in Ref.~\cite{Nieves:2012tt}, the signs of the
off-diagonal terms in the leading order contact terms of the potential $V^{\rm CT}_{\rm LO}[0^{++}]$ given below differ from those in the cited work. 
}
\begin{widetext}
\bea
&&{\cal L}^{(0)}_{HH}={\rm Tr}\left[H^\dagger_a \left(i \partial_0 +\frac{\vec\nabla^2}{2 \bar{M}}\right)_{ba} H_b\right]
+{\rm Tr}\left[\Hb^\dagger_a\left(i\partial_0+\frac{\vec\nabla^2}{2\bar{M}}\right)_{ab}\Hb_b\right]
+\frac{\delta}{4}{\rm Tr}[H^\dagger_a\sigma^i H_a \sigma^i]
+\frac{\delta}{4}{\rm Tr}[\Hb^\dagger_a\sigma^i\Hb_a \sigma^i]\nn \\
&&-\frac{C_{10}}{8}{\rm Tr}[\bar{H}^\dagger_a\tau^A_{aa^\prime}H^\dagger_{a^\prime}H_b\tau_{bb^\prime}^A\bar{H}_{b^\prime}] 
-\frac{C_{11}}{8}{\rm Tr}[\bar{H}^\dagger_a\tau^A_{aa^\prime}\sigma^i H^\dagger_{a^\prime}H_b\tau_{bb^\prime}^A\sigma^i\bar{H}_{b^\prime}],\label{Lag0}
\eea
\end{widetext}
where $a$ and $b$ are the isospin indices, $\sigma$'s and $\tau$'s are the spin and isospin Pauli matrices, respectively, and the trace is taken in the spin space. The 
isospin matrices are normalised as $\tau^A_{ab} \tau^B_{ba} =2\delta^{AB}$. The mass $\bar{M}$ in the kinetic terms is the spin-averaged $B$ meson mass, 
$\bar{M}=(3m_*+m)/4$, and $\delta=m_*-m\approx 45$~MeV.

The terms in the first line in Eq.~\eqref{Lag0} stand for the leading heavy and anti-heavy meson chiral perturbation theory Lagrangian of 
Refs.~\cite{Wise:1992hn,Burdman:1992gh,Yan:1992gz},
written in the two-component notation of Ref.~\cite{Hu:2005gf}. The terms proportional to the
potentials $C_{10}$ and $C_{11}$ correspond to the ${\cal O}(Q^0)$ $S$-wave contact interactions~\cite{Mehen:2011yh,AlFiky:2005jd}.
The superfields $H_a$ and $\bar{H}_a$ are defined in Eq.~(\ref{Hs}).

The effective Lagrangian at NLO derived in Ref.~\cite{Wang:2018jlv} reads
\begin{widetext}
\bea
&&{\cal L}_{HH}^{(2)}=
-\frac{D_{10}}8\left\{{\rm Tr}[\nabla^i\bar{H}^\dagger_a\tau^A_{aa^\prime}\nabla^i H^\dagger_{a^\prime}H_b\tau_{bb^\prime}^A\bar{H}_{b^\prime} 
+{\rm Tr}[\bar{H}^\dagger_a\tau^A_{aa^\prime}H^\dagger_{a^\prime}\nabla^i H_b\tau_{bb^\prime}^A\nabla^i\bar{H}_{b^\prime}]\right\} \nonumber\\
&&-\frac{D_{11}}{8}\left\{{\rm Tr}[\nabla^i\bar{H}^\dagger_a\tau^A_{aa^\prime}\sigma^j\nabla^i H^\dagger_{a^\prime}H_b\tau_{bb^\prime}^A\sigma^j \bar{H}_{b^\prime}+
{\rm Tr}[\bar{H}^\dagger_a\tau^A_{aa^\prime}\sigma^j H^\dagger_{a^\prime}\nabla^i H_b\tau_{bb^\prime}^A\sigma^j\nabla^i\bar{H}_{b^\prime}]\right\}\nn\\[-2mm]
\label{Lag2}\\[-2mm]
&&-\frac{D_{12}}{8}\left\{{\rm Tr}[(\nabla^i\bar{H}^\dagger_a \tau^A_{aa^\prime}\sigma^i \nabla^j H^\dagger_{a^\prime}+
\nabla^j\bar{H}^\dagger_a \tau^A_{aa^\prime}\sigma^i\nabla^i H^\dagger_{a^\prime}-\frac23\delta^{ij}\nabla^k\bar{H}^\dagger_a\tau^A_{aa^\prime}\sigma^i \nabla^k H^\dagger_{a^\prime})H_b
\tau_{bb^\prime}^A \sigma^j \bar{H}_{b^\prime}]\right. \nn \\
&&+\left.{\rm Tr}[\bar{H}^\dagger_a \tau^A_{aa^\prime} \sigma^i H^\dagger_{a^\prime} (
 \nabla^i H_b \tau_{bb^\prime}^A \sigma^j \nabla^j \bar{H}_{b^\prime} 
+\nabla^j H_b \tau_{bb^\prime}^A \sigma^j \nabla^i \bar{H}_{b^\prime} 
- \frac23\delta^{ij} \nabla^k H_b \tau_{bb^\prime}^A \sigma^j \nabla^k \bar{H}_{b^\prime} ) ] 
\right \}\nn,
\eea
\end{widetext}
where the contact terms proportional to the potentials $D_{10}$ and $D_{11}$ contribute to $S$-wave interactions 
while the term $D_{12}$ gives rise to the $S$-$D$ transitions. 
As explained in Ref.~\cite{Wang:2018jlv}, we are only interested in the $S$-$S$ and $S$-$D$ transitions for the $B^{(*)} \bar{B}^{(*)}$ scattering, so that all terms of the kind $\propto \nabla^i 
H^\dagger \nabla^j H$ 
contributing to $P$ waves are dropped.

Further, we define the combinations
\bea
&\mathcal{C}_d=\ds \frac18 (C_{11}+C_{10}), \quad \mathcal{C}_f= \frac18 (C_{11}-C_{10}),&\nonumber\\
&\mathcal{D}_d=\ds \frac18 (D_{11}+D_{10}), \quad \mathcal{D}_f= \frac18 (D_{11}-D_{10}), &\label{Ds}\\
&\mathcal{D}_{SD}=\ds \frac{2\sqrt{2}}{3}D_{12}, &\nonumber
\eea
where the subindex $d$ ($f$) labels the diagonal (off-diagonal) terms.
These are the parameters used in the main text.

\section{Partial wave projectors}
\label{app:projectors}

A complete set of the relevant projectors $\{P(\alpha,\ven)\}$, 
with $\alpha$ indicating an elastic channel as given in Eq.~(\ref{basisvec}), used to arrive at the partial-wave-projected potentials 
(\ref{VPWA}) reads (for simplicity, the unit vector $\ven$ is omitted in the argument)
\begin{eqnarray}
&&P\Bigl(B\bar{B}({}^1S_0)\Bigr)=1,\label{BB1S0}\\
&&P\Bigl(B^*\bar{B}^*({}^1S_0)\Bigr)=\sqrt{\frac{1}{3}}\varepsilon_{1i}\varepsilon_{2i},\label{BsBs1S0}\\
&&P\Bigl(B^*\bar{B}^*({}^5D_0)\Bigr)=-\sqrt{\frac{3}{8}}S_{ij}v_{ij},\label{BsBs5D0}\\
&&P\Bigl(B\bar{B}({}^1D_2)\Bigr)_{ij}=-\sqrt{\frac{15}{2}}v_{ij},\label{BB1B2}\\
&&P\Bigl(B\bar{B}^*({}^3S_1)\Bigr)_i=\varepsilon_i,\label{BBs3S1}\\
&&P\Bigl(B\bar{B}^*({}^3D_1)\Bigr)_i=-\frac{3}{\sqrt{2}}\varepsilon_j v_{ij},\label{BBs3B1}\\
&&P\Bigl(B\bar{B}^*({}^3D_2)\Bigr)_{ij}=-\frac{\sqrt{5}}{2}\varepsilon_k\Bigl(
i \varepsilon_{ikl}v_{lj}+i \varepsilon_{jkl} v_{li}\Bigr),\label{BBs3B2}
\end{eqnarray}
\begin{eqnarray}
&&P\Bigl(B^*\bar{B}^*({}^3S_1)\Bigr)_i=A_i,\label{BsBs3S1}\\
&&P\Bigl(B^*\bar{B}^*({}^3D_1)\Bigr)_i=-\frac{3}{\sqrt{2}}A_j v_{ij},\label{BsBs3B1}\\
&&P\Bigl(B^*\bar{B}^*({}^5S_2)\Bigr)_{ij}=\frac12S_{ij},\label{BsBs5S2}\\
&&P\Bigl(B^*\bar{B}^*({}^1D_2)\Bigr)_{ij}=-\sqrt{\frac{5}{2}}(\veep_1\cdot\veep_2)v_{ij},\label{BsBs1B2}\\
&&P\Bigl(B^*\bar{B}^*({}^5D_1)\Bigr)_{i}=-\frac{\sqrt{3}}{2} i \varepsilon_{ijk}S_{jm}v_{km},\label{BsBs5B1}\\
&&P\Bigl(B^*\bar{B}^*({}^5D_2)\Bigr)_{ij}=-\sqrt{\frac{45}{56}}\Bigl(
S_{ik}v_{kj}+S_{jk}v_{ki}\nonumber\\
&&\hspace*{45mm}-\frac23\delta_{ij}S_{kl}v_{kl}\Bigr),\label{BsBs5B2}\\
&&P\Bigl(B^*\bar{B}^*({}^5G_2)\Bigr)_{ij}=\sqrt{\frac{175}{32}}\, S_{kl}\, v_{ijkl},
\end{eqnarray}
where
\begin{eqnarray}
v_{ij}&=&n_i n_j-\frac13\delta_{ij},\label{vijdef}
\eea
\bea
v_{ijkl}&=&n_i n_j n_k n_l-\frac17(n_i n_j \delta_{kl} 
+ n_i n_k \delta_{jl} +n_i n_l \delta_{jk}
\nonumber\\ \nonumber
&+& n_j n_k \delta_{il} + n_j n_l \delta_{ik} 
+ n_k n_l \delta_{ij} )\label{vijkl}\\
&+&\frac1{35}\left (\delta_{ij}\delta_{kl} + \delta_{ik}\delta_{jl} +\delta_{il}\delta_{jk} \right),
\eea
\bea
A_i&=&\frac{i}{\sqrt{2}}\varepsilon_{ijk}\varepsilon_{1j}\varepsilon_{2k},\label{Aidef}\\
S_{ij}&=&\varepsilon_{1i}\varepsilon_{2j}+\varepsilon_{1j}\varepsilon_{2i}-\frac23\delta_{ij}(\veep_1\cdot\veep_2).
\label{Sijdef}
\end{eqnarray}

All projectors above are normalised as
\begin{equation}
\frac1{2J+1}\int\frac{d\Omega_n}{4\pi}P(\alpha,\ven)P^{\dagger}(\alpha,\ven)=1.
\end{equation}

\end{document}